\documentclass[aps,prc,reprint,superscriptaddress]{revtex4-1}

\usepackage{graphicx}

\usepackage{hyperref}

\usepackage{multirow}
\usepackage{lineno}
\usepackage{comment}
\usepackage{amsmath}
\usepackage{enumitem} 
\modulolinenumbers[5]
\hypersetup{ 
 colorlinks=true, 
 citecolor=blue, 
 filecolor=blue, 
 linkcolor=blue, 
 urlcolor=blue 
}

\begin{document}


\title{Measurement and analysis of the $^{246}$Cm and $^{248}$Cm neutron capture cross-sections at the EAR2 of the n\_TOF facility at CERN}

\date{\today}
\author{V.~Alcayne} \affiliation{Centro de Investigaciones Energ\'{e}ticas Medioambientales y Tecnol\'{o}gicas (CIEMAT), Spain} %
\author{A.~Kimura} \affiliation{Japan Atomic Energy Agency (JAEA), Tokai-Mura, Japan} %
\author{E.~Mendoza} \affiliation{Centro de Investigaciones Energ\'{e}ticas Medioambientales y Tecnol\'{o}gicas (CIEMAT), Spain} %
\author{D.~Cano-Ott} \affiliation{Centro de Investigaciones Energ\'{e}ticas Medioambientales y Tecnol\'{o}gicas (CIEMAT), Spain} %
\author{O.~Aberle} \affiliation{European Organization for Nuclear Research (CERN), Switzerland} %
\author{F.~Álvarez-Velarde} \affiliation{Centro de Investigaciones Energ\'{e}ticas Medioambientales y Tecnol\'{o}gicas (CIEMAT), Spain} %

\author{S.~Amaducci} \affiliation{INFN Laboratori Nazionali del Sud, Catania, Italy} \affiliation{Dipartimento di Fisica e Astronomia, Universit\`{a} di Catania, Italy} %
\author{J.~Andrzejewski} \affiliation{University of Lodz, Poland} %
\author{L.~Audouin} \affiliation{Institut de Physique Nucl\'{e}aire, CNRS-IN2P3, Univ. Paris-Sud, Universit\'{e} Paris-Saclay, F-91406 Orsay Cedex, France} %
\author{V.~B\'{e}cares} \affiliation{Centro de Investigaciones Energ\'{e}ticas Medioambientales y Tecnol\'{o}gicas (CIEMAT), Spain} %
% ...
%


\collaboration{The n\_TOF Collaboration (www.cern.ch/ntof)} \noaffiliation


\begin{abstract}
The $^{246}$Cm(n,$\gamma$) and $^{248}$Cm(n,$\gamma$) cross-sections have been measured at the Experimental Area 2 (EAR2) of the n\_TOF facility at CERN with three C$_6$D$_6$ detectors. This measurement is part of a collective effort to improve the capture cross-section data for Minor Actinides (MAs), which are required to estimate the production and transmutation rates of these isotopes in light water reactors and innovative reactor systems. In particular, the neutron capture in $^{246}$Cm and $^{248}$Cm open the path for the formation of other Cm isotopes and heavier elements such as Bk and Cf and the knowledge of (n,$\gamma$) cross-sections of these Cm isotopes plays an important role in the transport, transmutation and storage of the spent nuclear fuel. The reactions $^{246}$Cm(n,$\gamma$) and $^{248}$Cm(n,$\gamma$) have been the two first capture measurements analyzed at n\_TOF EAR2. 
Until this experiment and two recent measurements performed at J-PARC, there was only one set of data of the capture cross-sections of $^{246}$Cm and $^{248}$Cm, that was obtained in 1969 in an underground nuclear explosion experiment.
In the measurement at n\_TOF a total of 13 resonances of $^{246}$Cm between 4 and 400 eV and 5 of $^{248}$Cm between 7 and 100 eV have been identified and fitted. The radiative kernels obtained for $^{246}$Cm are compatible with JENDL-5, but some of them are not with JENDL-4, which has been adopted by JEFF-3.3 and ENDF/B-VIII.0. The radiative kernels obtained for the first three $^{248}$Cm resonances are compatible with JENDL-5, however, the other two are not compatible with any other evaluation and are 20\% and 60\% larger than JENDL-5.
\end{abstract}

\maketitle 
\section{Introduction}\label{sec:Introduction}

Accurate neutron capture cross-section data for Minor Actinides (MAs) are required for accurate neutronic calculations of nuclear systems (light water reactors, fast reactors, and accelerator-driven systems) and for determining quantities relevant to the transport, transmutation, and storage of the nuclear fuel, such as the radiotoxicity, the decay heat, or the neutron emission rates. In particular, the capture cross-sections of $^{246}$Cm and $^{248}$Cm are important for determining the content of these isotopes in the spent nuclear fuel and because they open the path for the formation of heavier Cm isotopes and elements such as Bk and Cf. In addition, $^{246}$Cm is important since it is one of the main neutron emitters after 300 years of cooling down \cite{Aliberti_TargetAcurracy_2006}.

The neutron capture cross-sections of $^{246}$Cm and $^{248}$Cm ($^{246, 248}$Cm) have been measured at n\_TOF Experimental Area 2 (EAR2) \cite{Weis_EAR2_2015,Sabate_FluxEar2_2017} with three C$_6$D$_6$ detectors \cite{Plag_C6D6Detectors_2003} and a dedicated sample. In addition to the $^{246, 248}$Cm sample, another one was measured in the same experimental campaign at n\_TOF to obtain the capture cross-section of $^{244}$Cm \cite{Alcayne_Cm244_246WONDER_2019,Alcayne_NDCm244_2019,Alcayne_CmND_23}. In this case, the measurement was performed in two areas of the n\_TOF facility: Experimental Area 1 (EAR1) \cite{Guerrero_EAR1_2013} and EAR2. The final results of the $^{244}$Cm measurements will be published in a future article. The samples used in the experiments were provided by Japan Atomic Energy Agency (JAEA), which used material from the same batch to perform another two capture measurements with the ANNRI setup at J-PARC \cite{Kimura_Cm244_2012,Kawase_Cm244_2021}. The only other previous capture measurements of these isotopes were performed in 1969 with the neutrons produced by an underground nuclear explosion \cite{Moore_Cm244_246_1971}.

In this paper, the experimental setup is described in Sec. \ref{sec:exp_setup}, the procedure for obtaining the experimental capture yields is presented in Sec. \ref{sec:data_analysis}, the analysis of the yields for retrieving the capture cross-sections is detailed in Sec. \ref{sec:ResAnalysis} and the conclusions are presented in Sec. \ref{sec:Conclusions}.

\section{Experimental setup}\label{sec:exp_setup}

\subsection{n\_TOF EAR2}
Neutrons at n\_TOF are produced by spallation reactions with a 20 GeV/c pulsed proton beam that impinges on a lead block. The pulses have a nominal intensity of $7\times10^{12}$ protons and a time spread of 7 ns Root Mean Square (RMS) \cite{Guerrero_EAR1_2013}. There are two experimental areas at n\_TOF: the EAR1 located at $\sim$185 m \cite{Guerrero_EAR1_2013} in the horizontal direction, and the EAR2 \cite{Weis_EAR2_2015}, located at $\sim$20 m in the vertical direction. 

The neutron fluence in EAR2 is $\sim$40 times larger than in EAR1 \cite{Sabate_FluxEar2_2017}. In addition, EAR2 is $\sim$10 times closer to the target than EAR1, so the signal-to-background ratio produced by the natural radioactivity of the samples is improved by more than two orders of magnitude compared to EAR1. These features have made possible new measurements with shorter half-lives isotopes (i.e. highly radioactive), smaller capture cross-sections and/or smaller samples \cite{Barbagallo_Be7_2016,Sabate_33Sn_2017,Damone_Be7_2018,Stamatopoulos_Pu240_2020}.

\subsection{Detection setup}
\begin{figure}[htb]
\begin{center}
\includegraphics[width=1.0\linewidth]{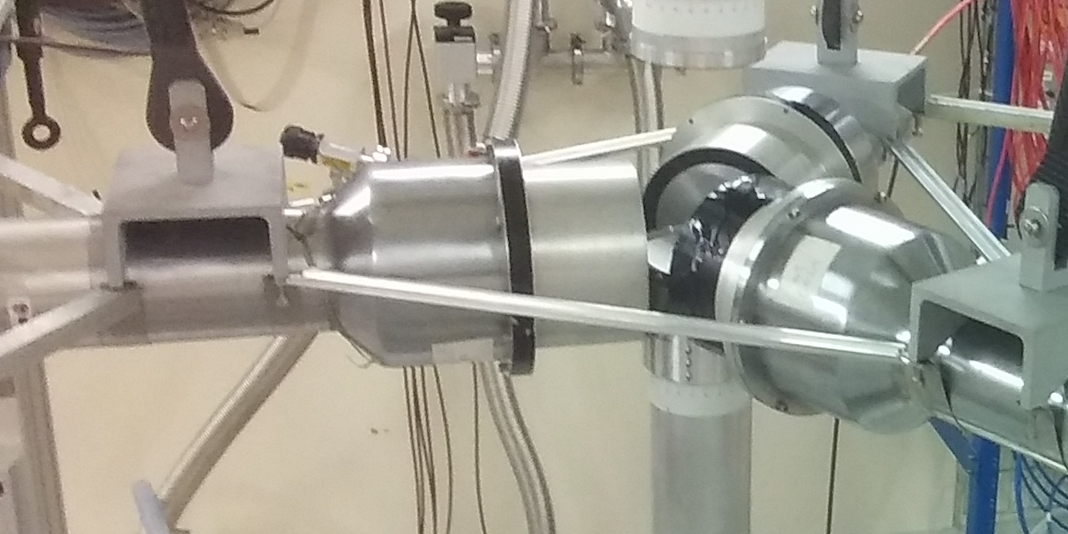} 
\includegraphics[width=1.0\linewidth]{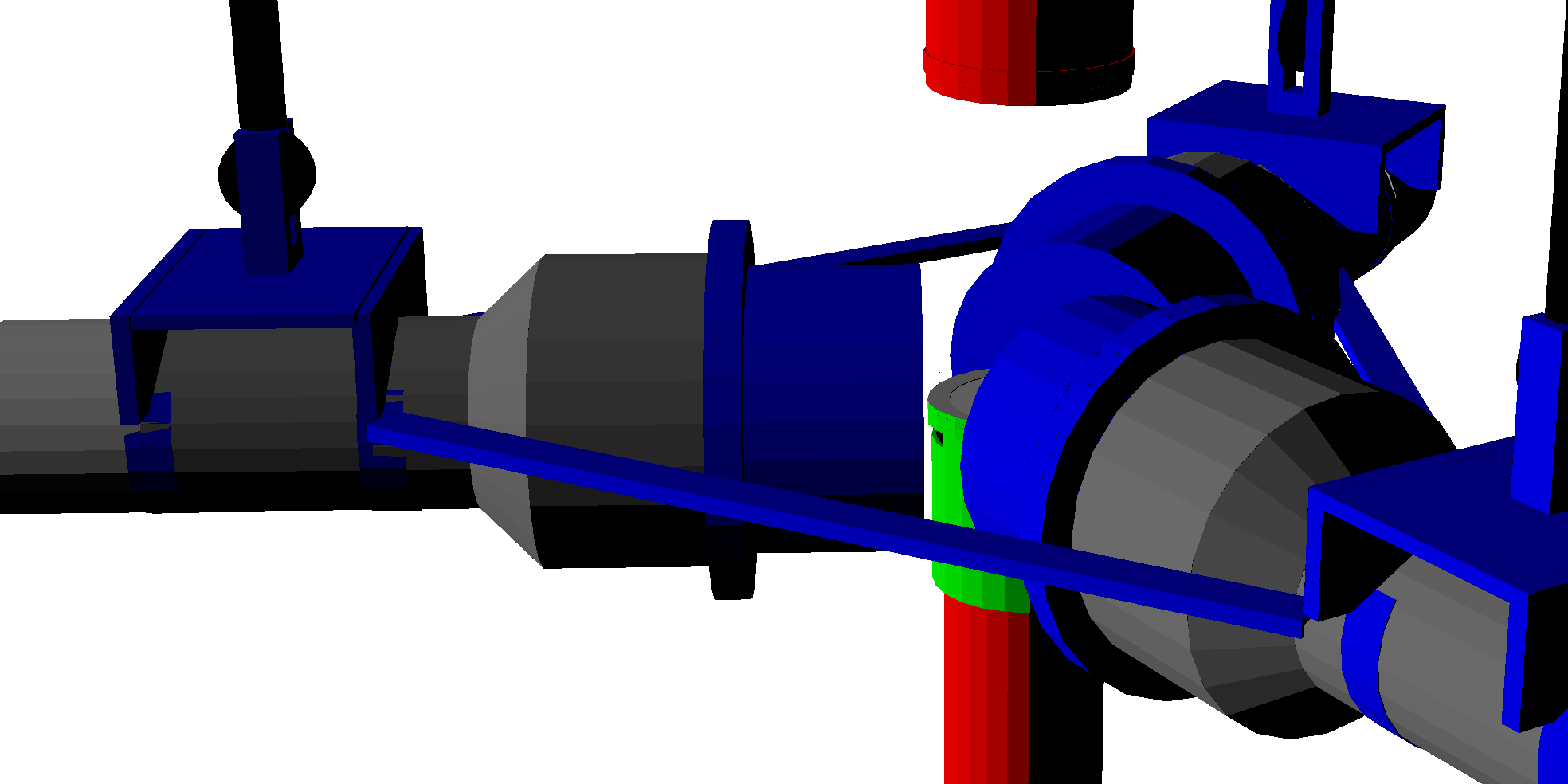} 
 \caption{At the top panel, a picture of the experimental setup used in the measurement consisting of three C$_6$D$_6$ detectors. In the bottom panel, the geometry implemented in Geant4 \cite{Agostinelli_GEANT4_2003} to simulate the detection system response.}
\label{fig:Experimental_setup}
\end{center}
\end{figure}

The prompt $\gamma$-rays emitted in the (n,$\gamma$) reactions were detected with three BICRON detectors filled with C$_6$D$_6$ liquid \cite{Plag_C6D6Detectors_2003}, whose front face is located at 5 cm from the center of the sample. A general view of the experimental setup is shown in the top panel of Fig. \ref{fig:Experimental_setup}. C$_6$D$_6$ detectors have been used since 2006 for many capture measurements at the n\_TOF facility \cite{Marrone_Sm151_2006,Domingo_Bi209_2006,Casanovas_204Tl_2020}, including also measurements of actinides 
\cite{Gunsing_Th232_2012,Fraval_Am241_2014,Mingrone_U238_2017,Mastromarco_U236_2017,Lerendegui_Pu242_2018}. The main advantages of these detectors are the low neutron sensitivity and good time resolution, with pulses of $\sim$10 ns Full Width Half Maximum. 

Two additional detectors have been used to monitor the neutron beam. A wall current monitor detector measures the proton current in each pulse before impinging on the lead block. The other detector is a silicon monitor (SiMon2) \cite{Consentino_Simon_2015} used to measure the neutron fluence at EAR2. It consists of four silicon pad detectors with a surface of $3\times3$ cm$^2$ and a thickness of 300 $\mu$m. The four detectors are placed outside the beam looking to an in-beam $^6$LiF foil for detecting the products of the $^6$Li(n,t) reaction. The cross-section for this reaction is standard from thermal to 1 MeV \cite{iaea_StandartLibraires_2007}.

The signals from all the individual detectors have been recorded by the n\_TOF Digital Acquisition System (DAQ) \cite{Abbondano_DAQ_2005}, based on SPDevices ADQ412DC-3G cards with 1 GHz sampling rate and 14-bit resolution. The data from the silicon monitor and the C$_6$D$_6$ detectors are analyzed using a pulse shape routine to extract the amplitude and time of the signals \cite{Zugec_DataProcessing_2016,Guerrero_SignalBC501A_2008}. This information and the detector numbers have been stored in ROOT files \cite{Brun_ROOT_1997} for their posterior analysis.

\subsection{Cm sample}\label{sec:TheSample}
The Cm sample used in the experiment was acquired by the JAEA from the Russia Research Institute of Atomic Research in 2007. The sample consists of a Cm oxide pellet of dimensions 2.5 mm (radius) x 0.5 mm (height) enclosed in an aluminum casing of dimensions 4.5 mm (radius) x 1.2 mm (height). Pellets created with the same batch material were used in the Kimura \textit{et al.} \cite{Kimura_Cm244_2012} and Kawase \textit{et al.} \cite{Kawase_Cm244_2021} experiments. 

The pellet contains approximately 1.1 mg of $^{246}$Cm and 0.2 mg of $^{248}$Cm. The absolute masses of the different isotopes are not accurately known. However, their relative fractions are precisely determined since molar isotopic enrichment was well-established in 2010 by thermal ionization mass spectrometry and $\alpha$-particle spectrometry using a small amount of the same batch used to prepare the samples \cite{Kimura_Cm244_2012}. The mole fractions of the isotopes at the time when the n\_TOF measurement was performed (summer 2017) are presented in Table \ref{tab:IsotopicAbundace}. These values have been obtained from the isotopic evolution of the values measured in 2010, taking into account the radioactive decays.
\begin{table}[htb]
\begin{center}
 \begin{tabular}{c c }\hline \hline 
 Isotope& Mole fraction (\%)\\ \hline
 $^{240}$Pu&\hspace{0.075cm}9.2 \hspace{0.075cm}$\pm$ 0.2 \\
 $^{243}$Am&\hspace{0.075cm}1.2 \hspace{0.075cm}$\pm$ 0.2 \\
 $^{244}$Cm&20.1 $\pm$ 0.4 \\
 $^{245}$Cm&\hspace{0.075cm}1.0 \hspace{0.075cm}$\pm$ 0.3 \\
 $^{246}$Cm&57.0 $\pm$ 1.3 \\
 $^{247}$Cm&\hspace{0.075cm}2.8 \hspace{0.075cm}$\pm$ 0.4 \\
 $^{248}$Cm&\hspace{0.075cm}8.7 \hspace{0.075cm}$\pm$ 0.2 \\
 \hline \hline
 \end{tabular}
\end{center}
\caption{Isotopic abundances in the sample, at the time of the measurement at n\_TOF (summer 2017).}
\label{tab:IsotopicAbundace}
\end{table}
 
As can be seen in Table \ref{tab:IsotopicAbundace}, there is also a significant amount of $^{244}$Cm in the sample. As mentioned above, the results for this isotope will be presented in a future work, which would also include measurements performed with another sample performed at EAR1 and EAR2 \cite{Alcayne_Cm244_246WONDER_2019,Alcayne_NDCm244_2019,Alcayne_CmND_23}. 

 Together with Cm isotopes, the sample contains also $^{240}$Pu, which is the daughter of $^{244}$Cm (T$\mathrm{_{1/2}}$ ($^{244}$Cm): 18.11 years \cite{Chechev_Cm244HalfLife_2006}). As explained in detail in Section \ref{sec:NorPu240} $^{240}$Pu is used for the normalization of the capture cross-sections of $^{246}$Cm and $^{248}$Cm.

\begin{figure}[htb]
\begin{center}
 \includegraphics[width=0.9\linewidth]{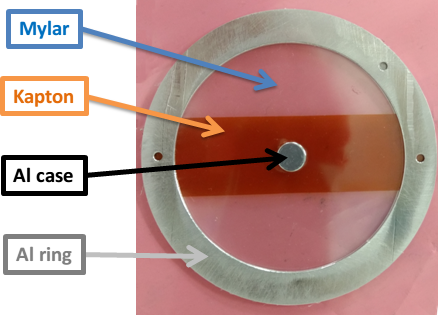} 
 \caption{Picture including the Cm sample (Al case), the Kapton strip (Kapton), the Mylar foil (Mylar) and the outside aluminum ring (Al ring). The Al ring inner diameter of 7 cm is considerably larger than the beam diameter, which is less than 6 cm.} 
\label{fig:Target}
\end{center}
\end{figure}

In order to reduce the in-beam material the Cm sample was placed in the center of the beam using a ring made with aluminum, Mylar and Kapton (Fig. \ref{fig:Target}). 

\section{Data analysis}\label{sec:data_analysis}

\subsection{Data Reduction} \label{sec:DataReduction}
\begin{figure}[htb] 
\begin{center}
\includegraphics[width=\columnwidth]{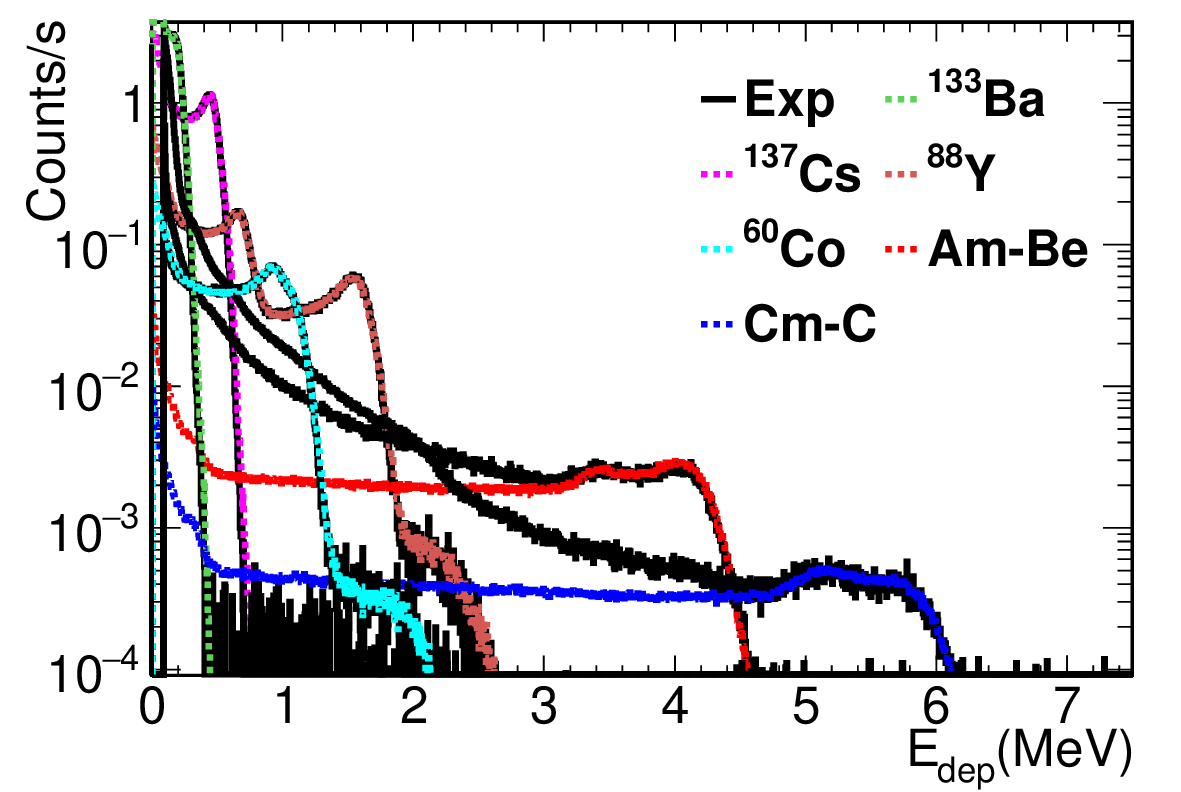} 
\caption{Simulated detector responses (dashed lines with different colors) compared to measured spectra from calibration sources (in black).}
 \label{fig:calibration}
\end{center}
\end{figure}

The energy calibration and energy resolution of each C$_6$D$_6$ detector were determined by comparing the detector response to several $\gamma$-ray sources obtained experimentally and via Monte Carlo simulations. The detailed geometry implemented in Geant4 \cite{Agostinelli_GEANT4_2003} for the simulations is shown in Fig. \ref{fig:Experimental_setup}. After the calibration and energy resolution determination, the detector response to the six $\gamma$-ray sources is accurately reproduced with the simulations as shown in Fig. \ref{fig:calibration}. The calibration procedure was repeated every week during the 3 months of the experiment to monitor and correct the small gain shifts (less than 7\% along the entire experiment).

After a careful analysis, we noticed that the detectors exhibit changes in gain as a function of the time-of-flight (i.e. neutron energy ($E_n$)), requiring a correction of the energy calibration. More details concerning this effect are provided in \cite{Alcayne_sted_2024}. We attribute the origin of the gain shifts to the effect of the EAR2 particle flash (consisting of relativistic charged particles, high-energy neutrons, and prompt $\gamma$-rays with very short time-of-flight, less than 1 $\mu s$) on the C$_6$D$_6$ photomultipliers. The gain of the detectors suddenly changes just after the particle flash and it takes about 10 $\mu s$ to recover the values obtained with the calibration sources. These 10 $\mu s$ corresponds, approximately, to time-of-flight neutron energies of $\sim$0.02 eV. 

\begin{figure}[htb] 
\begin{center}
\includegraphics[width=\columnwidth]{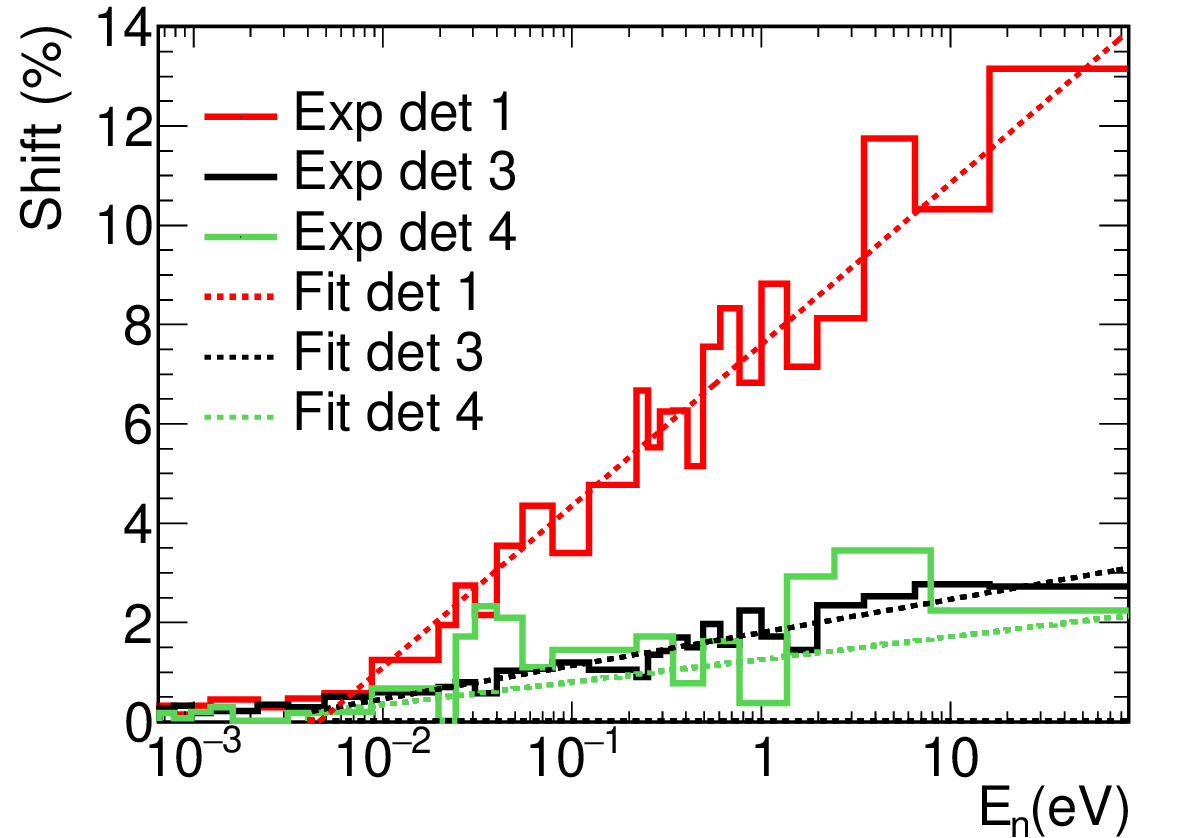} 
\caption{Experimental gain shift for each detector (Exp.) obtained with the $^{88}$Y measurement as a function of the neutron energy, i.e. time-of-flight. The dashed lines are the fits of each detector response with a logarithmic function.}
 \label{fig:GainShiftY88}
\end{center}
\end{figure}

The gain shift effect has been characterized for each detector as a function of the neutron energy using an $^{88}$Y $\gamma$-ray source placed close to the detectors in measurements with the neutron beam, as shown in Fig. \ref{fig:GainShiftY88}. The obtained shift values have been then fitted with a logarithmic function, shown in Fig. \ref{fig:GainShiftY88}, which was used to correct the energy calibration of each detector as a function of the neutron energy. As described in detail in references \cite{Alcayne_sted_2024,Alcayne_Thesis_2022} these gain shifts can not be attributed to pile-up effects.

The DAQ system used at n\_TOF records all the signals. However, the analysis routine may fail when fitting two consecutive pulses located very close one to another. To correct for these pile-up effects, the paralyzable method described in \cite{Knoll_2000} has been used, with a constant dead-time ($\tau$) of 20 ns. The maximum corrections applied to the yields due to the pile-up effects are smaller than 2\%.

\subsection{Background} \label{sec:Background}

The background is defined as the events detected in the C$_6$D$_6$ detectors originated from reactions other than (n,$\gamma$) in $^{246}$Cm or $^{248}$Cm. The background can be divided into various components:

\begin{enumerate}[label=(\roman*)]
 \item Beam-off background produced by the activity of the sample. As presented in Fig \ref{fig:CountinRateBKG}, the beam-off background is considerably low compared to other background sources, due to the high instantaneous fluence of the EAR2.
 \item Beam-on background determined with a Dummy sample identical to the one of Cm but without any actinide (Fig. \ref{fig:CountinRateBKG}). The characterization of the complete background not related to reactions on the actinides is obtained with this measurement. The subtraction of this background is one of the main sources of uncertainty in this experiment. To address this, we meticulously examined regions between resonances, where actinide influence on the time-of-flight spectrum is minimal. We ensured compatibility of counts between the dummy and sample after subtracting the beam-off background in these regions. The resulting average scaling factor of 1.0035(40) was utilized to scale the background appropriately.

\begin{figure}[htb] 
\begin{center}
 \includegraphics[width=\columnwidth]{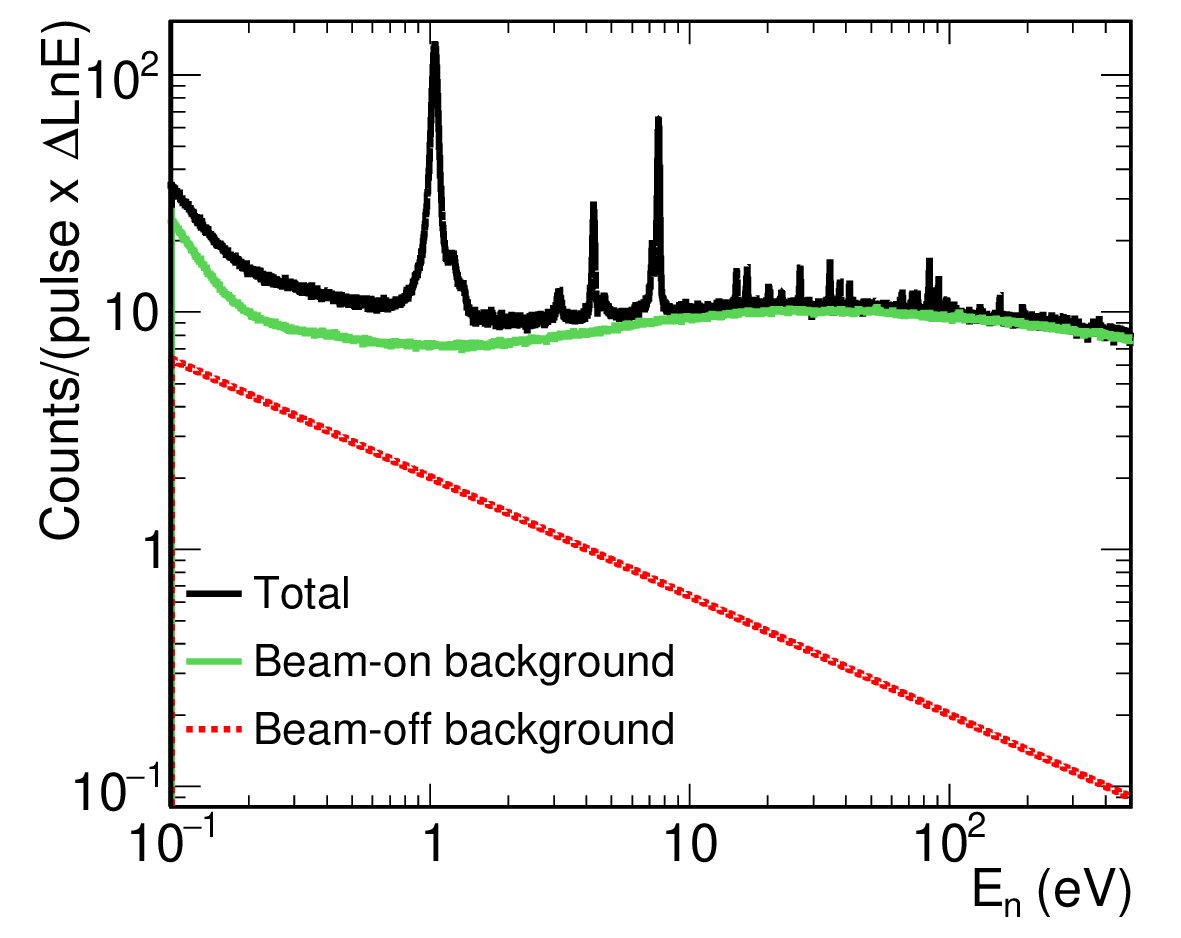} 
 \includegraphics[width=\columnwidth]{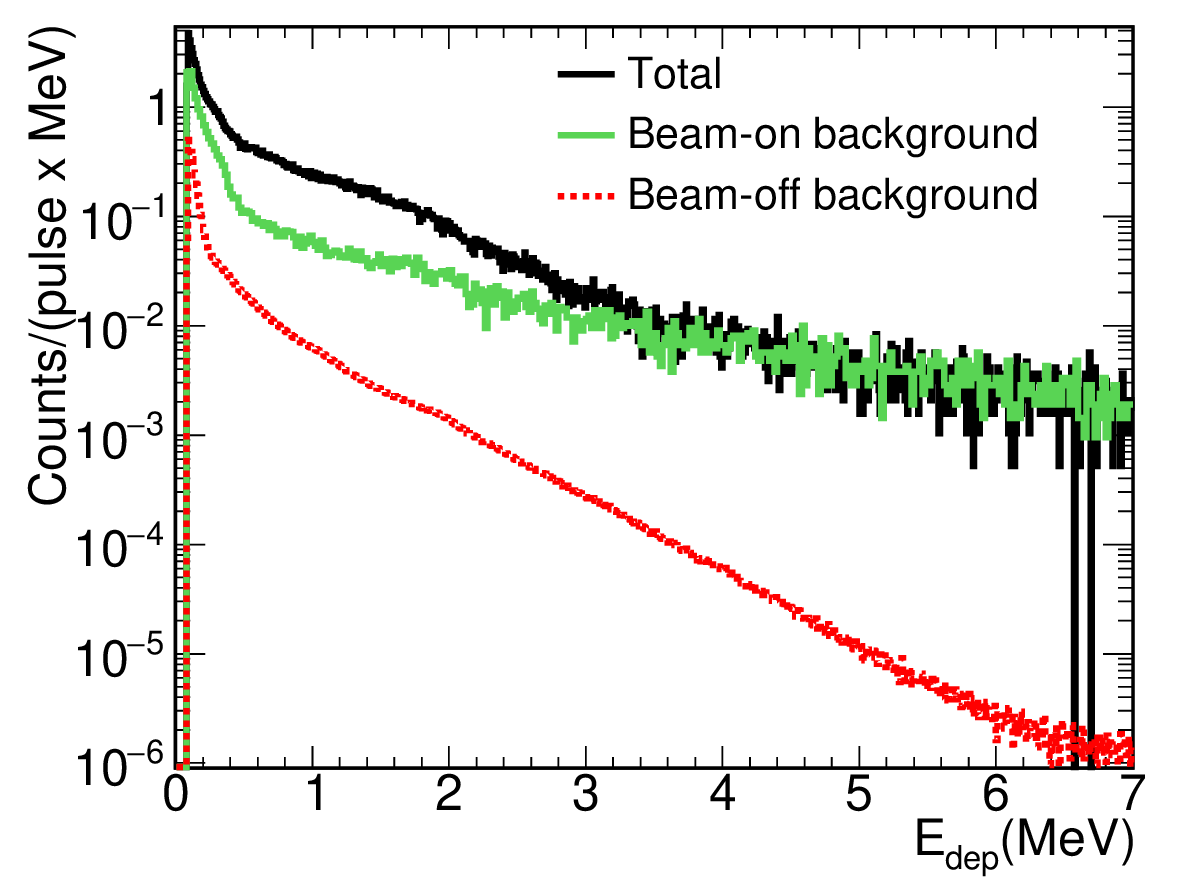} 
 \caption{Top panel: Measured counting rate (0.12 $<$ E$\mathrm{_{dep}}$ $<$ 6.0 MeV) as a function of the neutron energy for the Cm sample (Total), together with the beam-on and beam-off backgrounds (see the text for details). The units are counts per pulse and per unit lethargy ($\Delta$LnE, which means that the bin contents have been divided by the natural logarithm of the ratio between the upper and lower bin limits). Bottom panel: measured deposited energy spectra for the first resonance of $^{246}$Cm, between 4.2 and 4.35 eV, together with the beam-on and beam-off backgrounds.}
 \label{fig:CountinRateBKG}
\end{center}
\end{figure}
 \item Regarding the background due to the interaction of the neutron beam with the actinides in the sample the different reactions have been separated:
 \begin{enumerate}[]

 \item Capture: in most cases, the resonances of $^{240}$Pu, $^{243}$Am, $^{244}$Cm, $^{245}$Cm, and $^{247}$Cm present in the sample (see Tab. \ref{tab:IsotopicAbundace}), are sufficiently separated from those of $^{246}$Cm and $^{248}$Cm, so they do not have any significant impact to the analysis of the Resonance Parameters (RP) in the Resolved Resonance Region (RRR). In any case, uncertainties in the RP parameters due to the presence of these resonances are negligible.
 \item Elastic scattering: this component is negligible in the analyzed region, due to the low neutron sensitivity of the detectors \cite{Plag_C6D6Detectors_2003} and the capture to elastic scattering cross-section ratio of the actinides in the sample.
 \item Fission: the background due to fission reactions has been subtracted from the capture yield in the resonance analysis, as described in Sec. \ref{sec:ResAnalysis}. This background component has been determined from the fission detection efficiency ($\varepsilon_{fis}$) and the fission cross-section present in the JENDL-4.0 evaluation. The JEFF-3.3 \cite{JEFF3.3_2017}, JENDL-5 \cite{JENDL5} and ENDF/B-VIII.0 \cite{Brown_ENDF8_2018} libraries have adopted the JENDL-4.0 for the fission of these isotopes. The value of $\varepsilon_{fis}$ has been determined by comparing the experimental yield for the first (2.94 eV) and second (3.17 eV) resonances of $^{247}$Cm, which has a significant fission reaction channel, and the evaluated cross-section from JENDL-4.0. It has been then assumed that the $\varepsilon_{fis}$ values are the same for all the Cm isotopes. The fit to the two $^{247}$Cm resonances to obtain $\varepsilon_{fis}$ is presented in Fig. \ref{fig:FissionEff}. The derived value of $\varepsilon_{fis}$ is 0.085(22), which is about three times larger than the capture detection efficiency of the actinides of the measurement (Table \ref{tab:Ef_noWF}). 
\begin{figure}[htb] 
\begin{center}
 \includegraphics[width=\columnwidth]{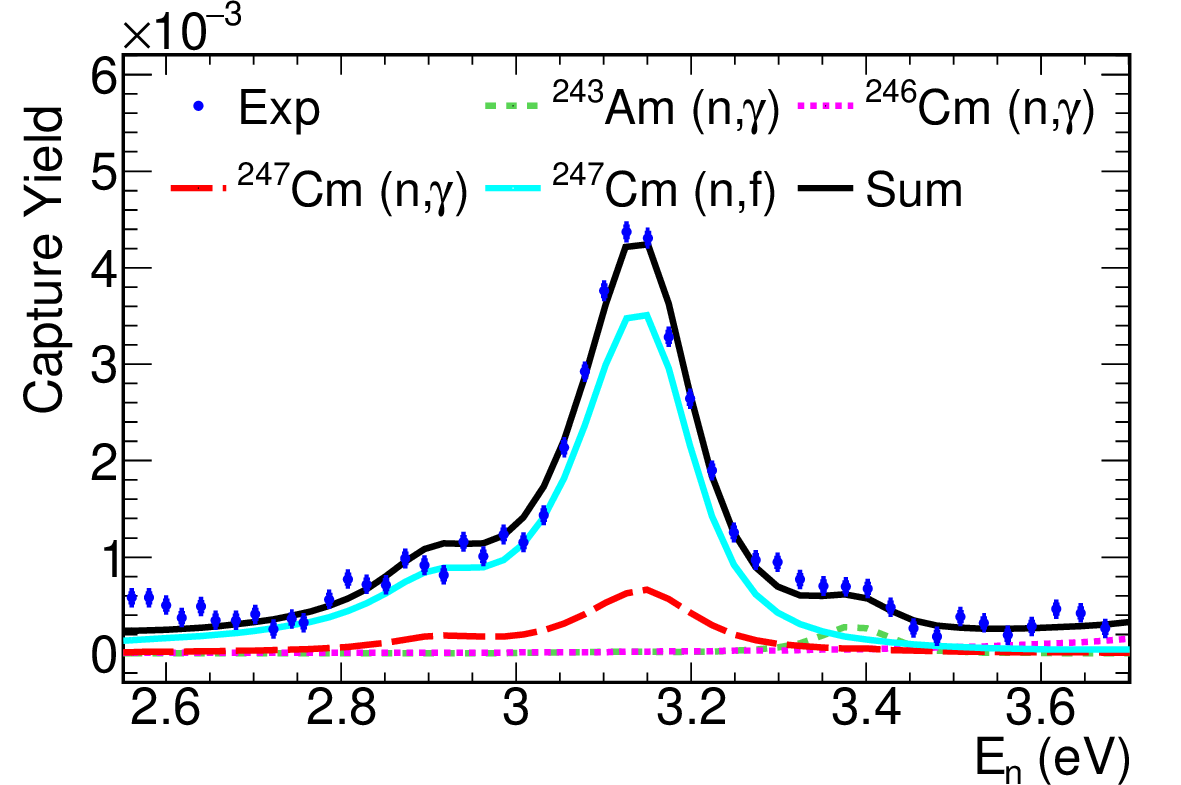} 
\caption{Experimental yield (Exp) of the Cm sample together with the partial contribution of the different isotopes ($^{243}$Am, $^{246}$Cm and $^{247}$Cm) calculated with JENDL-4.0. The contribution of the $^{247}$Cm fission yield has been re-scaled to fit the experimental data, thus obtaining $\varepsilon_{fis}$. The black line (Sum) is the sum of the contributions of the different yields.}
 \label{fig:FissionEff}
\end{center}
\end{figure}

The uncertainty in the determination of the fission detection efficiency (25\%) has been estimated by considering 20 \% uncertainty due to the $^{247}$Cm fission cross-section in JENDL-4.0, a 15\% uncertainty due to the uncertainty in the abundance of $^{247}$Cm in the sample (Table \ref{tab:IsotopicAbundace}), and a 5\% uncertainty due to counting statistics.

 \end{enumerate}

\end{enumerate}

\subsection{Pulse Height Weighting Technique (PHWT)}\label{sec:Data_PWHT}
The efficiency to detect a capture cascade ($\varepsilon_c$) depends in principle on the $\gamma$-cascade deexcitation pattern that may also depend on the neutron energy. To avoid this dependency, the spectra from C$_6$D$_6$ detectors have been traditionally analyzed with the Pulse Height Weighting Technique (PHWT) \cite{Mackin_PHWT_1967,Tain_PHWT_2004,Mendoza_PHWT_23}. That is mainly based on the fulfillment of these two conditions:
\begin{enumerate}[label=(\roman*)]
 \item The individual detector $\gamma$-ray efficiency ($\varepsilon_\gamma$) 
 has to be much smaller than one ($\varepsilon_\gamma$ $\ll$ 1), so that at most one of the $\gamma$-rays of the cascade is detected in each detector.
 \item $\varepsilon_\gamma$ has to be proportional to the energy of the $\gamma$-ray ($E_\gamma$), $\varepsilon_\gamma$= $k\cdot E_\gamma$.
\end{enumerate}
Assuming these conditions, $\varepsilon_c$ becomes almost proportional to the sum energy of the $\gamma$-rays of the cascade ($E_c$), so it is independent of the deexcitation pattern:
\begin{equation}\label{Eq:PHWT}
\varepsilon_c=1-\prod_{j}( 1-\varepsilon _{\gamma j}) \approx \sum_{j} \varepsilon_{\gamma j}\approx k \cdot E_c
\end{equation}
In this expression the $j$ index loops through all the $\gamma$-rays of the cascade. For the actinides measured in this experiment, the neutron separation energies ($S_n$) are of the order of 6 MeV, which is more than four orders of magnitude larger than the maximum incident neutron analyzed energy of $\sim$400 eV. Therefore, the sum energy of the resulting capture cascade can be approximated as the neutron separation energy of the isotope ($E_c= S_n+E_n\cong S_n$).

The detector itself does not satisfy condition (ii) of the PHWT. Therefore, a mathematical manipulation of the detector response is performed to establish a proportional relationship between the detection efficiency and the energy of the $\gamma$-rays. The counts registered at each deposited energy are weighted by a factor depending on its energy (pulse height), given by the so-called Weighting Function (WF). 

Following the same procedure as in \cite{Tain_PHWT_2004}, the WF for the detection setup was calculated by simulating the detector response to 30 monoenergetic $\gamma$-rays from 0.1 to 10 MeV using Geant4. For practical purposes, the WF was assumed to have the polynomial dependence (of 5th degree) on the deposited energy and its final version was obtained from fitting individual points with such a polynomial. Applying the obtained WF to the detector response, the proportionality condition between the weighted counts and E$_{\gamma}$ is satisfied with an RMS deviation better than 0.5\% for $\gamma$-rays with $E_\gamma$ between 0.1 and 10 MeV.

\subsection{Corrections to the PHWT}\label{sec:CorrectionsMC}
In practice, there are different experimental effects, which may disturb the straightforward application of the PHWT technique. In our case, they are:

\begin{enumerate}[label=(\roman*)]
 \item 
 The 0.12 MeV threshold set in each detector, which makes condition (ii) of the PHWT technique unfulfilled for low energy $\gamma$-rays.
 \item
Effect of $\gamma$-ray summing, produced when two or more $\gamma$-rays are detected in one detector simultaneously.

\item
The internal electron conversion, leading to the emission of non-detectable electrons instead of $\gamma$-rays.

\end{enumerate}
 
To account for the deviations produced by these effects, correction factors (F$\mathrm{_{PHWT}}$) are calculated for each isotope by performing Monte Carlo simulations of the detection of (n,$\gamma$) cascades.
 The (n,$\gamma$) cascades of $^{240}$Pu, $^{246}$Cm and $^{248}$Cm necessary to determine the F$\mathrm{_{PHWT}}$ have been obtained with the NuDEX code \cite{Mendoza_PHWT_23}. This code generates the full level scheme of the nucleus, together with the branching ratios of each level, to compute the (n,$\gamma$) cascades using models of photon strength functions (PSFs) and level density (LD). The data are taken from the RIPL-3 \cite{Capote_RIPL_2007} and ENSDF \cite{ENSDF} databases, among others.

\begin{figure}[htb]
\begin{center}
 \includegraphics[width=\columnwidth]{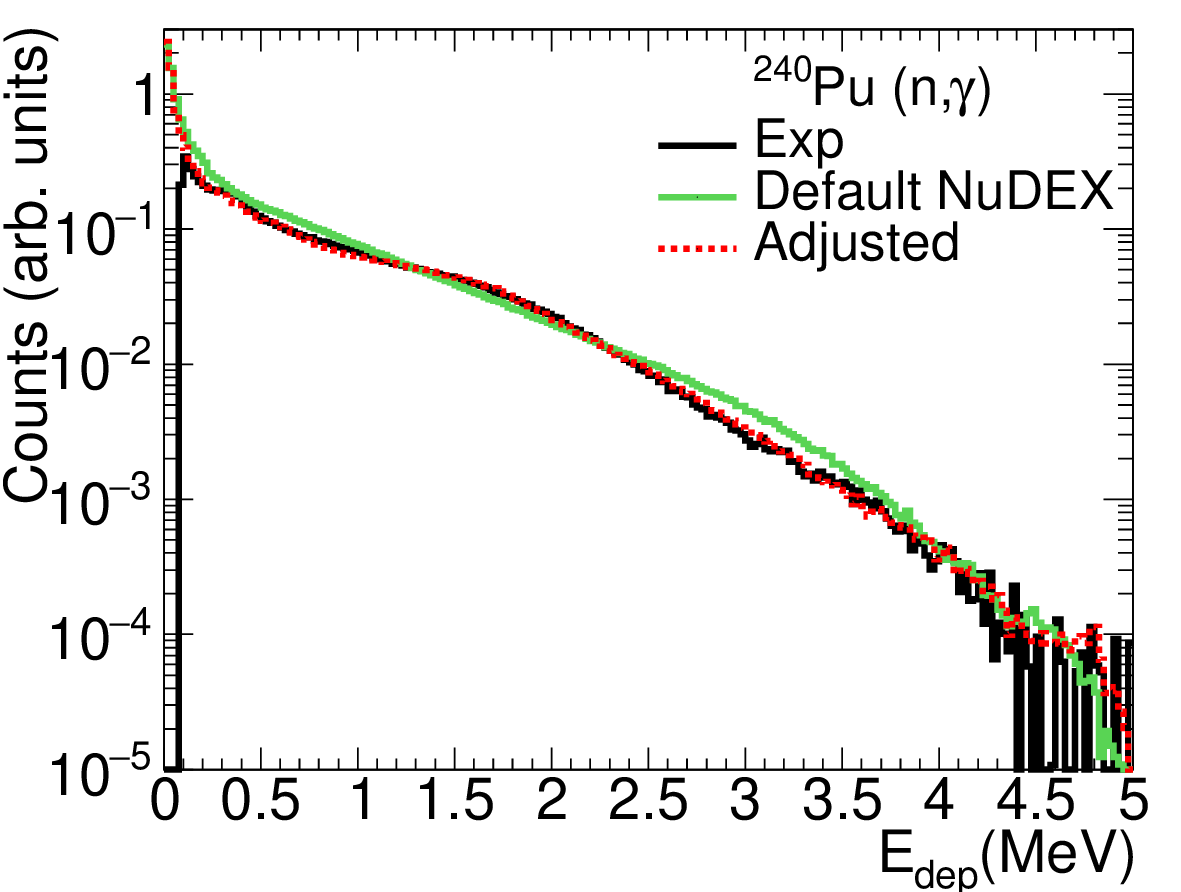} 
 \includegraphics[width=\columnwidth]{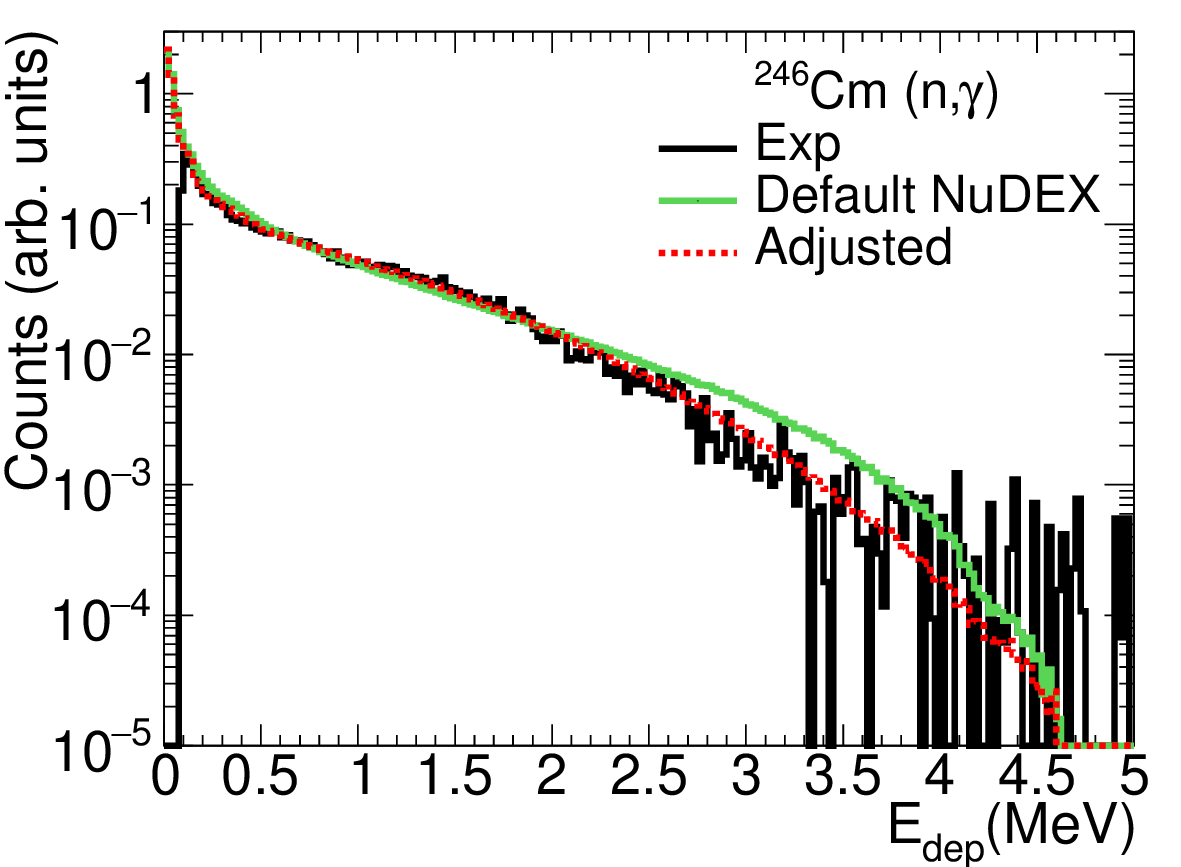} 
\caption{Comparison of experimental and simulated deposited energy spectra of $^{240}$Pu(n,$\gamma$) (0.9 $<E_n<$ 1.1 eV)(top) and $^{246}$Cm(n,$\gamma$) (4.2 $<E_n<$ 4.35 eV) (bottom). The experimental spectra (Exp) are compared with the simulated spectra obtained with the default NuDEX cascades (Default NuDEX) and with the cascades obtained after adjusting the PSF parameters to reproduce the experimental spectra (Adjusted). This plot shows data only for one of the detectors, hereafter the same procedure would be followed for simplicity. }
 \label{fig:Cascades}
\end{center}
\end{figure}

 The (n,$\gamma$) cascades for $^{240}$Pu and $^{246}$Cm have been firstly obtained with the default PSF and LD values given by NuDEX. As can be seen in Fig. \ref{fig:Cascades} the simulation of the detector response with the default NuDEX cascades does not exactly reproduce the experimental spectra obtained for this reaction. Therefore adjusted PSF and LD have been obtained and used for the cascades of $^{240}$Pu and $^{246}$Cm. The adjusted cascades of $^{240}$Pu have been obtained reproducing the experimental data obtained with the Total Absorption Calorimeter (TAC) \cite{Guerrero_TAC_2009}, the process is described in detail in \cite{Mendoza_PSF_2020,Alcayne_Thesis_2022}. The (n,$\gamma$) cascades of $^{246}$Cm have been obtained by fitting the PSF and LD to the deposited energy spectrum obtained for the first resonance at 4.3 eV. As shown in Fig. \ref{fig:Cascades}, the adjusted cascades for $^{240}$Pu and $^{246}$Cm reproduce the experimental data significantly more accurately than the default cascades generated by NuDEX.

\begin{figure}[htb]
\begin{center}
 \includegraphics[width=\columnwidth]{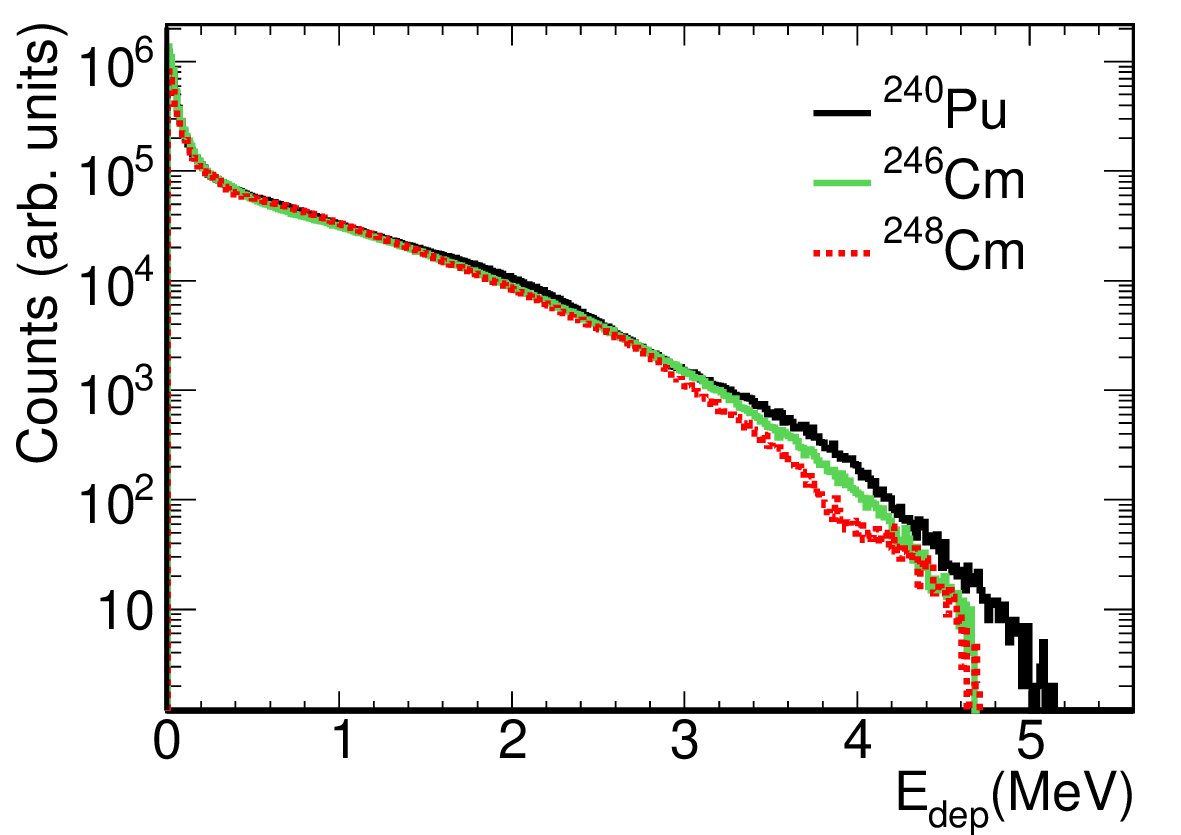} 
\caption{Comparison between the simulated deposited energy spectra in the 
detector due to $^{240}$Pu(n,$\gamma$), $^{246}$Cm(n,$\gamma$) and $^{248}$Cm (n,$\gamma$) cascades. These cascades were used in the analysis.}
 \label{fig:ALLCascades}
\end{center}
\end{figure}

Unfortunately, we were unable to extract any spectrum for $^{248}$ due to the lack of statistics, so the default NuDEX cascades are used for this isotope. As shown in Fig. \ref{fig:ALLCascades}, the shapes of the simulated deposited energy spectra of the three cascades are similar.

The adopted (n,$\gamma$) cascades have been used to calculate the F$\mathrm{_{PHWT}}$ parameters for each isotope. It could be that for other resonances of the same isotope, we had different cascades. However, according to the values obtained in Table 3 of \cite{Mendoza_PHWT_23} the change in F$\mathrm{_{PHWT}}$ for other resonances are of the order of 0.5\% for all the nuclei of this work.

The yields of Cm have been normalized to the strongest resonance of $^{240}$Pu at 1.056 eV. In consequence, the calculated Cm cross-sections will depend not on the individual F$\mathrm{_{PHWT}}$ values, but on the F$\mathrm{_{PHWT}}$($^{246}$Cm)/F$\mathrm{_{PHWT}}$($^{240}$Pu) and F$\mathrm{_{PHWT}}$($^{248}$Cm)/F$\mathrm{_{PHWT}}$($^{240}$Pu) ratios. Table \ref{tab:FPHWTThresholds} shows the F$\mathrm{_{PHWT}}$ values and their ratios for different thresholds. In the analysis a threshold of 0.12 MeV is used, however, there could be some uncertainty in the exact threshold value used due to the detector calibration. Nevertheless, the differences in the ratios for detection thresholds varying between 0.10 and 0.15 MeV are less than 0.3\%.

 \begin{table}[htb]
\begin{center}
 \begin{tabular}{l c c c c } \hline\hline 
 &0.1 MeV&0.12 MeV&0.15 MeV&0.30 MeV \\\hline
$^{240}$Pu&1.0925(4)&1.0977(4)&1.1056(4)&1.1505(4)\\
$^{246}$Cm&1.1641(4)&1.1705(4)&1.1805(4)&1.2334(4)\\
$^{248}$Cm&1.0831(4)&1.0886(4)&1.0971(4)&1.1446(4)\\
$^{240}$Pu/$^{246}$Cm&0.9385(5)&0.9378(5)&0.9365(5)&0.9327(5)\\
$^{240}$Pu/$^{248}$Cm&1.0087(5)&1.0084(5)&1.0078(5)&1.0052(5)\\ \hline\hline
 \end{tabular}
\end{center}
\caption{F$\mathrm{_{PHWT}}$ values and ratios between them for different detection thresholds, and the ratios between them. The uncertainties in the table are due to counting statistics in the Monte Carlo simulations.}
\label{tab:FPHWTThresholds}
\end{table}

In practice, it is necesary to propagate the uncertainties in the cascades to the F$\mathrm{_{PHWT}}$ values. If we compare the F$\mathrm{_{PHWT}}$($^{246}$Cm)/F$\mathrm{_{PHWT}}$($^{240}$Pu) ratio obtained with the fitted cascades and with the default NuDEX cascade the difference is lower than 1.5\% \cite{Alcayne_Thesis_2022}. Taking this into account, in addition to the uncertainties in the Geant4 simulations, we have estimated the uncertainty in the F$\mathrm{_{PHWT}}$($^{246}$Cm)/F$\mathrm{_{PHWT}}$($^{240}$Pu) ratio at 1\% and in the F$\mathrm{_{PHWT}}$($^{248}$Cm)/F$\mathrm{_{PHWT}}$($^{240}$Pu) ratio at 2\%. The second is higher due to not having adjusted the cascades of $^{248}$Cm.

\subsection{Capture yield}\label{sec:CaptureYield}
The theoretical neutron capture yield of an isotope $i$, $Y_{i}(E_{n})$, is defined as the fraction of incident neutrons that induce (n,$\gamma$) reaction in the sample in that isotope. The yield is related to the capture cross-section of the isotope ($\sigma_{\gamma i}$) with:
\begin{equation}\label{Eq:YieldTheory}
Y_{i}(E_n)=F_{m}(E_n)(1-e^{-n\cdot\sigma_{tot}(E_n)})\cdot\frac{A_i \cdot \sigma_{\gamma i}(E_n)}{\sigma_{tot}(E_n)} 
\end{equation}
where $n$ is the sample areal density, $A_{i}$ is the atom fraction of that isotope, $\sigma_{tot}$ is the sum of the total cross-sections of all the isotopes present in the sample, weighted by their isotopic abundances, and $F_{m}$ is a factor to correct for the multiple interactions. The calculation of $F_{m}$ is described in Sec. \ref{sec:ResAnalysis}. This theoretical yield is then compared with the experimental yield to obtain the capture cross-section of the isotope. The experimental yield can be determined after applying the PHWT, described in Section \ref{sec:Data_PWHT}, as follows:
 \begin{equation}\label{Eq:yield_WF}
Y_{PHWT,exp,i}=F_{PHWT,i}\frac{C_w-B_w}{S_{n,i}\cdot\phi_n}
\end{equation}
where $C_w$ is the total weighted counting rate, $B_w$ is the background weighted counting rate, $S_{n,i}$ is the total energy of the cascade for the isotope $i$ and $\phi_n$ is the number of neutrons impinging on the sample per unit time.

As a consequence of the pulse-height weighting procedure, the statistical fluctuations obtained in the weighted yield are larger than with the standard counting technique, i.e. the uncertainties due to counting statistics are larger. It was then proposed in \cite{Lerendegui_Pu242_2018,Lerendegui_Thesis_2018,Mendoza_PHWT_23} to obtain the yield without weighting the counts and then to normalize this yield to the weighted yield obtained with the PHWT. By doing so the uncertainties can be reduced. To use this approach, the efficiency to detect the cascades ($\varepsilon_{c,i}$) must be equal for all the measured resonances in the respective isotope. The actinides of the sample ($^{240}$Pu, $^{246}$Cm and $^{248}$Cm) are 0$^+$ heavy nuclei with more than a million levels below the neutron separation energy. Therefore, as calculated in \cite{Mendoza_PHWT_23}, the relative variation in detection efficiency between each resonance is to be at most $\sim$1\% for these isotopes. This 1\% relative uncertainty in efficiency is significantly smaller than the increase of the counting statistics uncertainty due to the pulse-height weighting procedure. We thus analyzed the data without the application of pulse-height weights, then the yield is calculated as:
 \begin{equation}\label{Eq:yield_noWF}
Y_{exp,i}=\frac{ C-B }{\varepsilon_{c,i} \cdot \phi_n}
\end{equation}
where $C$ is the total counting rate and $B$ is the background counting rate. The counting rate was calculated with an energy deposited threshold of 0.12 MeV and an upper limit of 6 MeV.

The number of neutrons impinging on the sample per unit time ($\phi_n$) has been calculated multiplying the total fluence by the fraction of neutrons intercepted by the sample (Beam Intersection Factor, BIF). The BIF is not expected to change significantly in the energy range of this measurement. The energy dependence of the total fluence during the experiment has been obtained with the SiMon2 neutron detector \cite{Consentino_Simon_2015}. This fluence is in excellent agreement with the evaluated fluence \cite{Sabate_FluxEar2_2017}, which was obtained with a combination of various measurements with different detectors in 2015 and 2016. Therefore, the SiMon2 fluence has been used for the analysis with a 1\% uncertainty in its energy dependence. The BIF of the Cm sample has been obtained by using the saturated resonance method \cite{Macklin_Saturated_1979,Borella_C6D6_2007} with the first resonance of $^{197}$Au at 4.9 eV. For this purpose, a gold sample with the same radius as the Cm pellet has been measured obtaining a BIF of 0.0314(7). For the time-to-energy conversion, the time-of-flight effective distance was calculated to reproduce the energies of the resonances of $^{197}$Au in the JEFF-3.3 library until 400 eV. The distance obtained was 19.425(1) m.

 The detection efficiency ($\varepsilon_{c,i}$) has to be calculated to determine the non-weighted yield ($Y_{exp,i}$). The efficiency values of each isotope have been calculated normalizing to the weighted yield by comparing the yield integrals of the resonances with weight ($Y_{PHWT,exp,i}$) and without ($Y_{exp,i}$). The capture efficiency values obtained doing the weighted average over all the resonances for each isotope are presented in Table \ref{tab:Ef_noWF}.

\begin{table}[!ht]
\begin{center}
\begin{tabular}{lccc}
\hline\hline
 &$^{240}$Pu&$^{246}$Cm&$^{248}$Cm\\\hline
$\varepsilon_{c}$&0.03023(2) &0.02898(18)&0.02685(50)\\\hline\hline
\end{tabular}
\caption{Efficiency to detect the capture cascades for one detector. The uncertainties in the table are only due to counting statistics.}
\label{tab:Ef_noWF}
\end{center}
\end{table}

\begin{figure}[htb]
\begin{center}
 \includegraphics[width=\columnwidth]{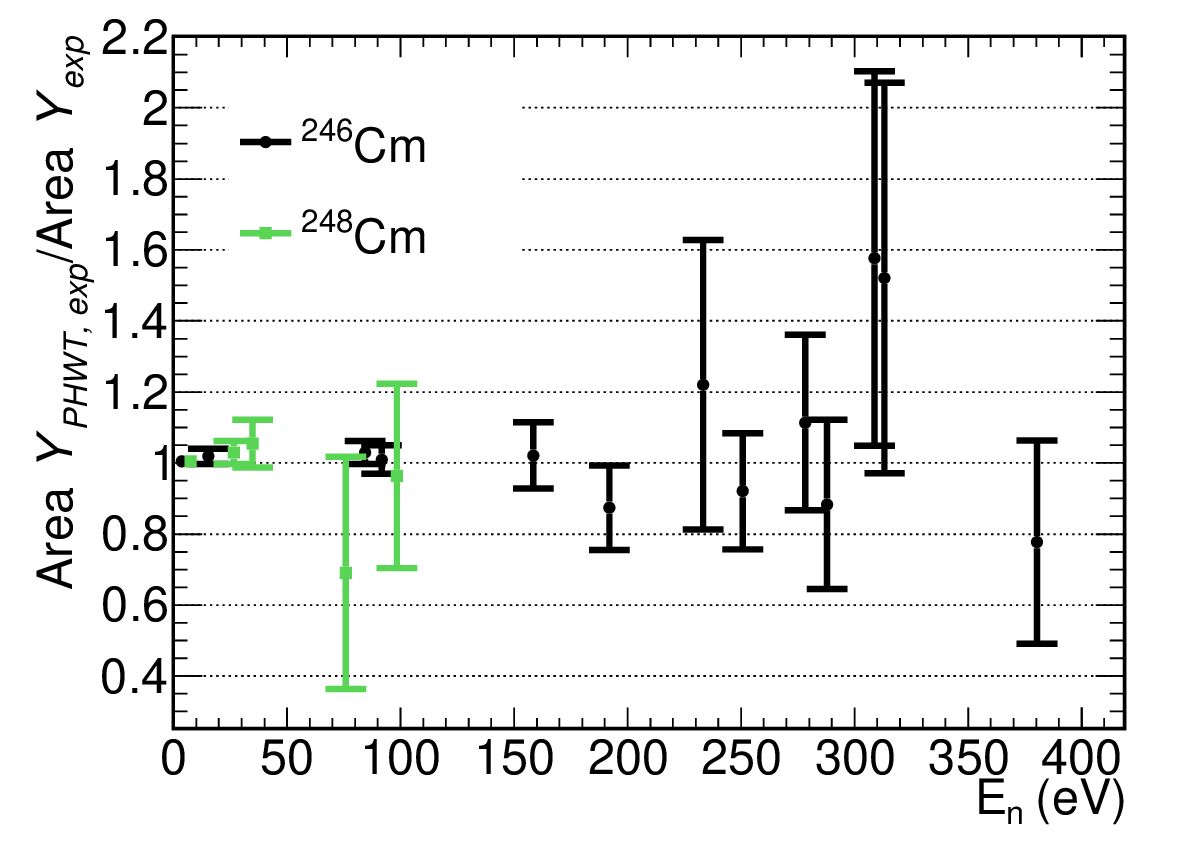} 
\caption{Ratio between the areas of the resonances of the yields obtained with and without the PHWT (i.e. from Eqs. \ref{Eq:yield_WF} and Eq. \ref{Eq:yield_noWF}) for $^{246}$Cm and $^{248}$Cm resonances. The uncertainties of the data are only due to counting statistics.}
 \label{fig:Yielk_WF_NOWF}
\end{center} 
\end{figure}

The areas of the resonances obtained with the weighted yield ($Y_{PHWT,exp,i}$) and non-weighted yield ($Y_{exp,i}$) are compatible as presented in Fig. \ref{fig:Yielk_WF_NOWF}.
 
 \subsection{Uncertainties}\label{sec:UncerYield}
The uncertainties considered in the yield calculation, in addition to those due to counting statistics, are the following:
\begin{enumerate}
\item 
Uncertainty in the normalization. The capture yield has been normalized to the first resonance of $^{240}$Pu at 1.056 eV, as described in Sec. \ref{sec:NorPu240}. The uncertainties associated with the normalization procedure are presented in Table \ref{tab:UncerNorma}. In addition to these uncertainties, the 2.75\% uncertainty in the capture cross-section of $^{240}$Pu at $\sim$1 eV in JEFF-3.3 should be considered.
\begin{table}[!ht]
\begin{center}
\begin{tabular}{lcc}
\hline\hline
Uncertainty &$^{246}$Cm&$^{248}$Cm\\ \hline
Abundance $^{240}$Pu/$^{246-248}$Cm in the sample &2.7&3.3\\ 
$\mathrm{F_{PHWT,^{246-248}Cm}/F_{PHWT,^{240}Pu}}$&1.0&2.0\\ 
 $\mathrm{\varepsilon_{c,^{240}Pu}/\varepsilon_{c,^{246-248}Cm}}$&1.1&2.0\\ \hline
Total quadratic sum&3.0&4.3\\ \hline\hline
\end{tabular}
\caption{Uncertainties (in \%) in the normalization. In the last row, the uncertainties are added quadratically to obtain the total normalization uncertainty.}
\label{tab:UncerNorma}
\end{center}
\end{table}
\item
Uncertainty in the gain shift correction of the energy calibration. The uncertainty in the correction has been estimated in 2-4\%, depending on the energy range. This uncertainty is propagated to a different value in each resonance that goes from 2 to 6\%.

\item 
Uncertainty in the shape of the neutron fluence, which is 1\%.

\item 
Uncertainty in the beam-on background subtraction. The uncertainty due to systematic effects in the determination of the background has been estimated to be 0.4\%. The propagation of this uncertainty is presented in Figs. \ref{fig:Cm246_Uncer} and \ref{fig:Cm248_Uncer}. 

\item
Uncertainty in the fission background produced by the actinides in the sample. The estimated uncertainty in the determination of the fission detection efficiency is 25\%. The propagation of this uncertainty to the resonance depends on the ratio of the capture and fission cross-sections and goes from 2 to 15\%.
\end{enumerate}

\section{Resonance analysis}\label{sec:ResAnalysis}

\subsection{Methodology}\label{sec:Methodology}
The capture yields in the range of 1-400 eV (where individual resonances can be well resolved in our experiment) have been analyzed with the SAMMY code \cite{Larsson_SAMMY_2006} using the Reich-Moore approximation to the R-matrix theory \cite{Frohner_Res_2000}. The different effects of the experimental yield have been taken into account with models implemented in the SAMMY code: multiple interactions, Doppler broadening, and resolution broadening.

The resolution broadening is a consequence of the fact that neutrons arriving at the sample at a certain time do not have all the same energy. The time-to-energy distribution of the neutrons is given by the so-called Resolution Function (RF). At the EAR2, contrary to the EAR1 \cite{Lorusso_RFEAR1_2004}, the RF depends on the sample position and its dimensions \cite{Vlachoudis_RFEAR2_2021}. This is due to the non-optimized geometry of the spallation target. 

\begin{figure}[htb]
\begin{center}
\includegraphics[width=\columnwidth]{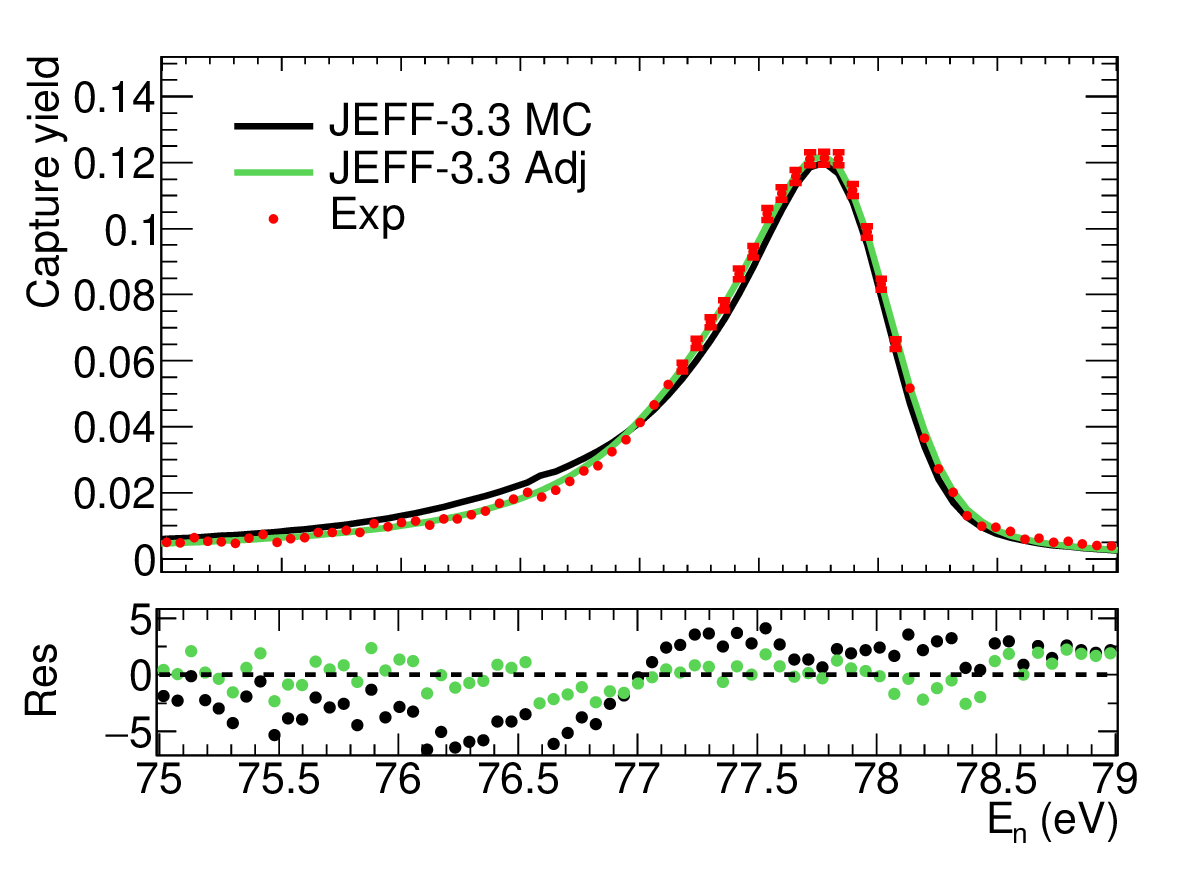}
\caption{Experimental $^{197}$Au capture yield near the 78.3 eV (Exp) for a 2.5 mm radius sample compared with two theoretical yields. Both have been obtained with the resonance parameters of JEFF-3.3 with two different RF. The first yield has been calculated with the RF obtained from Monte Carlo calculations (JEFF-3.3 MC), and the second (JEFF-3.3-Adj) uses the RF obtained after the adjustment. In the bottom panel of the figure, the residuals (Res) defined as the differences between the experimental data points and the theoretical JEFF-3.3 yield divided by the statistical uncertainties of the data points are plotted.}
\label{fig:RF_Fit}
\end{center}
\end{figure}

The RF at n\_TOF is normally calculated using simulations performed with FLUKA \cite{Ahdida_22,Lorusso_RFEAR1_2004}. In fact, at EAR2 this method works for samples covering all the neutron beam \cite{Vlachoudis_RFEAR2_2021}. However, as shown in Fig. \ref{fig:RF_Fit}, the code is not able to reproduce the shape of the yield for small samples. For this reason, the RF for the Cm campaign has been obtained by performing an adjustment. The parameters have been fitted to match the well-established resonances of $^{197}$Au measured in the range from 4 to 400 eV for a sample with the same dimensions (2.5 mm radius) as the Cm sample \cite{Alcayne_Thesis_2022}. Fig. \ref{fig:RF_Fit} shows that this RF reproduces the capture yield much more accurately than the RF obtained from Monte Carlo calculations with FLUKA. The resonance of Au presented in Fig. \ref{fig:RF_Fit} has a neutron energy of 78.27 eV and in the plot, the center of the resonance is at around 77.8 eV, this shift is due to the fact that the RF apart from changing the shape of the resonances also shifts its position. The uncertainties in the obtained RF have been quantified and propagated to the resonance analysis \cite{Alcayne_Thesis_2022}.

At n\_TOF EAR2 due to the considerable uncertainty in the RF, we are mostly sensitive to the area of the resonances and not to their shape. For this reason in the fit, we fitted only the neutron width ($\Gamma_n$), which for the resonances of the experiment is usually the most sensitive to the resonance area. Therefore for radiation and fission widths, respectively $\Gamma_\gamma$ and $\Gamma_f$, the values from JENDL-4.0 have been taken. The libraries JEFF-3.3 and ENDF/B-VIII.0 report the same RP for these isotopes. The energy of the resonances ($E_0$) has been also obtained on the fits.

\begin{figure*}[!htb]
\begin{center}
\includegraphics[width=0.49\textwidth]{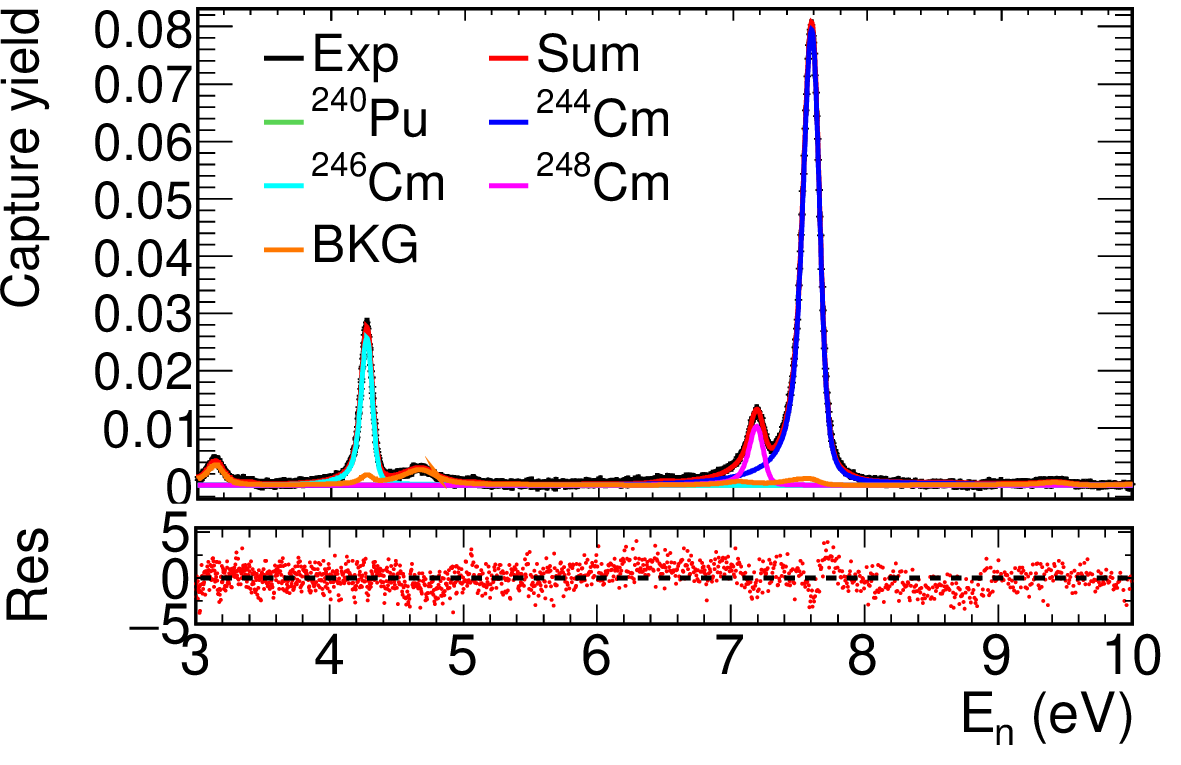}
\includegraphics[width=0.49\textwidth]{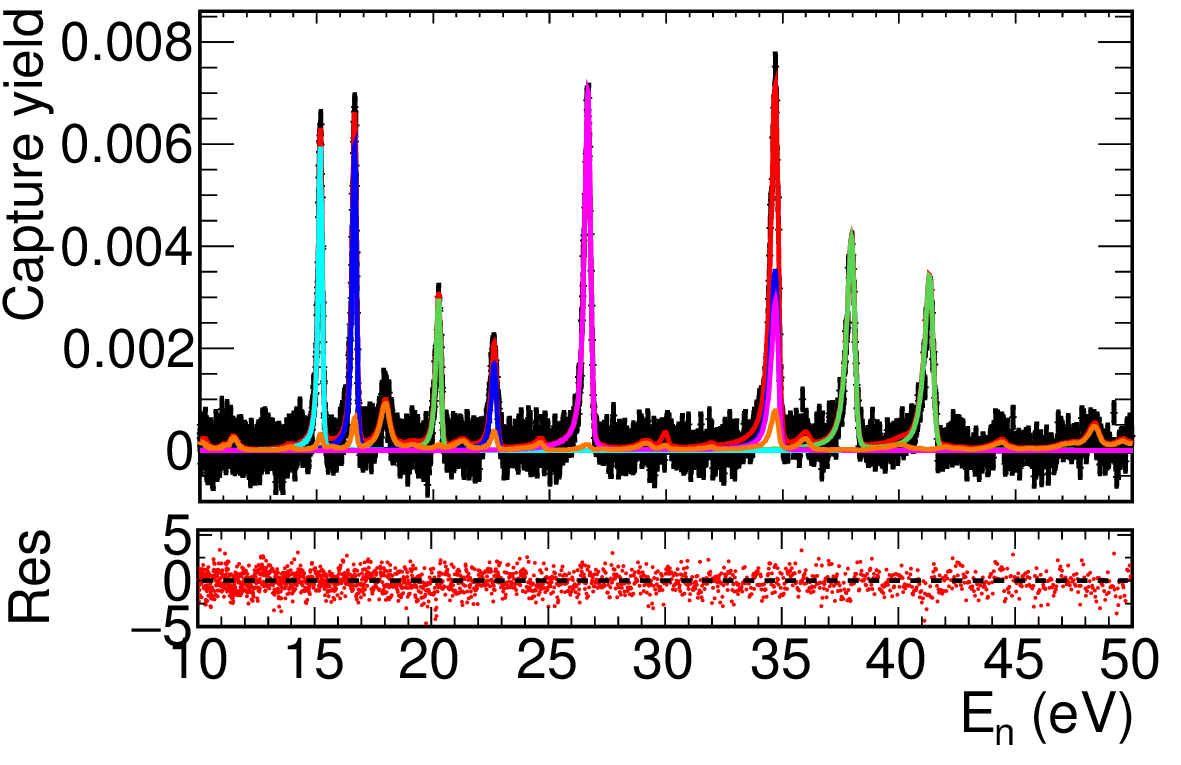}
\includegraphics[width=1.0\textwidth]{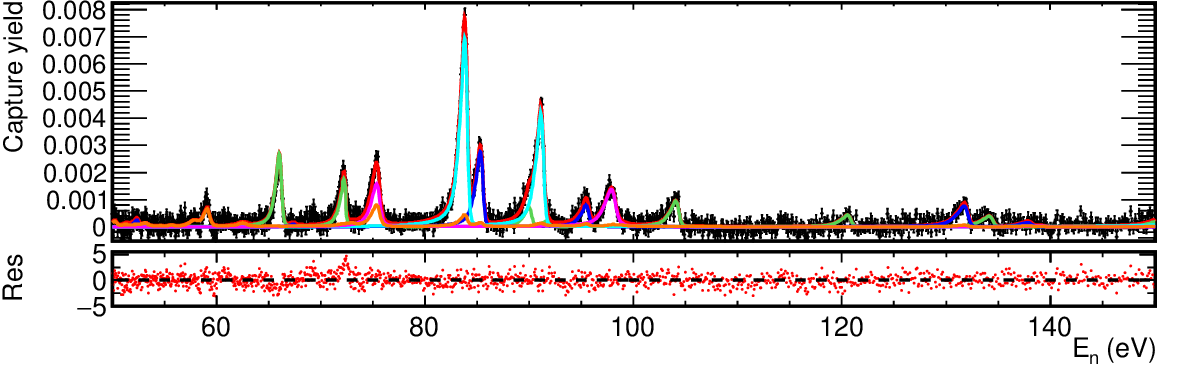}
\includegraphics[width=1.0\textwidth]{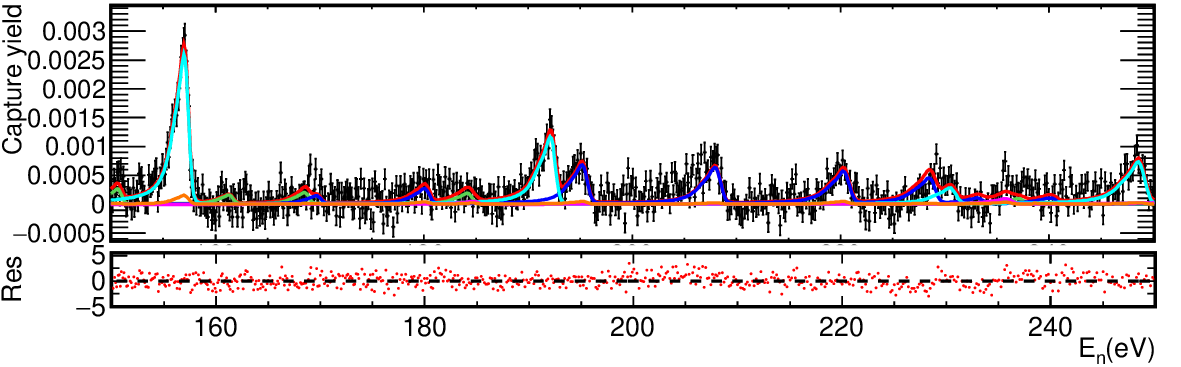}
\includegraphics[width=1.0\textwidth]{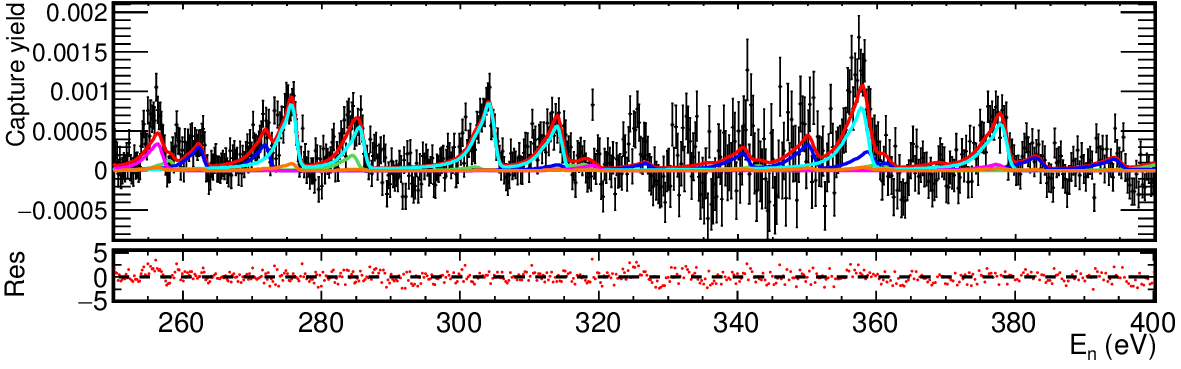}
\caption{Experimental capture yield measured in EAR2 for the Cm sample compared with the fitted yields. The experimental capture yield (Exp) includes the uncertainties due to counting statistics only. The green, blue and cyan lines correspond to the capture yield for each isotope. The orange line (BKG) corresponds to the background due to fission and capture events in the actinides. The red line (Sum) corresponds to the sum of all the capture and fission yields. It is important to notice that the resonances in the figures are shifted in energy due to the RF, see Section \ref{sec:Methodology} for more details. }
\label{fig:Cm246_EAR2_FitRes}
\end{center}
\end{figure*}

\begin{figure*}[!htb]
\includegraphics[width=0.49\textwidth]{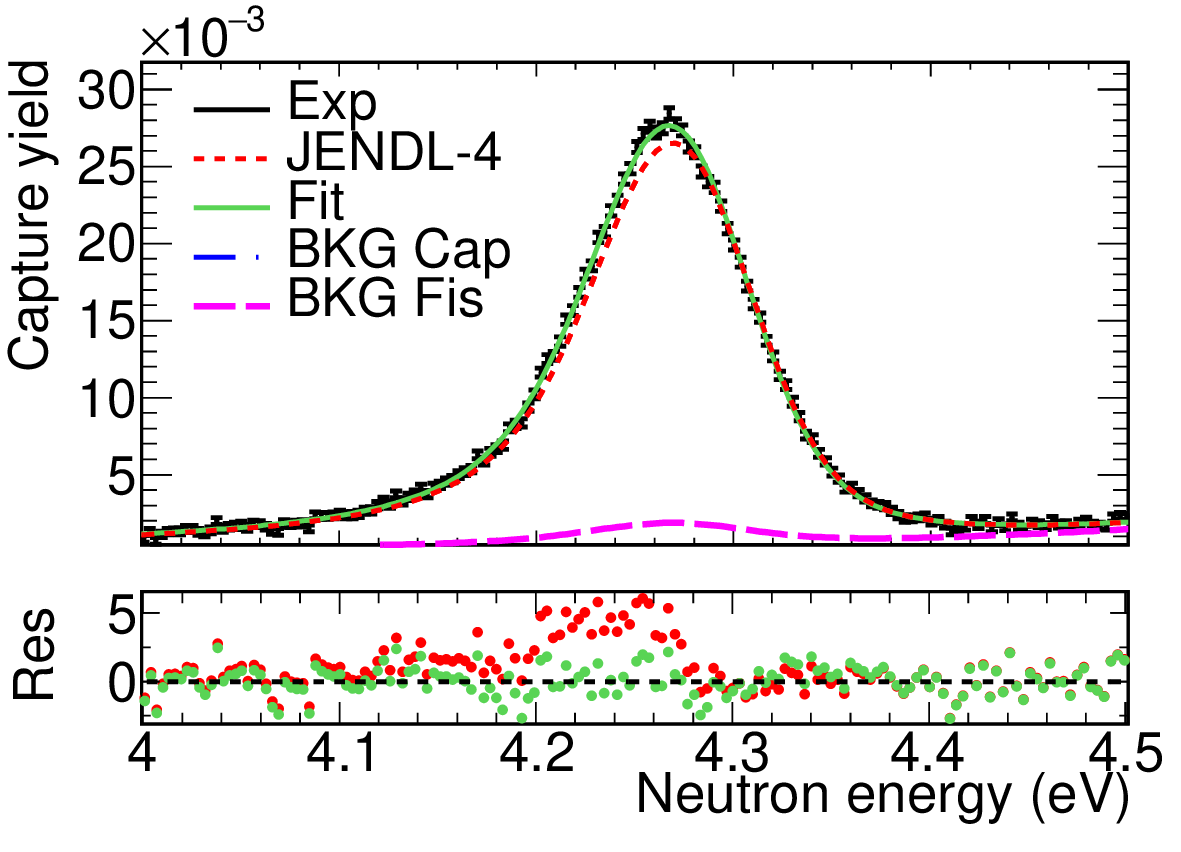}
\includegraphics[width=0.49\textwidth]{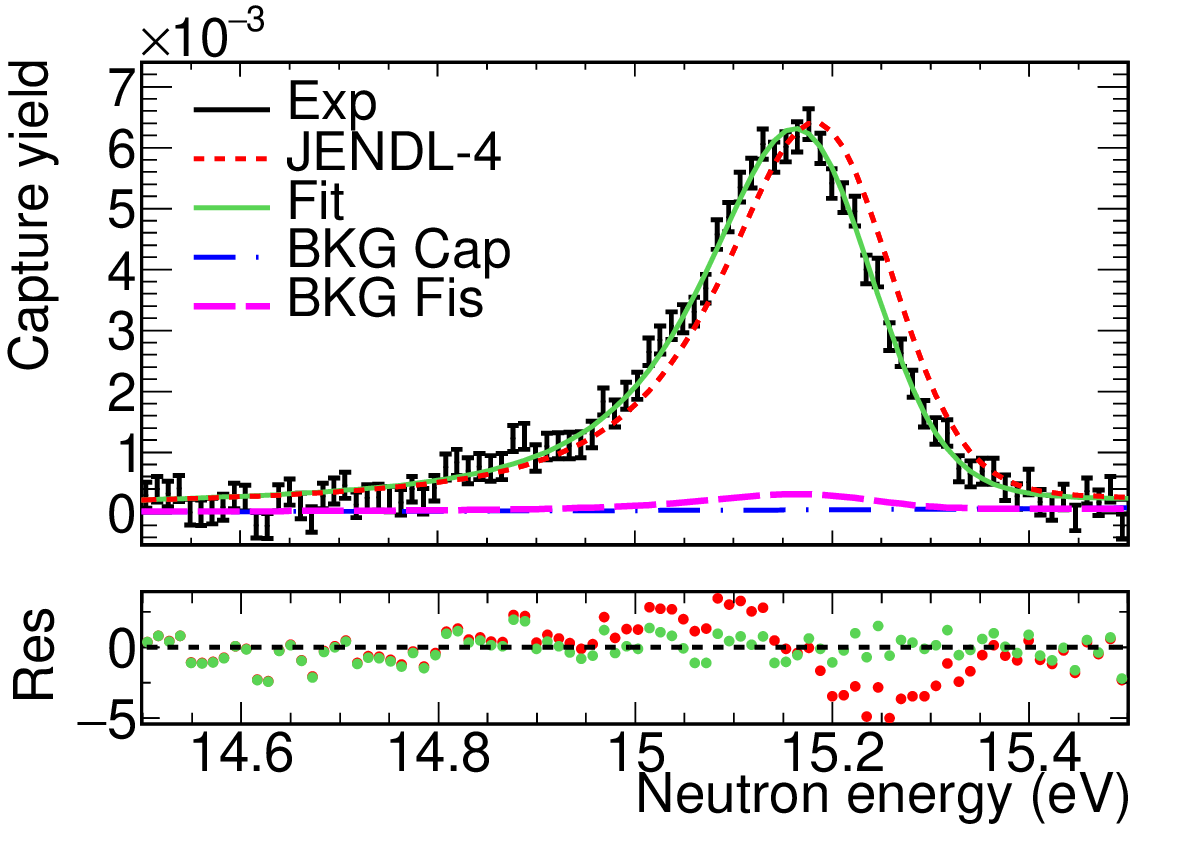}
\includegraphics[width=0.49\textwidth]{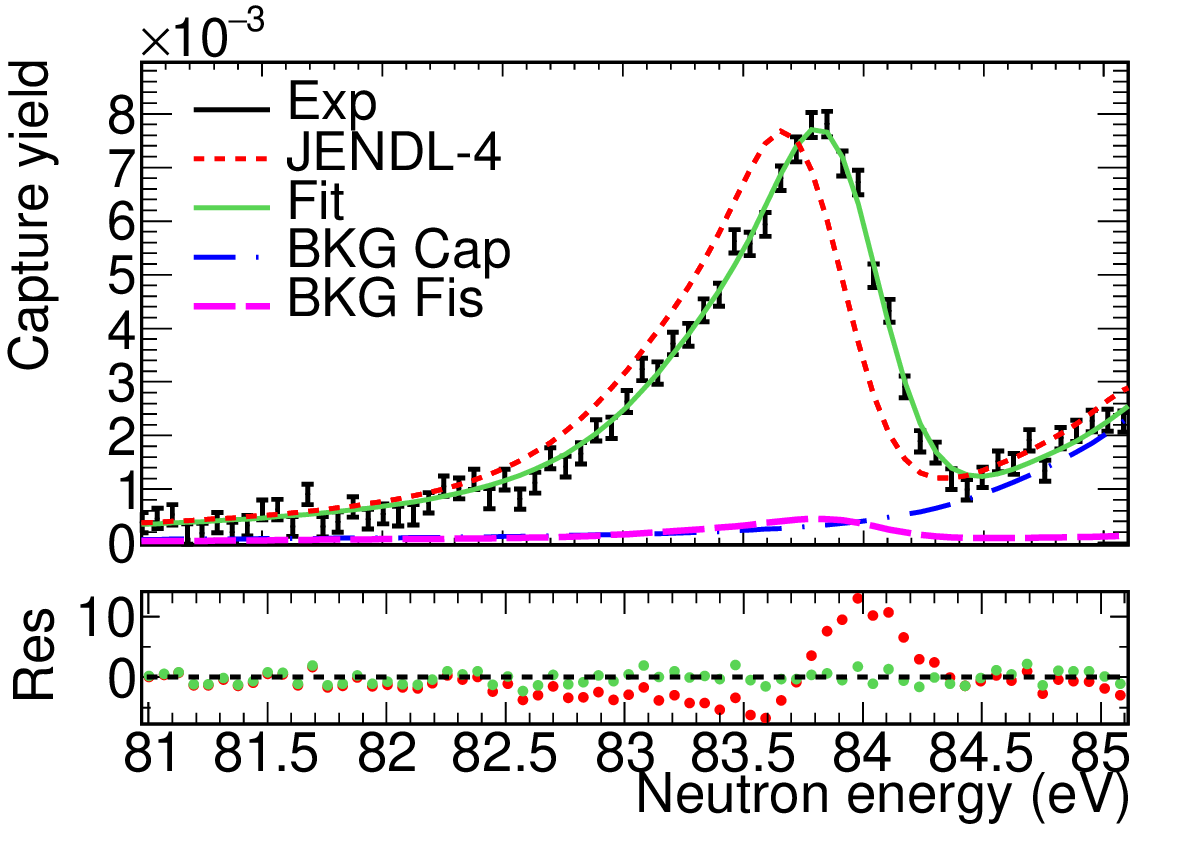}
\includegraphics[width=0.49\textwidth]{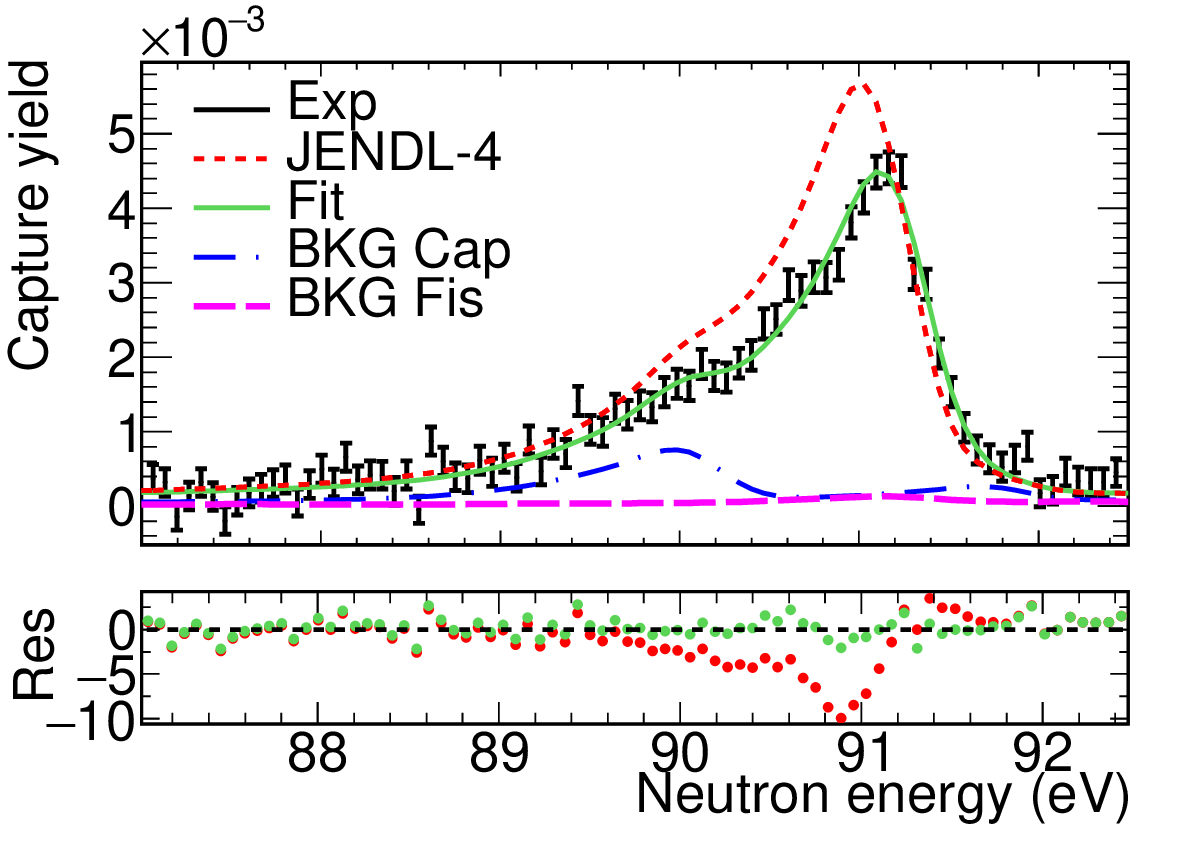}
\includegraphics[width=0.49\textwidth]{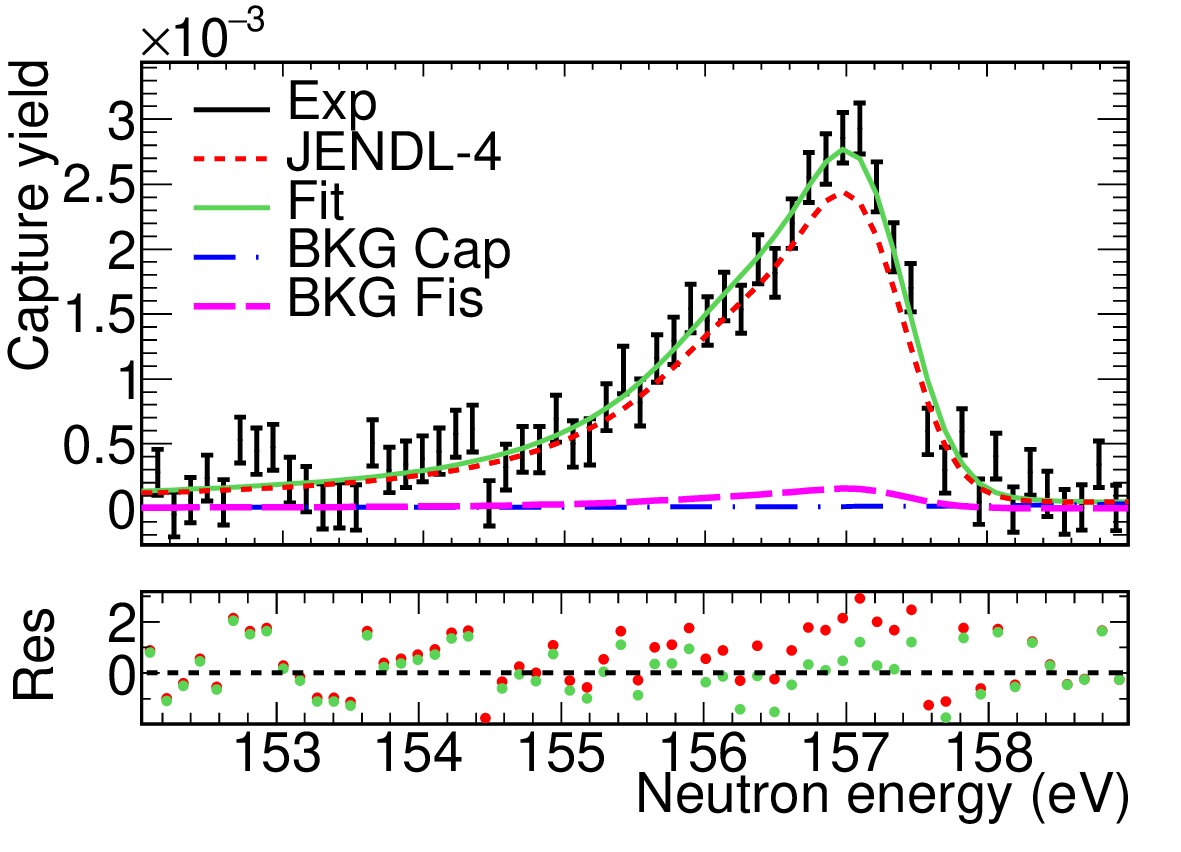}
\includegraphics[width=0.49\textwidth]{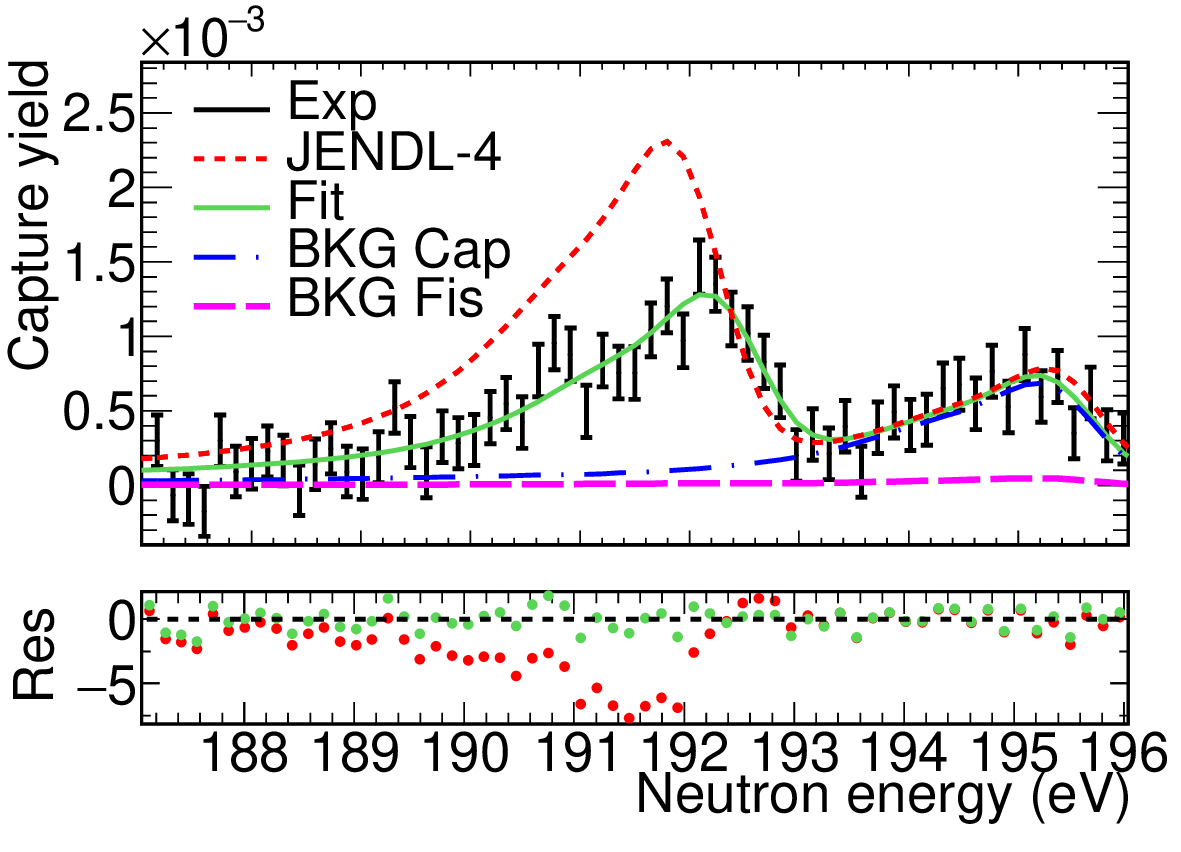}
\caption{Experimental $^{246}$Cm capture yields close to the first six resonances (black) compared with the yield obtained with the fit (green) and with the JENDL-4.0 data (red). The evaluation in JENDL-4.0 has also been used in JEFF-3.3 and ENDF/B-VIII.0. In blue, the calculation of the background due to the other actinides and in pink the fission background. It is important to notice that the resonances in the figures are shifted in energy due to the RF, see Section \ref{sec:Methodology} for more details.}
\label{fig:Cm246_Fit_4_193}
\end{figure*}

\begin{figure*}[!htb]
\includegraphics[width=0.49\textwidth]{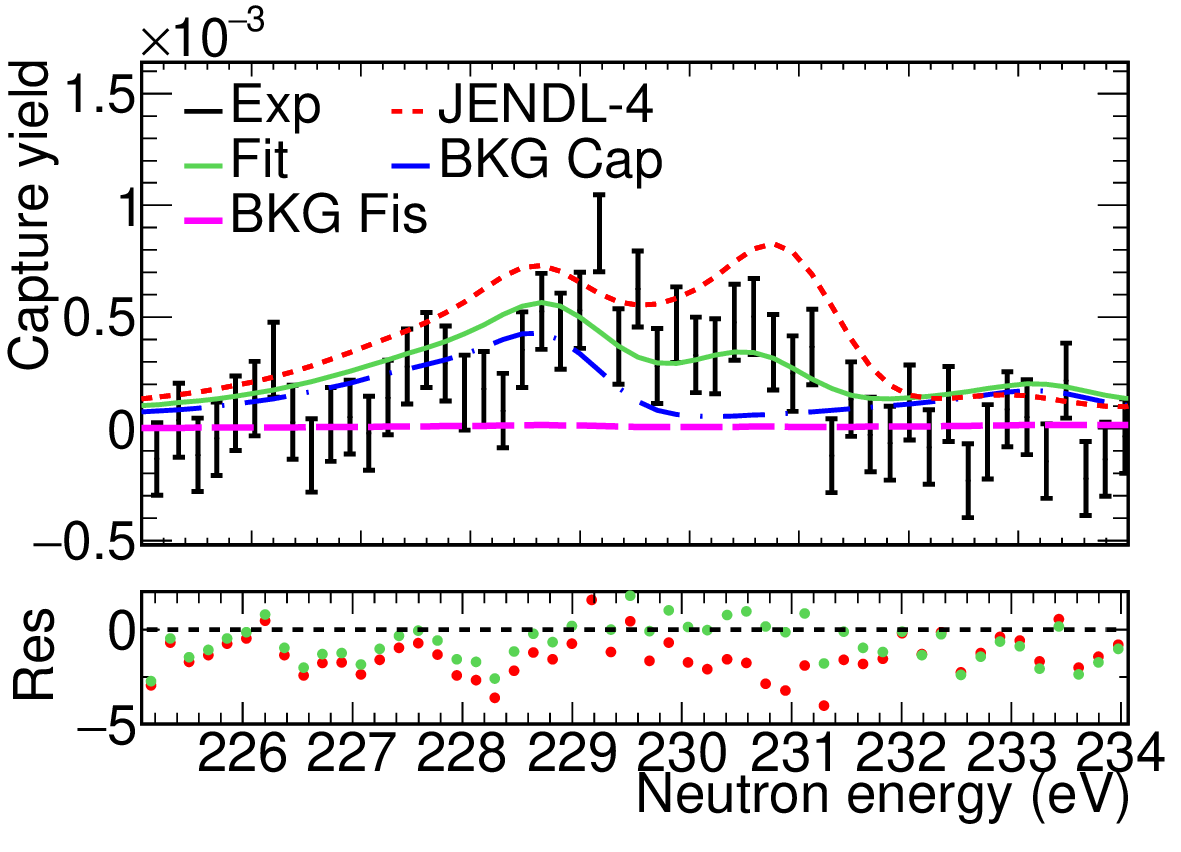}
\includegraphics[width=0.49\textwidth]{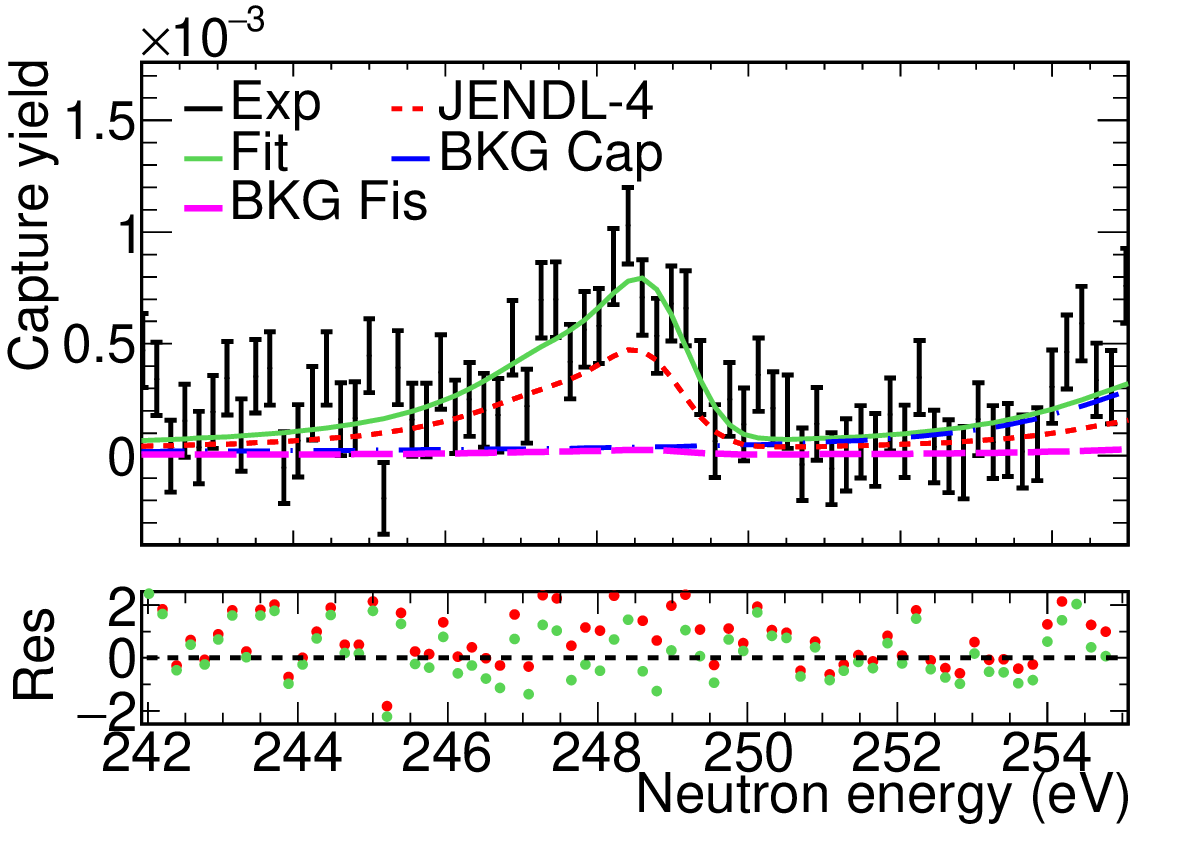}
\includegraphics[width=0.49\textwidth]{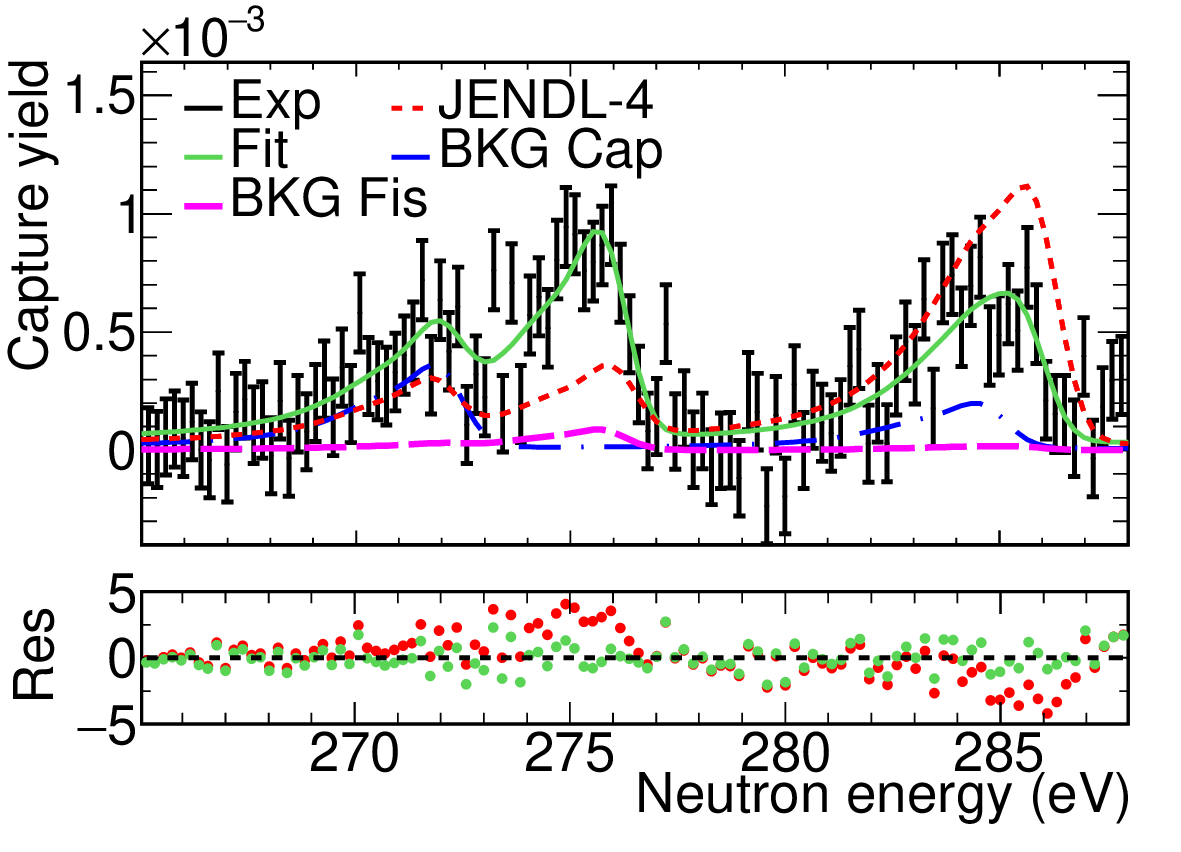}
\includegraphics[width=0.49\textwidth]{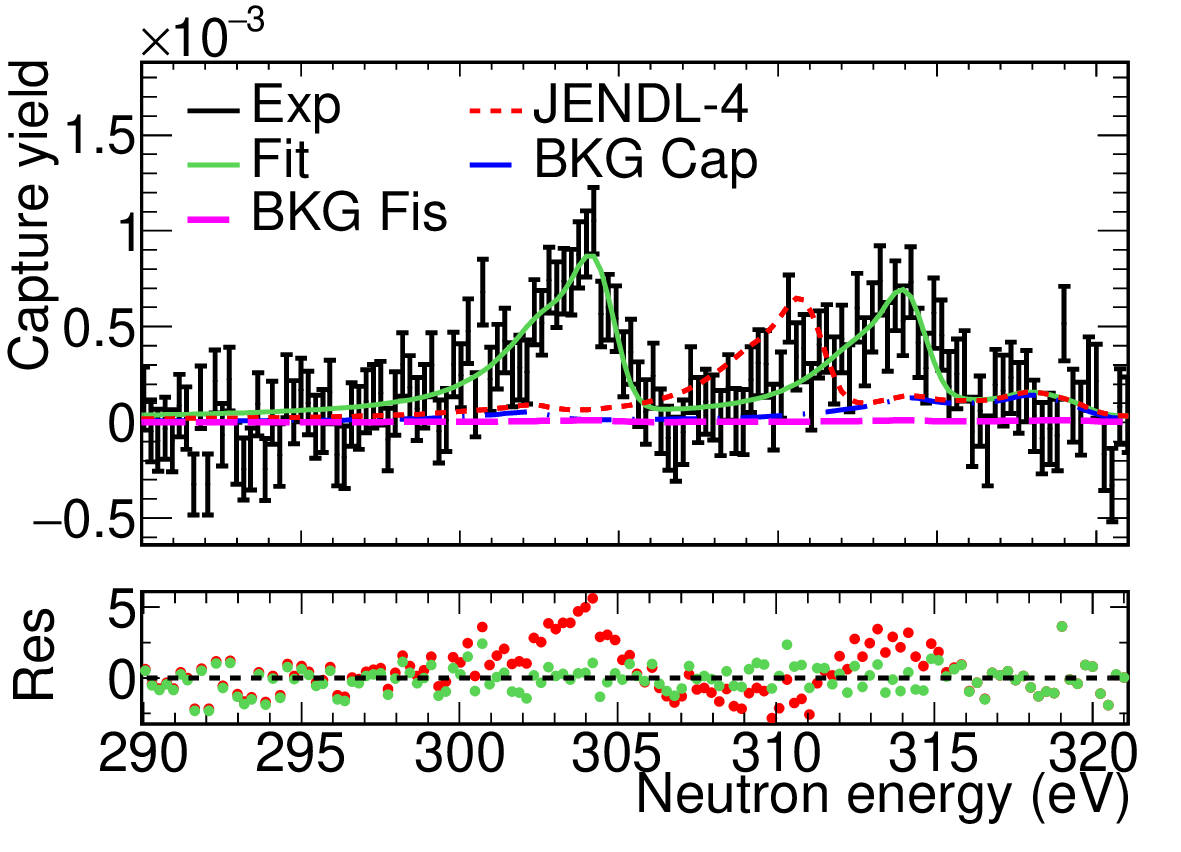}
\includegraphics[width=0.49\textwidth]{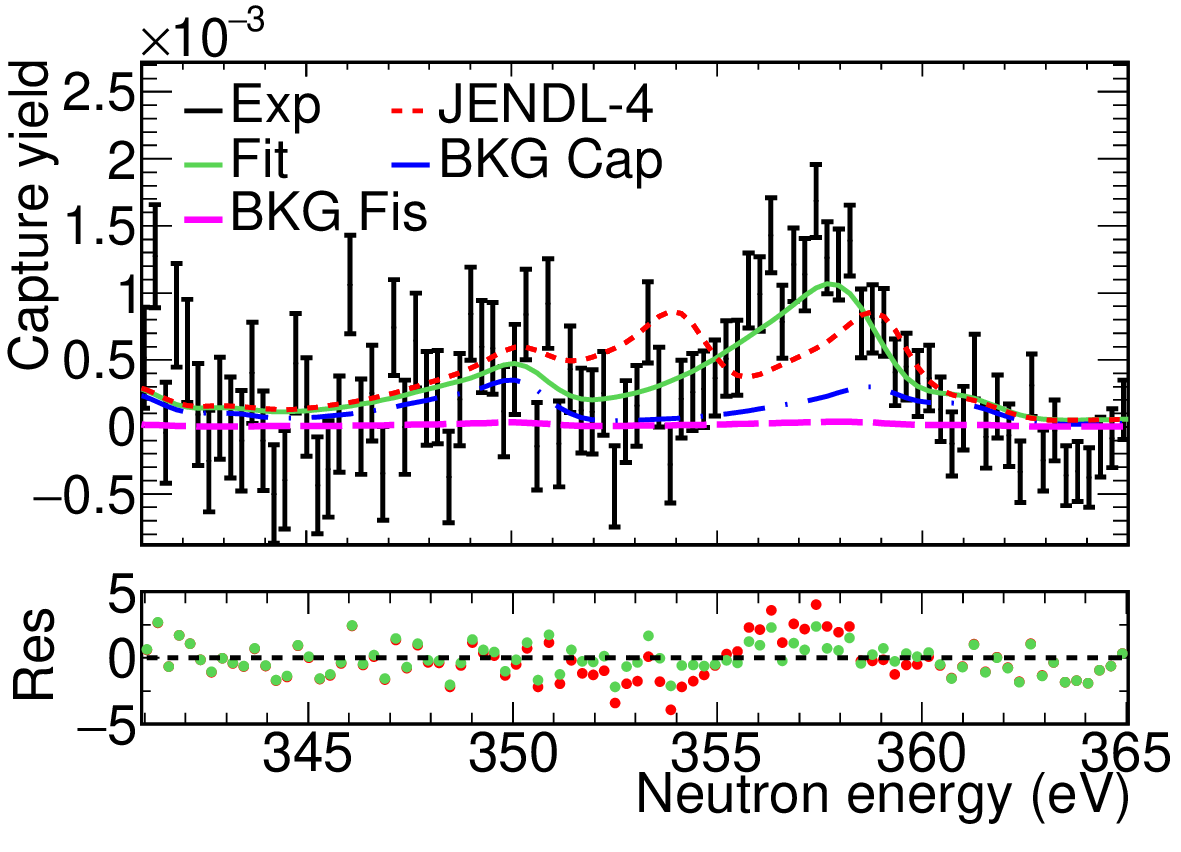}
\includegraphics[width=0.49\textwidth]{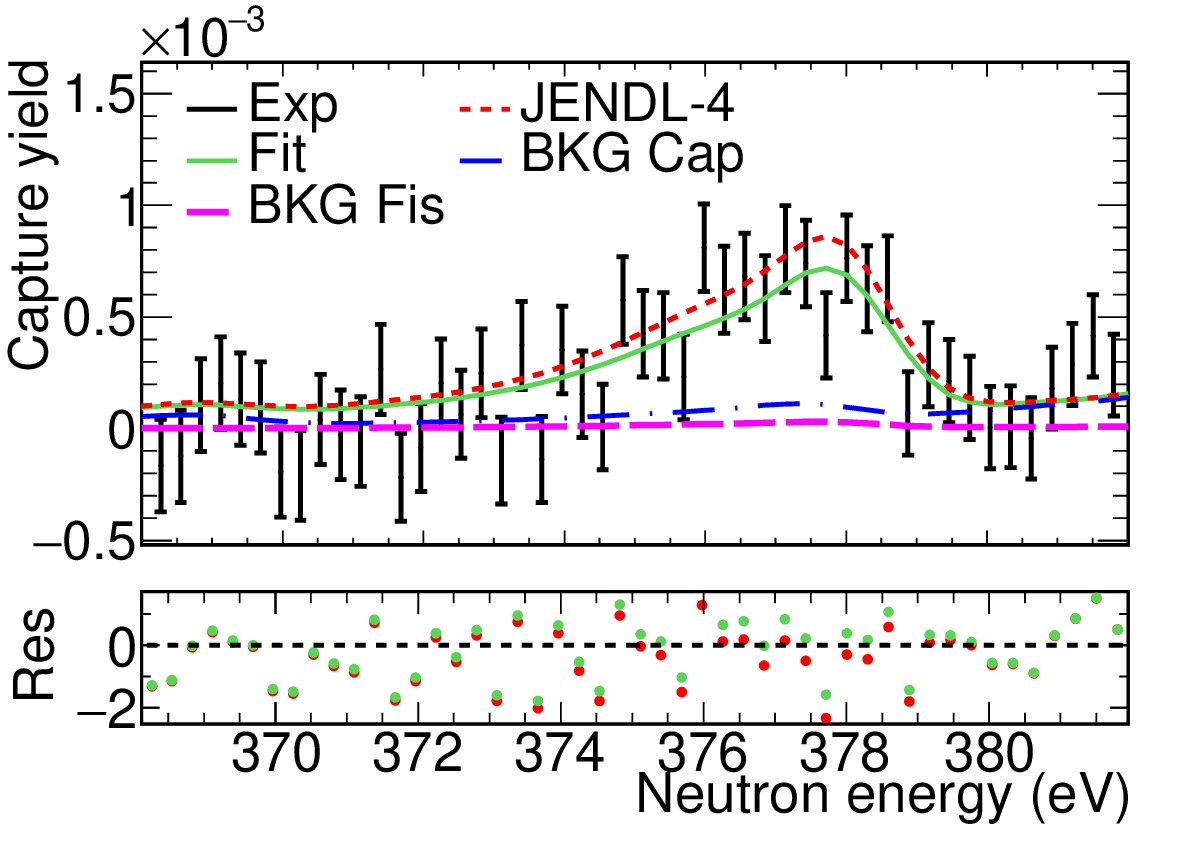}
\caption{Same as Fig. \ref{fig:Cm246_Fit_4_193} but for the resonances between 230 and 400 eV.}
\label{fig:Cm246_Fit_232_381}
\end{figure*}

\begin{figure*}[!htb]
\includegraphics[width=0.49\textwidth]{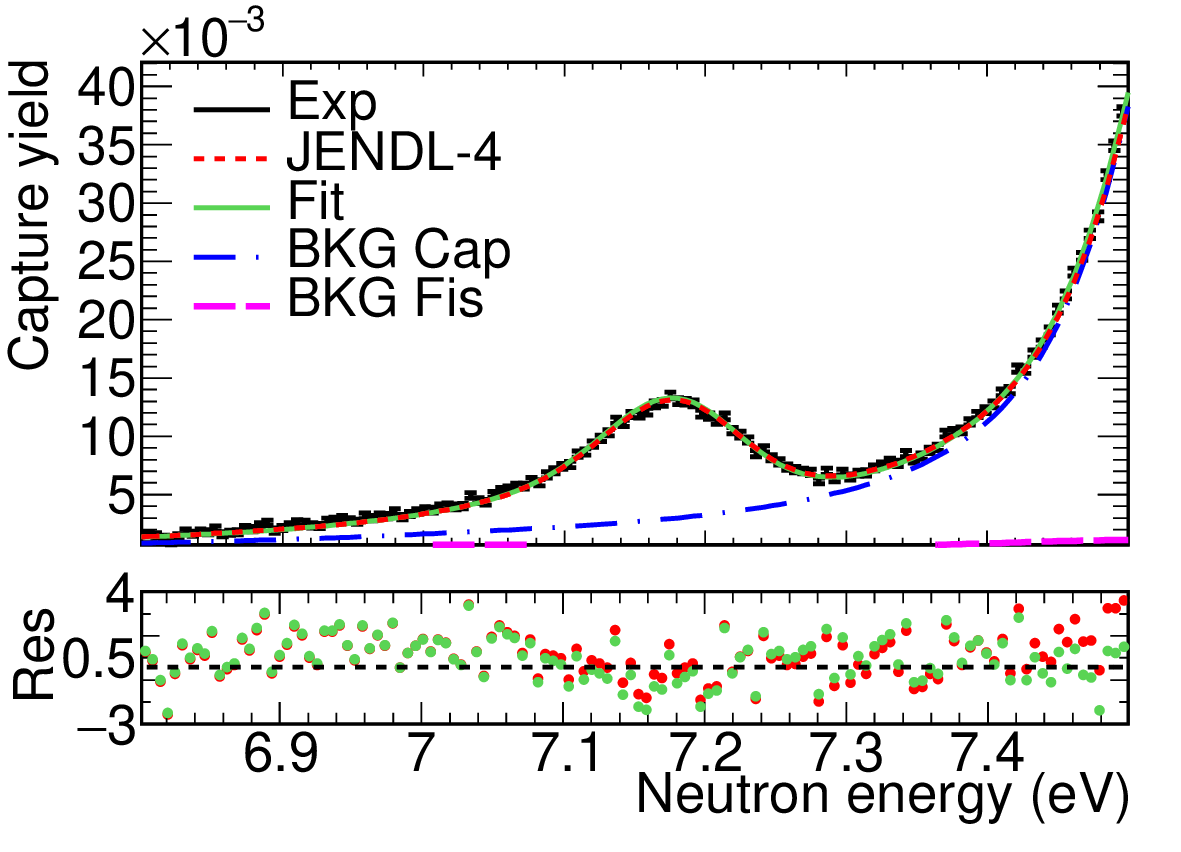}
\includegraphics[width=0.49\textwidth]{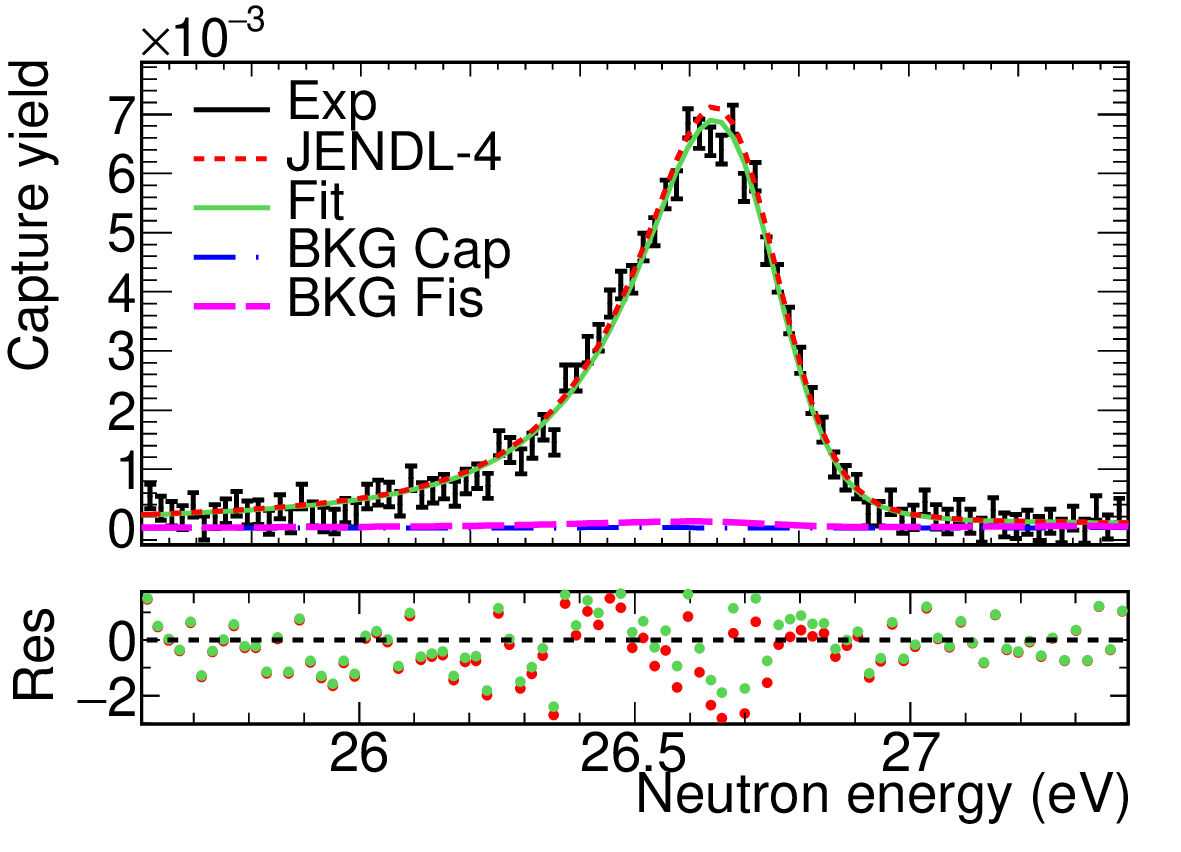}
\includegraphics[width=0.49\textwidth]{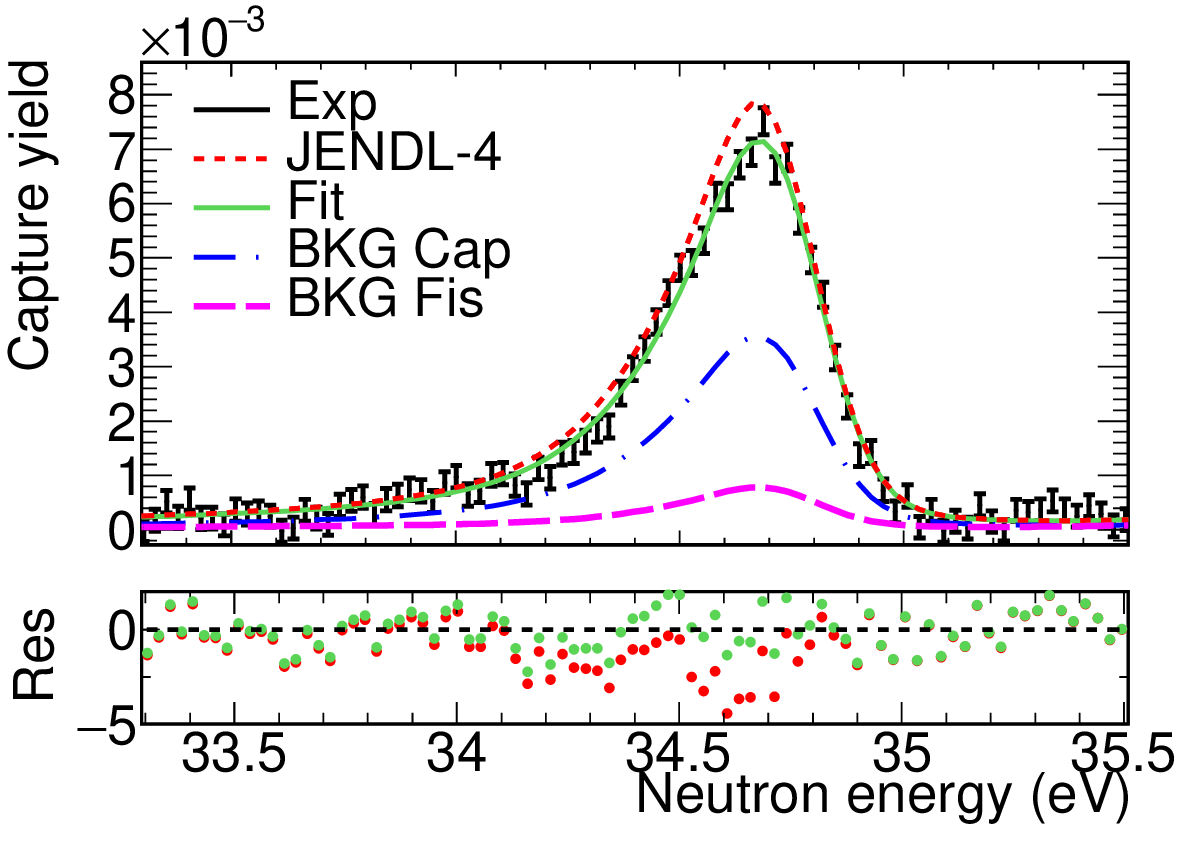}
\includegraphics[width=0.49\textwidth]{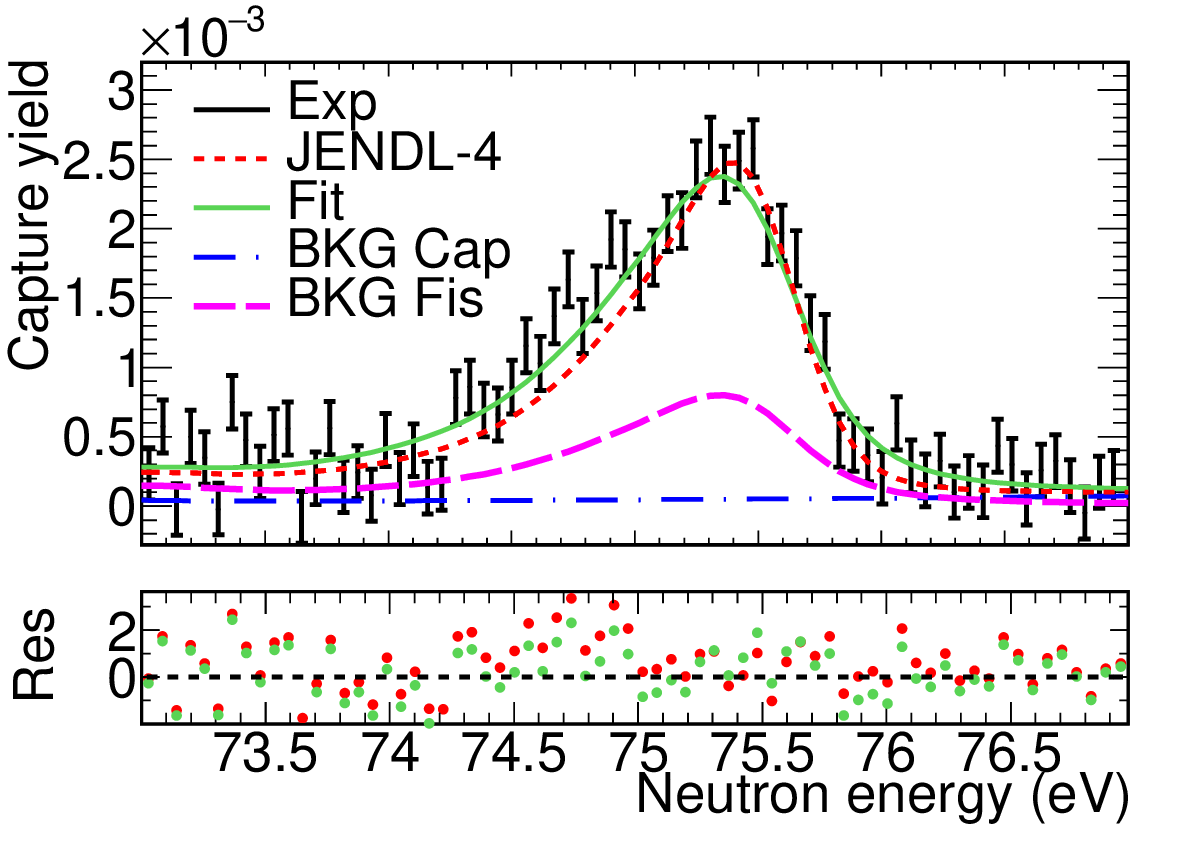}
\includegraphics[width=0.49\textwidth]{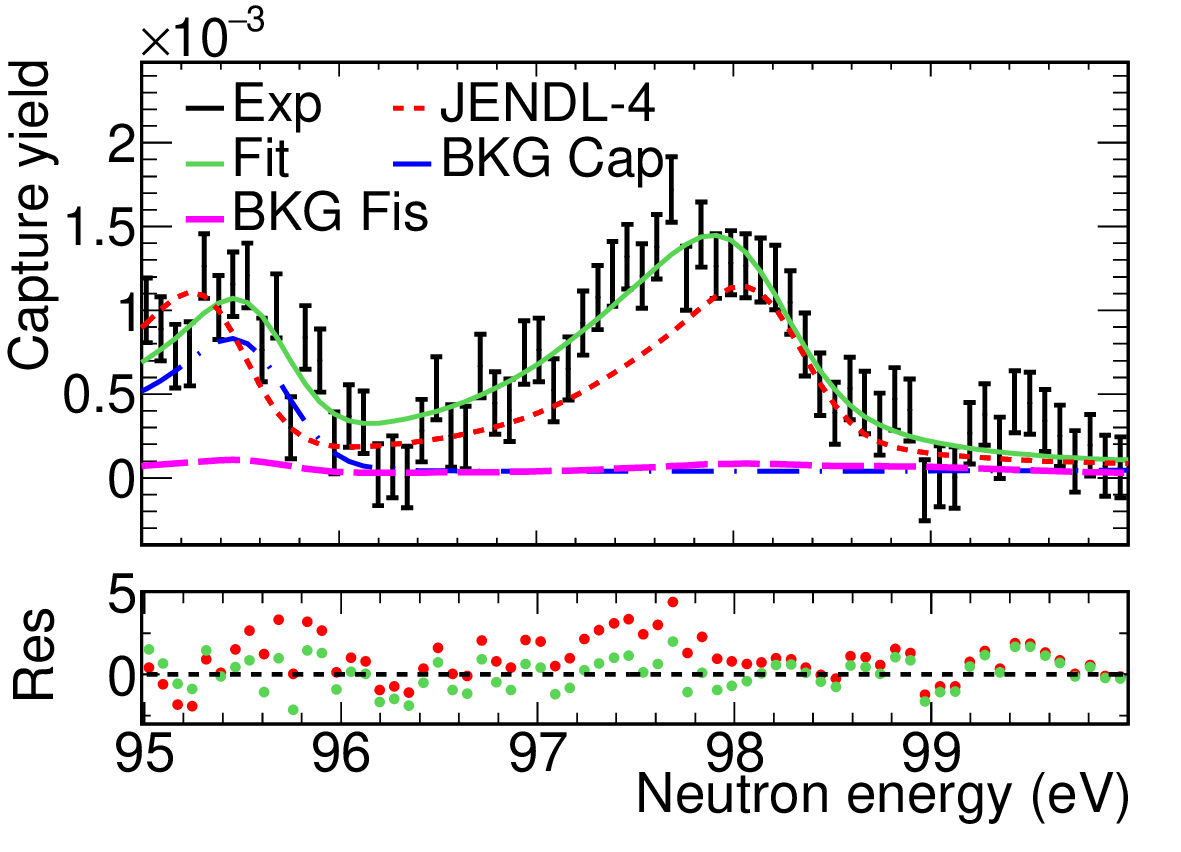}
\caption{Same as Fig. \ref{fig:Cm246_Fit_4_193} but for the resonances of $^{248}$Cm.}
\label{fig:Cm248_Fit_4_100}
\end{figure*}

All the resonances have been assumed to be s-wave with $J^\pi=1/2^+$ since no $p$-wave resonances should be observable in our experiment below 400 eV. The resonances of $^{240}$Pu and $^{244}$Cm have been also analyzed, but the results will be reported in a stand-alone publication, together with the results obtained from other measured Cm samples \cite{Alcayne_CmND_23}. In the SAMMY analysis, the isotopes $^{27}$Al and $^{16}$O, also mixed with Cm in the sample, have been included to properly account for their contributions to the self-shielding and multiple interactions corrections. The detection efficiencies for these two isotopes have been set to zero, as their contributions have already been subtracted with the dummy sample. The results of the fits are shown in Figs. \ref{fig:Cm246_EAR2_FitRes}, \ref{fig:Cm246_Fit_4_193}, \ref{fig:Cm246_Fit_232_381}, \ref{fig:Cm246_Fit_35_47_131} and \ref{fig:Cm248_Fit_4_100}.

The background resulting from fission reactions has been calculated using the fission detection efficiency ($\varepsilon_f$) and the fission yield of each isotope. As the fission yields depend on the RP, in particular in the fitted parameters E$_0$ and $\Gamma_n$, a recursive approach has been employed. This method involves initially subtracting the fission background estimated using the resonance parameters from the previous fitting iteration, followed by fitting the resonance parameters (E$_0$ and $\Gamma_n$). These two steps are iteratively repeated until convergence is reached.

\subsubsection{Normalization to \textsuperscript{240}Pu}\label{sec:NorPu240}

\begin{figure}[htb]
\includegraphics[width=\columnwidth]{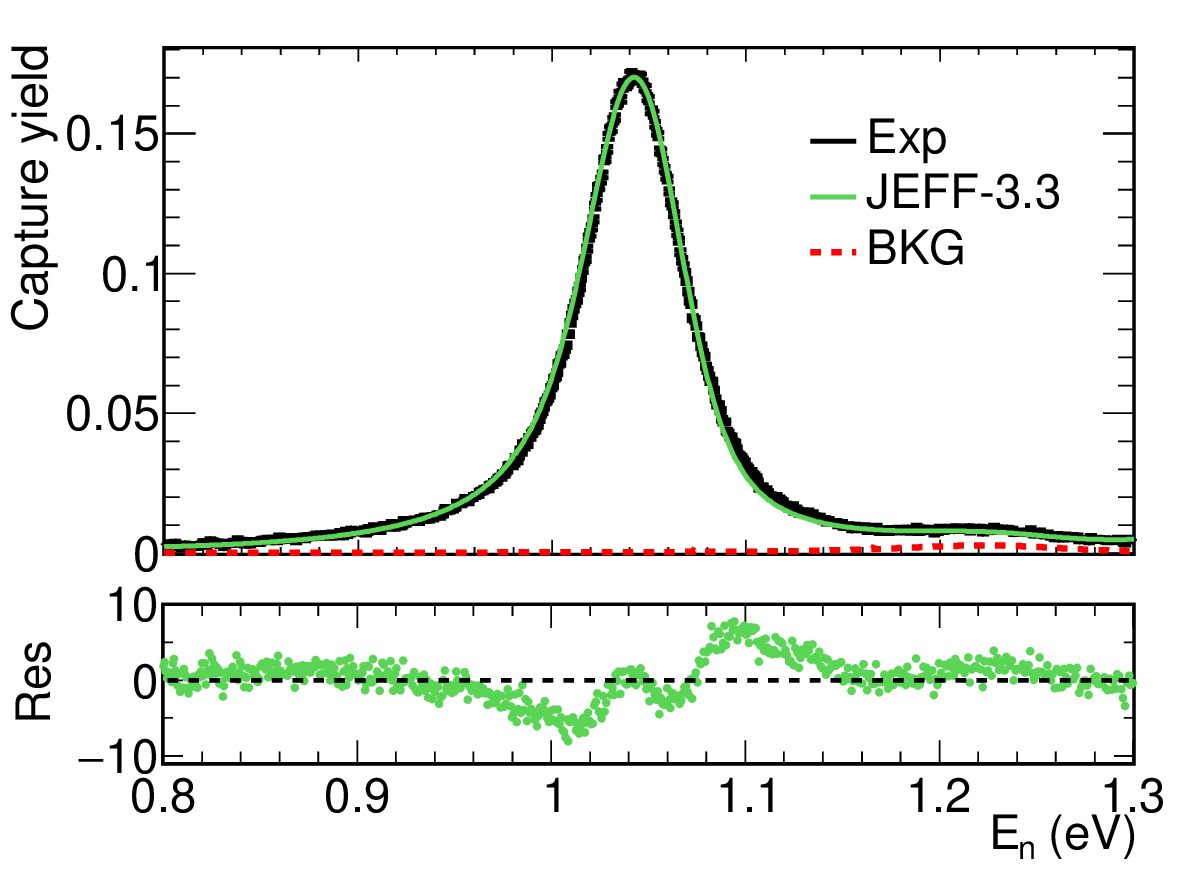}
\caption{$^{240}$Pu experimental capture yield (Exp) in the energy range of the strongest resonance compared to the yield obtained with the JEFF-3.3 resonance parameters (JEFF-3.3) used to perform the normalization. The background (BKG) is also included in the plot. In the bottom panel, the residuals are presented.}
\label{fig:Fit_Pu240}
\end{figure}

As already mentioned, the capture yield, and thus the capture cross-sections of $^{246}$Cm and $^{248}$Cm, have been normalized to the first resonance of $^{240}$Pu at 1.056 eV using the JEFF-3.3 resonance parameters. The experimental yield and the SAMMY fit of the Pu resonance are presented in Fig.\ref{fig:Fit_Pu240}. The uncertainties in the normalization, described in Table \ref{tab:UncerNorma}, are 3\% and 4.3\% for $^{246}$Cm and $^{248}$Cm, respectively. The mass of $^{240}$Pu in the sample which has been obtained from the fit is 0.159(4) mg (the uncertainty does not account for the uncertainty in the JEFF-3.3 cross-section).

\subsection{Uncertainties in the resonance parameters}\label{sec:UncerRP}
The uncertainties in the yield have been propagated to the E$_0$ and $\Gamma_n$ resonance parameters. In particular, the uncertainties due to counting statistics in the experimental yield have been propagated by the SAMMY code. However, the uncertainties due to systematic effects in the yield, summarized in Section \ref{sec:UncerYield}, have been propagated to the parameters with the method described in the following paragraphs. 

The different uncertainties have been divided into two groups, those that are considered to be fully uncorrelated between the fitted $\Gamma_n$ values of the resonances and those that are considered fully correlated. The uncertainties due to counting statistics, the resolution function, the fluence shape and the gain shifts of the deposited energy calibration have been considered as fully uncorrelated. On the other hand, the uncertainties due to the fission background, the beam-on background and the normalization have been considered as fully correlated.

Each of the uncertainties due to systematic effects have been propagated to E$_0$ and $\Gamma_n$ parameters by using two additional yields. If the nominal yield is described with a set of data points (E$_{j}$,Y$_{exp,j}$), with E$_{j}$ the neutron energy and Y$_{exp,j}$ the capture yield at the point $j$, then the two new yields have been constructed as (E$_{j}$,Y$_{exp,j}+\sigma_{k,j}$) and (E$_{j}$,Y$_{exp,j}-\sigma_{k,j}$), where $\sigma_{k,j}$ is the yield uncertainty due to the $k$-type of uncertainty at point $j$. The fitting process has been repeated for both yields, and two new sets of $E_0$ and $\Gamma_n$ parameters have been obtained for each $k$-type of uncertainty and for each resonance. The uncertainties in the fitted quantities have been obtained as half of the differences, in absolute value, between the two values obtained in the fits. The obtained uncertainties are given in Tables \ref{tab:RP_Cm246} and \ref{tab:RP_Cm248}, also in the tables the fully uncorrelated (U) and fully correlated (C) uncertainties were summed in squares. 

\subsection{Resonance parameters of \textsuperscript{246}Cm}
\begin{figure}[htb]
\includegraphics[width=\columnwidth]{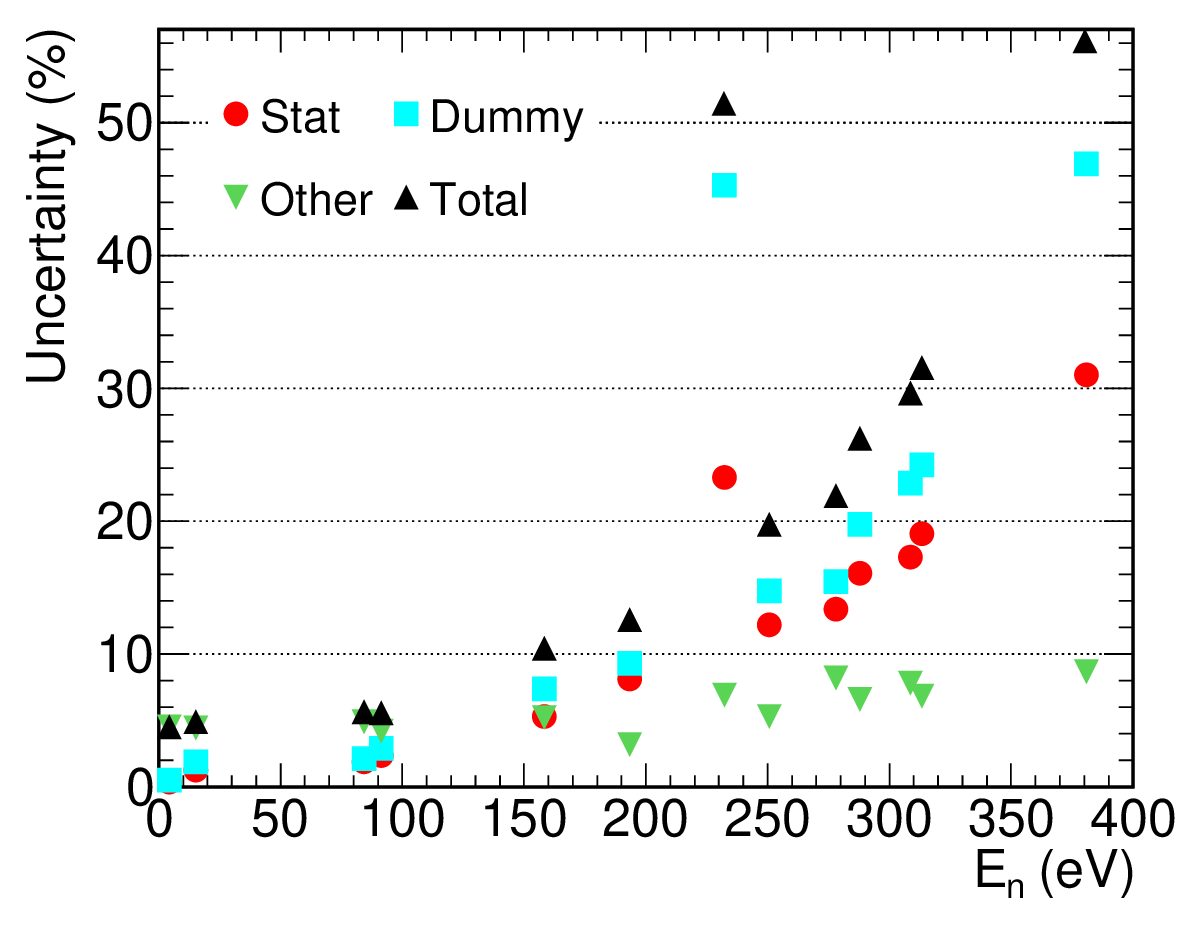}
\caption{Total uncertainties (in percentage) of the $\Gamma\mathrm{_{n}}$ parameters of $^{246}$Cm obtained in this work (Total). The uncertainties due to counting statistics (Stat), due to the subtraction of the beam-on background (Dummy) and the rest of the uncertainties (Other) are also presented for comparison.} 
\label{fig:Cm246_Uncer}
\end{figure}

\begin{figure}[!htb]

\includegraphics[width=\columnwidth]{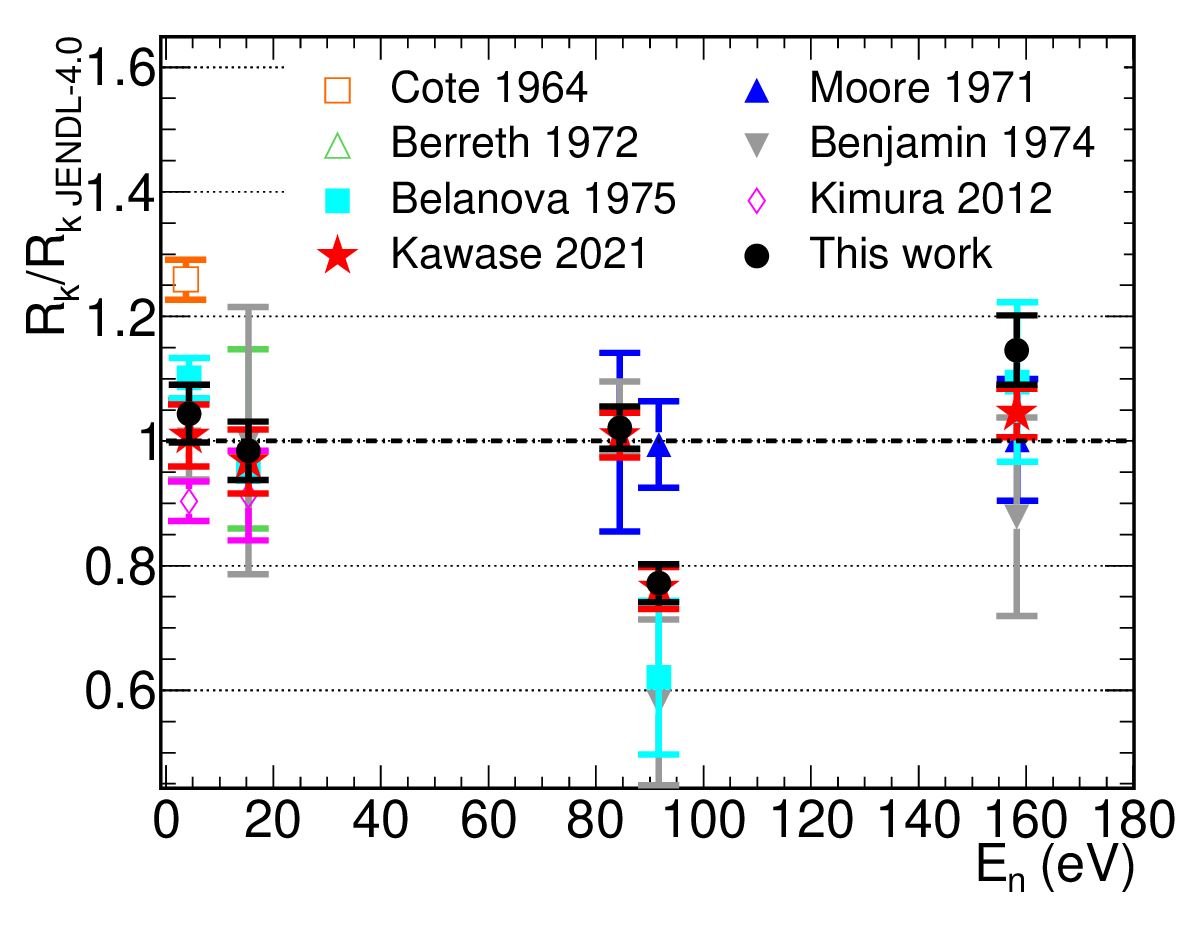}
\includegraphics[width=\columnwidth]{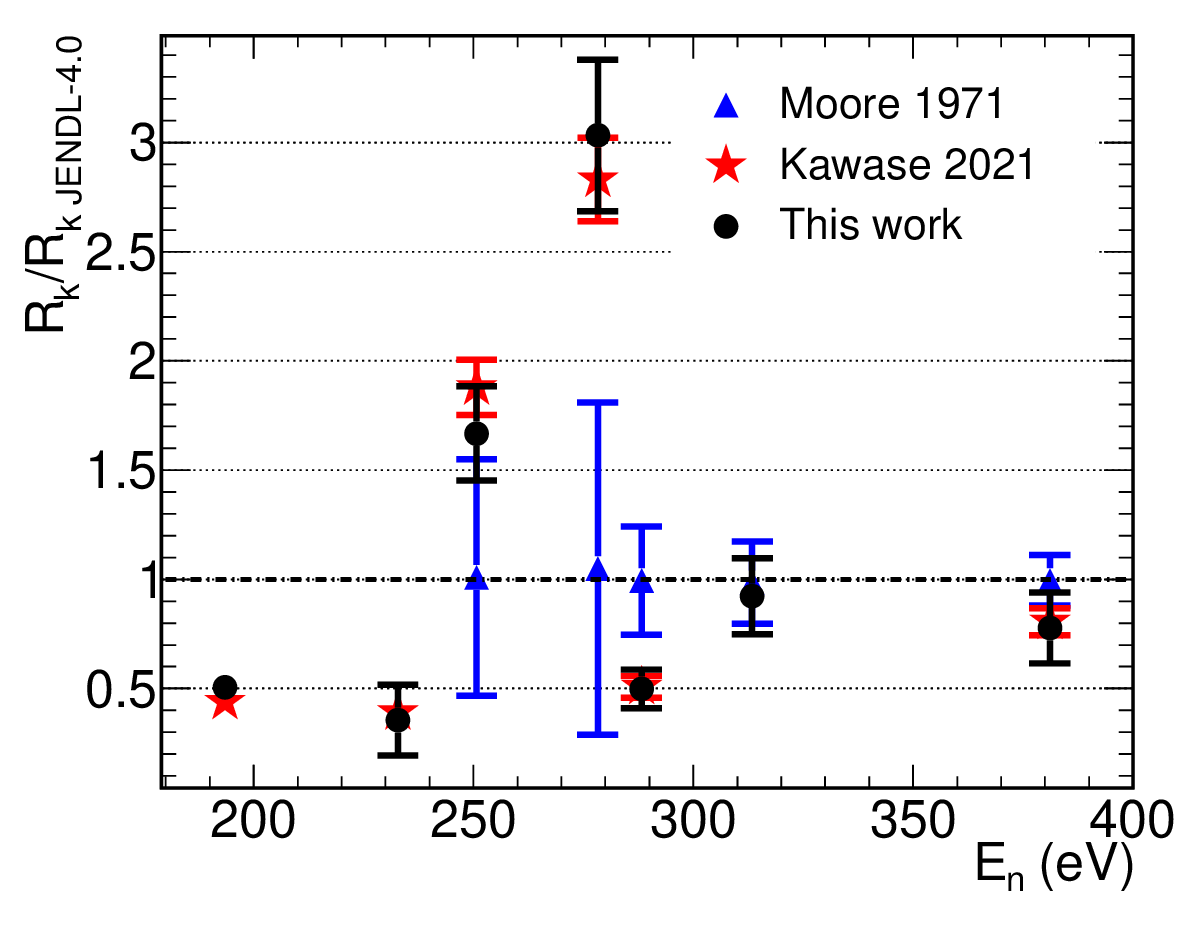}
\caption{Ratio between the $^{246}$Cm R$\mathrm{_k}$ values obtained in different experiments, including this work, and JENDL-4.0.}
\label{fig:Cm246_RK_PrevMeas}
\end{figure}

A total of 13 resonances of $^{246}$Cm have been fit, from 4 to 400 eV, as presented in Table \ref{tab:RP_Cm246}. The total uncertainties in the calculation of the $\Gamma_n$ parameters for the first four resonances are 4-6\% mainly due to the uncertainty in the normalization. The uncertainties at higher energies are dominated by the subtraction of the beam-on background and by counting statistics. The contribution of individual sources to the uncertainty in $\Gamma_n$ is presented in Fig. \ref{fig:Cm246_Uncer}.
 \begin{table*}[t]
\begin{center}
\begin{tabular}{c|c|cccccc|c|c|c|c|c|c}
\hline\hline
\multirow{2}{*}{E$\mathrm{_0}$}&\multirow{2}{*}{$\Gamma\mathrm{_{n}}$}&\multicolumn{10}{c|}{$\Gamma\mathrm{_{n}}$ uncertainty (meV)}&\multicolumn{2}{c}{R$\mathrm{_k}$ (meV)}\\ \cline{3-14}
&&Sta&Fis&Dummy&Gain&RF&Norm&Sum&Sum&\multirow{2}{*}{Total}&Total&\multirow{2}{*}{Value}&\multirow{2}{*}{Uncer.}\\ 
(eV)&(meV)&(U)&(C)&(C)&(U)&(U)&(C)&(C)&(U)&&(\%)&&\\ \hline 
4.3129(2)	&	0.324	&	0.001	&	0.009	&	0.001	&	0.003	&	0.005	&	0.009	&	0.013	&	0.006	&	0.015	&	4.5	&	0.315	&	0.016	\\
 15.310(1) 	&	0.541	&	0.006	&	0.009	&	0.009	&	0.004	&	0.015	&	0.015	&	0.020	&	0.017	&	0.026	&	4.8	&	0.524	&	0.029	\\
84.576(5)	&	23.4	&	0.4	&	0.8	&	0.5	&	0.3	&	0.4	&	0.7	&	1.1	&	0.7	&	1.3	&	5.6	&	13.82	&	0.52	\\
91.945(8) 	&	13.6	&	0.3	&	0.2	&	0.4	&	0.1	&	0.3	&	0.4	&	0.6	&	0.5	&	0.7	&	5.5	&	9.72	&	0.43	\\
 158.41(2) 	&	39.9	&	2.1	&	1.5	&	2.9	&	0.6	&	0.6	&	1.1	&	3.5	&	2.2	&	4.1	&	10	&	18.39	&	0.93	\\
 193.85(5)	&	17.7	&	1.4	&	0.1	&	1.6	&	0.2	&	0.1	&	0.5	&	1.7	&	1.4	&	2.2	&	13	&	11.7	&	1.0	\\
 232.5(2)	&	4.3	&	1.0	&	0.1	&	1.9	&	0.2	&	0.1	&	0.1	&	1.9	&	1.0	&	2.2	&	51	&	3.8	&	1.7	\\
 250.8(1) 	&	18.4	&	2.2	&	0.3	&	2.7	&	0.3	&	0.7	&	0.5	&	2.8	&	2.3	&	3.6	&	20	&	11.9	&	1.6	\\
 278.1(1)	&	31.6	&	4.2	&	1.9	&	4.9	&	0.7	&	1.4	&	0.9	&	5.3	&	4.4	&	6.9	&	22	&	16.2	&	1.9	\\
287.9(2) 	&	17.0	&	2.7	&	0.3	&	3.3	&	0.2	&	0.9	&	0.5	&	3.4	&	2.9	&	4.5	&	26	&	11.3	&	2.0	\\
 306.8(2) 	&	49.1	&	8.5	&	0.5	&	11.2	&	2.6	&	2.4	&	1.4	&	11.3	&	9.1	&	14.5	&	30	&	20.3	&	2.5	\\
 316.6(1) 	&	24.2	&	4.6	&	0.3	&	5.9	&	1.2	&	0.9	&	0.7	&	5.9	&	4.8	&	7.6	&	31	&	14.2	&	2.6	\\
																											
 381.1(2) 	&	59.9	&	18.5	&	2.9	&	28.0	&	1.6	&	3.6	&	1.7	&	28.2	&	18.9	&	34.0	&	57	&	21.9	&	4.6	\\

\hline\hline
 \end{tabular}
 \caption{$\Gamma_n$ and $E_0$ parameters of $^{246}$Cm obtained in this work. The different uncertainties for $\Gamma_n$ presented in the table are due to counting statistics (Stat), subtraction of the background produced by the fission events (Fis), subtraction of the beam-on background (Dummy), correction for the gain shifts of the energy calibration (Gain), determination of the RF (RF) and normalization (Norm). The uncertainty in the capture cross-section of $^{240}$Pu (2.75\%, according to JEFF-3.3) has not been included in the table. The different uncertainties are considered to be fully correlated (C) or fully uncorrelated (U) between the resonances. The quadratic sum of the different uncertainties has been also included. In addition, the radiative kernels with their uncertainties are also in the table.}
\label{tab:RP_Cm246}
\end{center}
\end{table*}
In the fits the $\Gamma_n$ and the energy of the resonances have been fit whereas $\Gamma_\gamma$ and $\Gamma_f$ were fixed. In order to compare our results with previous measurements obtained with different $\Gamma_\gamma$ and $\Gamma_f$ parameters the radiative kernels (R$_k = g_J \Gamma_\gamma\Gamma_n/\Gamma$) which are proportional to the area of the resonances have been calculated and compared in Fig. \ref{fig:Cm246_RK_PrevMeas}. The JENDL-4.0 values are obtained from the Maslov \textit{et al.} evaluation \cite{Maslov_EvalCm246_1996}. This evaluation, performed in 1996, is based on transmission measurements \cite{Cote_Cm244_1964,Berreth_Cm244_1972,Benjamin_Cm248_1974,Belanova_Cm_1975} and on the capture measurement by Moore \textit{et al.} \cite{Moore_Cm244_246_1971}. Details of all the measurements are in Table \ref{tab:PrevMeas}. The JEFF-3.3 and ENDF/B-VIII.0 \cite{Brown_ENDF8_2018} libraries have adopted the JENDL-4.0 values for this isotope. There are two other recent capture measurements by Kimura \textit{et al.} \cite{Kimura_Cm244_2012} and Kawase \textit{et al.} \cite{Kawase_Cm244_2021} performed at J-PARC with germanium detectors \cite{Tin_ANNRI_2009}. 
The latest JENDL-5 evaluation \cite{JENDL5} directly used the resonance parameters of the Kawase \textit{et al.} measurement.

\begin{table}[!ht]\begin{center}
\begin{tabular}{lccc}
\hline \hline 
\multirow{2}{*}{Experiment} & \multirow{2}{*}{Type}&\multicolumn{2}{c}{Energy range}\ \\
& & \textsuperscript{246}Cm& \textsuperscript{248}Cm\ \\
\hline
Cote \textit{et al.} (1964) \cite{Cote_Cm244_1964} & Transm.&4-27 eV &-\\
Moore \textit{et al.} (1971) \cite{Moore_Cm244_246_1971}& Capture &80-380 eV&26 eV\\
Berreth \textit{et al.} (1972) \cite{Berreth_Cm244_1972} & Transm.&4-16 eV &-\ \\
Benjamin \textit{et al.} (1972) \cite{Benjamin_Cm248_1974} & Transm.&4-160 eV &7-3000 eV\\
Belanova \textit{et al.} (1975) \cite{Belanova_Cm_1975} & Transm.&4-160 eV &7-100 eV\\
Kimura \textit{et al.} (2012) \cite{Kimura_Cm244_2012} & Capture &4-16 eV &7-27 eV\\
Kawase \textit{et al.} (2021) \cite{Kawase_Cm244_2021} & Capture &4-550 eV &7-100 eV\\
\hline \hline
\end{tabular}
\caption{Transmission (Transm.) and capture $^{246}$Cm and $^{248}$Cm measurements available in EXFOR in the RRR. Some of the measured yields extend up to higher energies, but the resonance parameters are only reported up to the energies presented in the table.}
\label{tab:PrevMeas}
\end{center}
\end{table}

Discrepancies have been found between the results obtained in this measurement and some of the previous ones. The observed resonance at 306.8 eV, assigned to $^{246}$Cm, was not present in JENDL-4. This assignment is based on the measurement with another Cm sample that contains a different ratio of $^{244}$Cm to $^{246}$Cm, more information on this sample can be found in \cite{Alcayne_Cm244_246WONDER_2019,Alcayne_NDCm244_2019,Alcayne_CmND_23}. In addition, this resonance is too big to be associated with $^{248}$Cm. This resonance was also observed in the recent measurement by Kawase \textit{et al.} with a neutron width of 34.6(31) meV, the value obtained in our measurement is 49(11) meV.

Furthermore, the resonance-like structure of $^{246}$Cm at 360.8 eV was not properly fitted with a radiative width of 34.7 meV, this value has been used to fit the rest of the resonances of $^{246}$Cm above 20 eV because these values were reported in JENDL-4.0. In order to correctly fit this resonance we had to use a value of at least 60 meV for $\Gamma_\gamma$. The reported radiative widths of all resonances were always between 25 and 35 meV and only fluctuations of a few percent for these values are expected. For this reason, we think that this structure is a doublet with a separation between the resonance of 2-3 eV. In the previous measurement by Moore \textit{et al.} the resonance was also considered as a doublet. In the Kawase \textit{et al.} measurement, the resonance is fitted with a radiative width of 34.7 meV.

\begin{figure*}[!htb]
\includegraphics[width=0.49\textwidth]{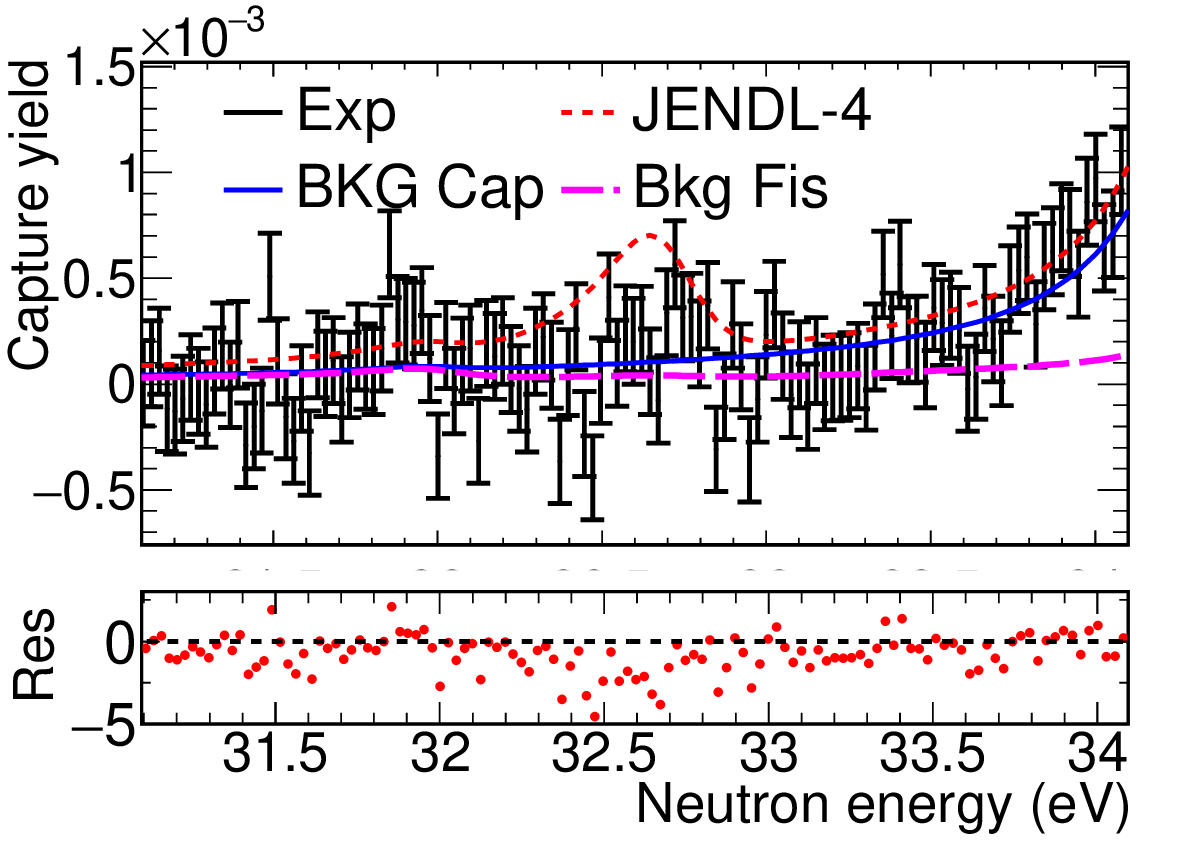}
\includegraphics[width=0.49\textwidth]{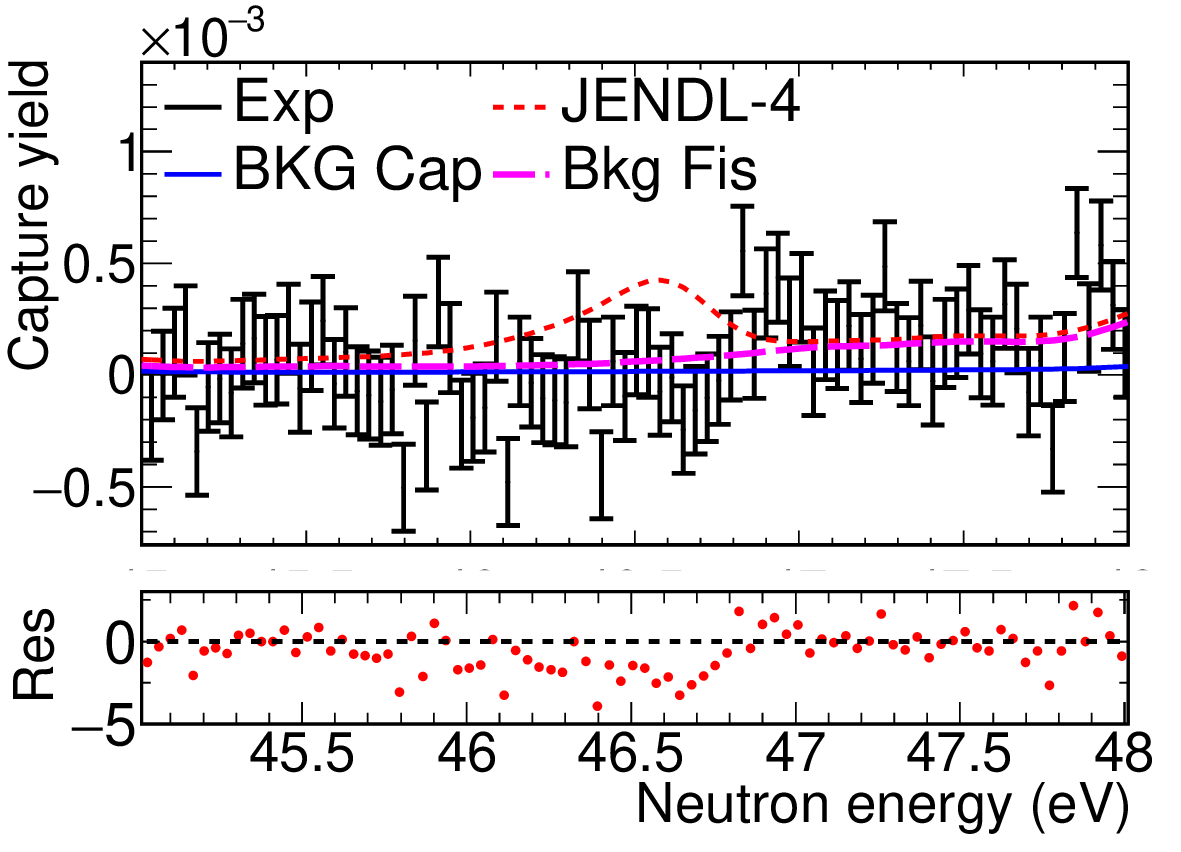}
\includegraphics[width=0.49\textwidth]{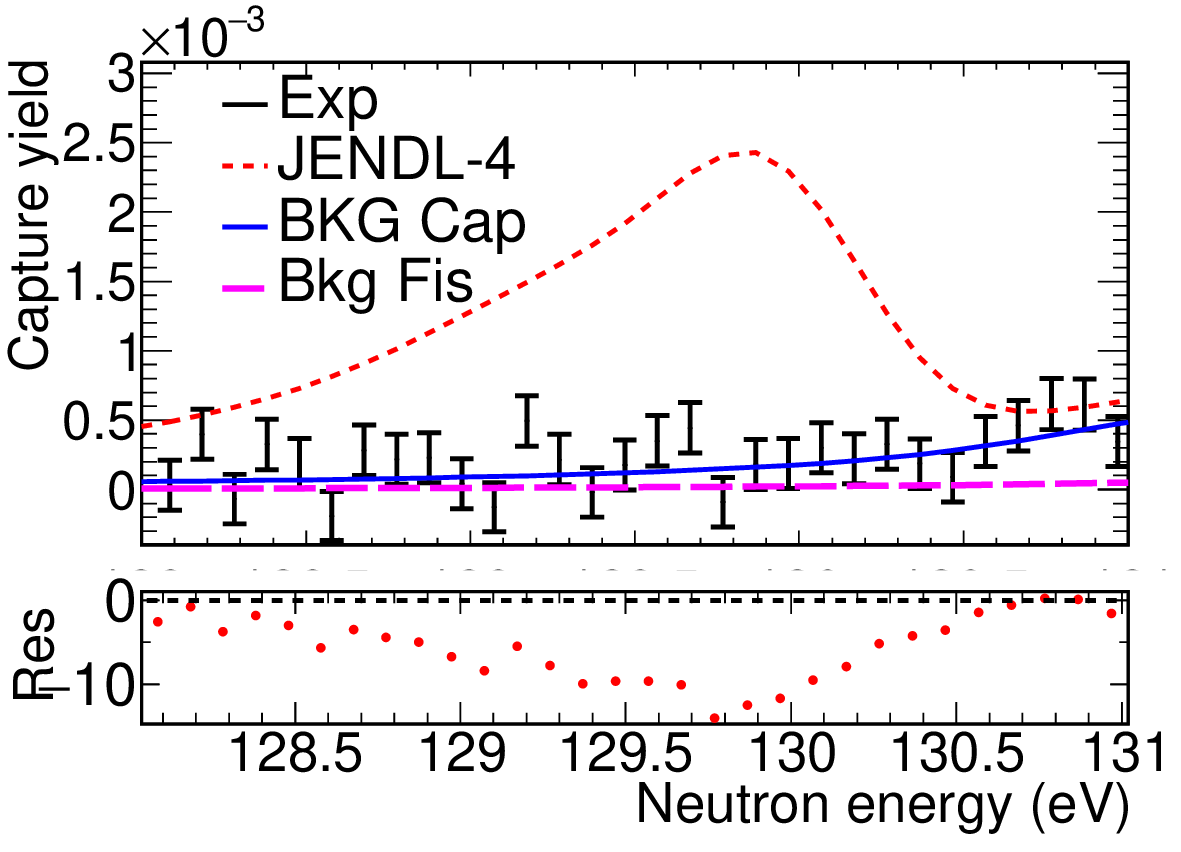}
\caption{Same as Fig. \ref{fig:Cm246_Fit_4_193} but for the resonances at 32.95, 47 and 131 eV.}
\label{fig:Cm246_Fit_35_47_131}
\end{figure*}

In addition, in the JENDL-4.0 library the RP for three weak resonances at 32.95, 47 and 131 eV were given. As shown in Fig. \ref{fig:Cm246_Fit_35_47_131}, our data are not compatible with the existence of the resonances with the areas given. The neutron width allowed by our data would need to be smaller by a factor of 3 to 10 depending on the resonance. Note that these resonances were included in the Maslov \textit{et al.} evaluation without being observed in any previous measurement.

The R$\mathrm{_k}$ values obtained for the first three resonances at 4.31, 15.31 and 84.58 eV are compatible with the previous measurements and with JENDL-4.0 and JENDL-5.
The remaining resonances are within one standard deviation not compatible with JENDL-4.0, but are compatible with the Kawase \textit{et al.} experiment, and thus with JENDL-5. For neutron energies below 160 eV, n\_TOF provided R$\mathrm{_k}$ values with uncertainties smaller than 5\%, which are similar to those of Kawase \textit{et al.} and considerably smaller than in the previous capture and transmission measurements. The uncertainties at higher energies are between 10 and 20\% for all but one resonance.

\subsection{Resonance parameters of \textsuperscript{248}Cm}
\begin{figure}[htb]
\includegraphics[width=\columnwidth]{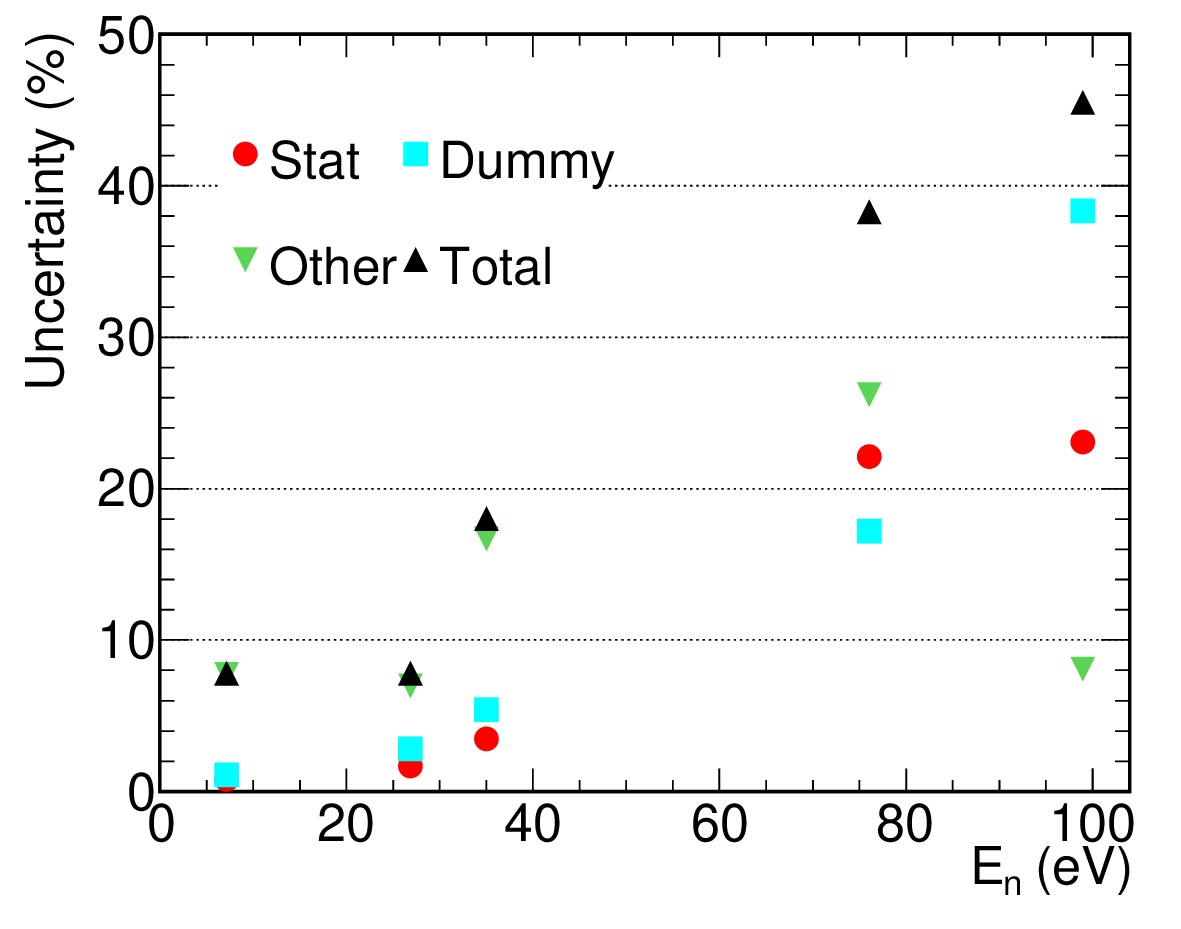}
\caption{Total uncertainties of the $\Gamma\mathrm{_{n}}$ parameters of $^{248}$Cm obtained in this work in percentage (Total). The uncertainties due to counting statistics (Stat), due to the subtraction of the beam-on background (Dummy) and the rest of the uncertainties (Other) are also presented for comparison.} 
\label{fig:Cm248_Uncer}
\end{figure}

A total of 5 resonances of $^{248}$Cm have been observed and fitted in the analysis from 7 to 100 eV, as presented in Table \ref{tab:RP_Cm248}. The uncertainties in the calculation of the $\Gamma_n$ parameters for the first two resonances are below 10\% and for the other three are below 50\%, as shown in Fig. \ref{fig:Cm248_Uncer}.
The JENDL-4.0, JEFF-3.3, ENDF/B-VIII.0 and JENDL-5 libraries adopt the same evaluation for this isotope, which is the one performed by Kikuchi \textit{et al.} \cite{Kikuchi_EvalCm248_1984}. This evaluation considers only the transmission data of Benjamin \textit{et al.} \cite{Benjamin_Cm248_1974} and the capture data of Moore \textit{et al.} \cite{Moore_Cm244_246_1971}. 

\begin{figure}[htb]
\includegraphics[width=\columnwidth]{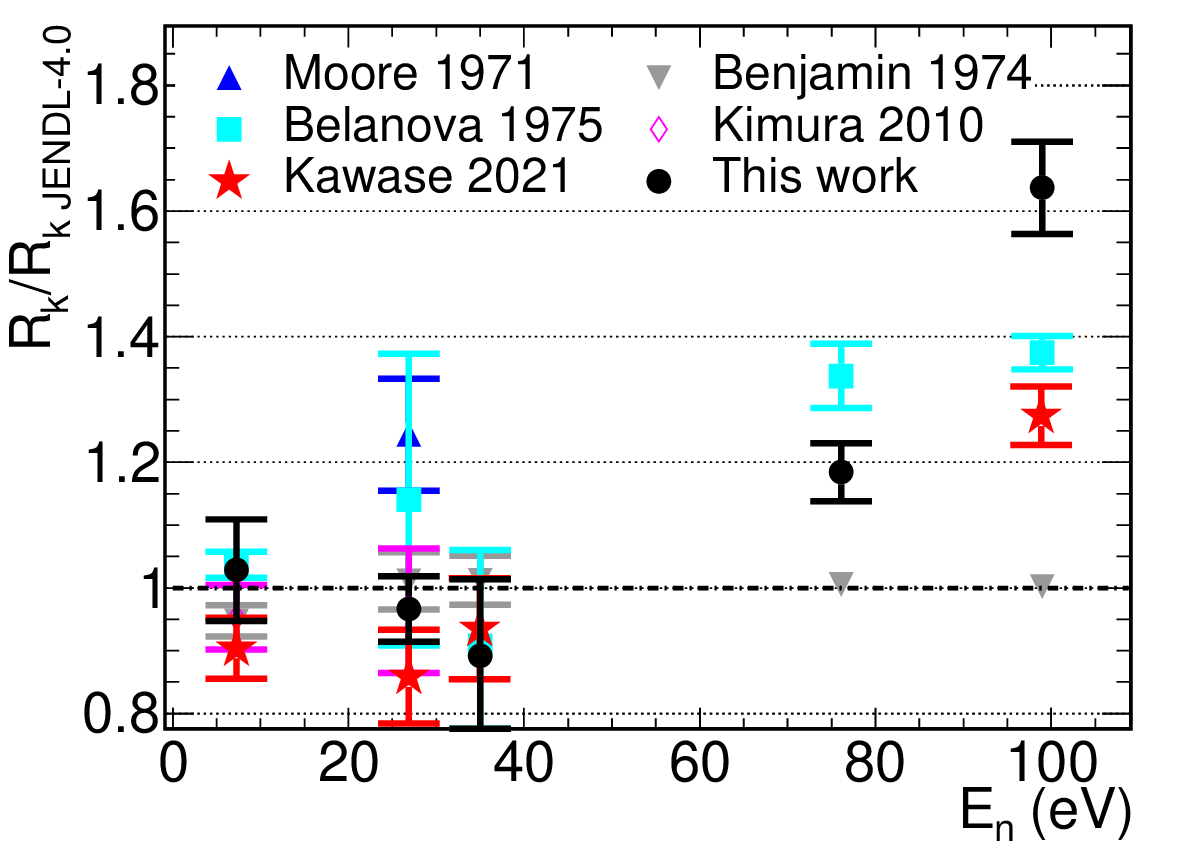}
\caption{Ratio between the $^{248}$Cm R$\mathrm{_k}$ values obtained in different experiments, including this work, and JENDL-4.0.}
\label{fig:Cm248_RK_PrevMeas}
\end{figure}

The values of the R$_k$ obtained for the first three resonances are compatible with the previous measurements and evaluations (Fig. \ref{fig:Cm248_RK_PrevMeas}). However, the values obtained for the other two resonances at 76 and 98.8 eV are 20\% and 65\% larger than the Kikuchi \textit{et al.} evaluation and are not compatible within one sigma with any other previous measurement. The uncertainties in the R$_k$ for the $^{248}$Cm resonances in this work are below 15\%.
 \begin{table*}[t]
\begin{center}
\begin{tabular}{c|c|cccccc|c|c|c|c|c|c}
\hline\hline
\multirow{2}{*}{E$\mathrm{_0}$}&\multirow{2}{*}{$\Gamma\mathrm{_{n}}$}&\multicolumn{10}{c|}{$\Gamma\mathrm{_{n}}$ uncertainty (meV)}&\multicolumn{2}{c}{R$\mathrm{_k}$ (meV)}\\ \cline{3-14}
&&Sta&Fis&Dummy&Gain&RF&Nor&Sum&Sum&\multirow{2}{*}{Total}&Total&\multirow{2}{*}{Value}&\multirow{2}{*}{Uncer.}\\ 
(eV)&(meV)&(U)&(C)&(C)&(U)&(U)&(C)&(C)&(U)&&(\%)&&\\ \hline 
7.246(1)&1.94&0.01&0.05&0.02&0.02&0.11&0.08&0.10&0.11&0.15&7.8&1.79&0.14 \\
 26.90(1)&18.2&0.3&0.2&0.5&0.2&1.0&0.8&0.9&1.0&1.4&7.7&11.56&0.60 \\
 35.01(1) &9.8&0.3&1.2&0.5&0.3&1.0&0.4&1.3&1.1&1.8&18&7.4&1.0 \\
 76.07(2) &268&59&61&46&7&31&11&77&67&102&36&23.37&0.91 \\
 98.84(4)&371&85&23&142&10&3&16&145&86&169&45&36.1&1.6 \\
\hline\hline \end{tabular}
 \caption{Same as Table \ref{tab:RP_Cm246} but for the $^{248}$Cm resonances.}
\label{tab:RP_Cm248}
\end{center}
\end{table*}

\section{Summary and conclusions}\label{sec:Conclusions}

The capture cross-sections of $^{246}$Cm and $^{248}$Cm have been measured at the EAR2 of the n\_TOF facility with three C$_6$D$_6$ detectors. This work is the first analysis of a capture measurement at n\_TOF EAR2, profiting from the high instantaneous fluence in this area to measure a sample with a very low mass ($\sim$1 mg of $^{246}$Cm and $\sim$0.2 mg of $^{248}$Cm ) and high radioactivity ($\sim$2 GBq). A total of 13 resonances of $^{246}$Cm between 4 and 400 eV and 5 resonances of $^{248}$Cm between 7 and 100 eV have been observed and analyzed. The radiative kernels of $^{246}$Cm have been obtained with an uncertainty of less than 5\% at energies below 160 eV. At higher energies, the uncertainties are between 10 and 20\% for the majority of the resonances. For $^{248}$Cm, the uncertainties in radiative kernels are around 10\% as a result of the increase of the uncertainties due to counting statistics. The radiative kernels of the first three resonances of $^{246}$Cm obtained in this work are compatible with all the previous measurements and evaluations. For the remaining $^{246}$Cm resonances, the radiative kernels are only compatible within one sigma with the recent measurement by Kawase \textit{et al}. In the case of $^{248}$Cm also the radiative kernels of the first three resonances are compatible with all previous measurements and evaluations, for the other two resonances the radiative kernels obtained are considerably higher than evaluations and previous experiments.


\begin{acknowledgments}
This work was supported in part by the I+D+i grant PGC2018-096717-B-C21 funded by MCIN/AEI/10.13039/501100011033, by project PID2021-123100NB-I00 funded by MCIN/AEI/10.13039/501100011033/FEDER, UE, by project PCI2022-135037-2 funded by MCIN/AEI/10.13039/501100011033/ and European Union NextGenerationEU/PRTR, by the European Commission H2020 Framework Programme project SANDA (Grant agreement ID: 847552), Croatian Science Foundation project IP-2022-10-3878 and by funding agencies of the n\_TOF participating institutions.

\end{acknowledgments}

\section*{APPENDIX: FITS OF THE RESONANCES}\label{Appen:Sec:Fits}

\bibliography{Cm246}

\begin{thebibliography}{59}%
\makeatletter
\providecommand \@ifxundefined [1]{%
 \@ifx{#1\undefined}
}%
\providecommand \@ifnum [1]{%
 \ifnum #1\expandafter \@firstoftwo
 \else \expandafter \@secondoftwo
 \fi
}%
\providecommand \@ifx [1]{%
 \ifx #1\expandafter \@firstoftwo
 \else \expandafter \@secondoftwo
 \fi
}%
\providecommand \natexlab [1]{#1}%
\providecommand \enquote  [1]{``#1''}%
\providecommand \bibnamefont  [1]{#1}%
\providecommand \bibfnamefont [1]{#1}%
\providecommand \citenamefont [1]{#1}%
\providecommand \href@noop [0]{\@secondoftwo}%
\providecommand \href [0]{\begingroup \@sanitize@url \@href}%
\providecommand \@href[1]{\@@startlink{#1}\@@href}%
\providecommand \@@href[1]{\endgroup#1\@@endlink}%
\providecommand \@sanitize@url [0]{\catcode `\\12\catcode `\$12\catcode
  `\&12\catcode `\#12\catcode `\^12\catcode `\_12\catcode `\%12\relax}%
\providecommand \@@startlink[1]{}%
\providecommand \@@endlink[0]{}%
\providecommand \url  [0]{\begingroup\@sanitize@url \@url }%
\providecommand \@url [1]{\endgroup\@href {#1}{\urlprefix }}%
\providecommand \urlprefix  [0]{URL }%
\providecommand \Eprint [0]{\href }%
\providecommand \doibase [0]{http://dx.doi.org/}%
\providecommand \selectlanguage [0]{\@gobble}%
\providecommand \bibinfo  [0]{\@secondoftwo}%
\providecommand \bibfield  [0]{\@secondoftwo}%
\providecommand \translation [1]{[#1]}%
\providecommand \BibitemOpen [0]{}%
\providecommand \bibitemStop [0]{}%
\providecommand \bibitemNoStop [0]{.\EOS\space}%
\providecommand \EOS [0]{\spacefactor3000\relax}%
\providecommand \BibitemShut  [1]{\csname bibitem#1\endcsname}%
\let\auto@bib@innerbib\@empty
\bibitem [{\citenamefont {Aliberti}\ \emph {et~al.}(2006)\citenamefont
  {Aliberti} \emph {et~al.}}]{Aliberti_TargetAcurracy_2006}%
  \BibitemOpen
  \bibfield  {author} {\bibinfo {author} {\bibfnamefont {G.}~\bibnamefont
  {Aliberti}} \emph {et~al.},\ }\href {\doibase 10.1016/j.anucene.2006.02.003}
  {\bibfield  {journal} {\bibinfo  {journal} {Ann. Nucl. Energy}\ }\textbf
  {\bibinfo {volume} {33}},\ \bibinfo {pages} {700} (\bibinfo {year}
  {2006})}\BibitemShut {NoStop}%
\bibitem [{\citenamefont {Weiss}\ \emph {et~al.}(2015)\citenamefont {Weiss}
  \emph {et~al.}}]{Weis_EAR2_2015}%
  \BibitemOpen
  \bibfield  {author} {\bibinfo {author} {\bibfnamefont {C.}~\bibnamefont
  {Weiss}} \emph {et~al.} (\bibinfo {collaboration} {The n\_TOF
  Collaboration}),\ }\href {\doibase 10.1016/j.nima.2015.07.027} {\bibfield
  {journal} {\bibinfo  {journal} {Nucl. Instrum. Methods A}\ }\textbf {\bibinfo
  {volume} {799}},\ \bibinfo {pages} {90} (\bibinfo {year} {2015})}\BibitemShut
  {NoStop}%
\bibitem [{\citenamefont {Sabat{\'e}-Gilarte}\ \emph
  {et~al.}(2017{\natexlab{a}})\citenamefont {Sabat{\'e}-Gilarte} \emph
  {et~al.}}]{Sabate_FluxEar2_2017}%
  \BibitemOpen
  \bibfield  {author} {\bibinfo {author} {\bibfnamefont {M.}~\bibnamefont
  {Sabat{\'e}-Gilarte}} \emph {et~al.} (\bibinfo {collaboration} {The n\_TOF
  Collaboration}),\ }\href {\doibase 10.1140/epja/i2017-12392-4} {\bibfield
  {journal} {\bibinfo  {journal} {Eur. Phys. J. A}\ }\textbf {\bibinfo {volume}
  {53}},\ \bibinfo {pages} {210} (\bibinfo {year}
  {2017}{\natexlab{a}})}\BibitemShut {NoStop}%
\bibitem [{\citenamefont {Plag}\ \emph {et~al.}(2003)\citenamefont {Plag} \emph
  {et~al.}}]{Plag_C6D6Detectors_2003}%
  \BibitemOpen
  \bibfield  {author} {\bibinfo {author} {\bibfnamefont {R.}~\bibnamefont
  {Plag}} \emph {et~al.} (\bibinfo {collaboration} {The n\_TOF
  Collaboration}),\ }\href {\doibase 10.1016/s0168-9002(02)01749-7} {\bibfield
  {journal} {\bibinfo  {journal} {Nucl. Instrum. Methods A}\ }\textbf {\bibinfo
  {volume} {496}},\ \bibinfo {pages} {425} (\bibinfo {year}
  {2003})}\BibitemShut {NoStop}%
\bibitem [{\citenamefont {Alcayne}\ \emph {et~al.}(2019)\citenamefont {Alcayne}
  \emph {et~al.}}]{Alcayne_Cm244_246WONDER_2019}%
  \BibitemOpen
  \bibfield  {author} {\bibinfo {author} {\bibfnamefont {V.}~\bibnamefont
  {Alcayne}} \emph {et~al.} (\bibinfo {collaboration} {The n\_TOF
  Collaboration}),\ }\href {\doibase 10.1051/epjconf/201921103008} {\bibfield
  {journal} {\bibinfo  {journal} {EPJ Web Conf.}\ }\textbf {\bibinfo {volume}
  {211}},\ \bibinfo {pages} {03008} (\bibinfo {year} {2019})}\BibitemShut
  {NoStop}%
\bibitem [{\citenamefont {Alcayne}\ \emph {et~al.}(2020)\citenamefont {Alcayne}
  \emph {et~al.}}]{Alcayne_NDCm244_2019}%
  \BibitemOpen
  \bibfield  {author} {\bibinfo {author} {\bibfnamefont {V.}~\bibnamefont
  {Alcayne}} \emph {et~al.} (\bibinfo {collaboration} {The n\_TOF
  Collaboration}),\ }\href {\doibase 10.1051/epjconf/202023901034} {\bibfield
  {journal} {\bibinfo  {journal} {EPJ Web Conf.}\ }\textbf {\bibinfo {volume}
  {239}},\ \bibinfo {pages} {01034} (\bibinfo {year} {2020})}\BibitemShut
  {NoStop}%
\bibitem [{\citenamefont {Alcayne}\ \emph {et~al.}(2023)\citenamefont {Alcayne}
  \emph {et~al.}}]{Alcayne_CmND_23}%
  \BibitemOpen
  \bibfield  {author} {\bibinfo {author} {\bibfnamefont {V.}~\bibnamefont
  {Alcayne}} \emph {et~al.} (\bibinfo {collaboration} {The n\_TOF
  Collaboration}),\ }\href {\doibase 10.1051/epjconf/202328401009} {\bibfield
  {journal} {\bibinfo  {journal} {EPJ Web of Conf.}\ }\textbf {\bibinfo
  {volume} {284}},\ \bibinfo {pages} {01009} (\bibinfo {year}
  {2023})}\BibitemShut {NoStop}%
\bibitem [{\citenamefont {Guerrero}\ \emph {et~al.}(2013)\citenamefont
  {Guerrero} \emph {et~al.}}]{Guerrero_EAR1_2013}%
  \BibitemOpen
  \bibfield  {author} {\bibinfo {author} {\bibfnamefont {C.}~\bibnamefont
  {Guerrero}} \emph {et~al.} (\bibinfo {collaboration} {The n\_TOF
  Collaboration}),\ }\href {\doibase 10.1140/epja/i2013-13027-6} {\bibfield
  {journal} {\bibinfo  {journal} {Eur. Phys. J. A}\ }\textbf {\bibinfo {volume}
  {49}},\ \bibinfo {pages} {27} (\bibinfo {year} {2013})}\BibitemShut {NoStop}%
\bibitem [{\citenamefont {Kimura}\ \emph {et~al.}(2012)\citenamefont {Kimura}
  \emph {et~al.}}]{Kimura_Cm244_2012}%
  \BibitemOpen
  \bibfield  {author} {\bibinfo {author} {\bibfnamefont {A.}~\bibnamefont
  {Kimura}} \emph {et~al.},\ }\href {\doibase 10.1080/00223131.2012.693887}
  {\bibfield  {journal} {\bibinfo  {journal} {J. Nucl. Sci. Technol.}\ }\textbf
  {\bibinfo {volume} {49}},\ \bibinfo {pages} {708} (\bibinfo {year}
  {2012})}\BibitemShut {NoStop}%
\bibitem [{\citenamefont {Kawase}\ \emph {et~al.}(2021)\citenamefont {Kawase}
  \emph {et~al.}}]{Kawase_Cm244_2021}%
  \BibitemOpen
  \bibfield  {author} {\bibinfo {author} {\bibfnamefont {S.}~\bibnamefont
  {Kawase}} \emph {et~al.},\ }\href {\doibase 10.1080/00223131.2020.1864492}
  {\bibfield  {journal} {\bibinfo  {journal} {J. Nucl. Sci. Technol.}\ }\textbf
  {\bibinfo {volume} {58}},\ \bibinfo {pages} {764} (\bibinfo {year}
  {2021})}\BibitemShut {NoStop}%
\bibitem [{\citenamefont {Moore}\ and\ \citenamefont
  {Keyworth}(1971)}]{Moore_Cm244_246_1971}%
  \BibitemOpen
  \bibfield  {author} {\bibinfo {author} {\bibfnamefont {M.}~\bibnamefont
  {Moore}}\ and\ \bibinfo {author} {\bibfnamefont {G.}~\bibnamefont
  {Keyworth}},\ }\href {\doibase 10.1103/PhysRevC.3.1656} {\bibfield  {journal}
  {\bibinfo  {journal} {Phys. Rev. C}\ }\textbf {\bibinfo {volume} {3}},\
  \bibinfo {pages} {1656} (\bibinfo {year} {1971})}\BibitemShut {NoStop}%
\bibitem [{\citenamefont {Barbagallo}\ \emph {et~al.}(2016)\citenamefont
  {Barbagallo} \emph {et~al.}}]{Barbagallo_Be7_2016}%
  \BibitemOpen
  \bibfield  {author} {\bibinfo {author} {\bibfnamefont {M.}~\bibnamefont
  {Barbagallo}} \emph {et~al.} (\bibinfo {collaboration} {The n\_TOF
  Collaboration}),\ }\href {\doibase 10.1103/PhysRevLett.117.152701} {\bibfield
   {journal} {\bibinfo  {journal} {Phys. Rev. Lett.}\ }\textbf {\bibinfo
  {volume} {117}},\ \bibinfo {pages} {152701} (\bibinfo {year}
  {2016})}\BibitemShut {NoStop}%
\bibitem [{\citenamefont {Sabat{\'e}-Gilarte}\ \emph
  {et~al.}(2017{\natexlab{b}})\citenamefont {Sabat{\'e}-Gilarte} \emph
  {et~al.}}]{Sabate_33Sn_2017}%
  \BibitemOpen
  \bibfield  {author} {\bibinfo {author} {\bibfnamefont {M.}~\bibnamefont
  {Sabat{\'e}-Gilarte}} \emph {et~al.} (\bibinfo {collaboration} {The n\_TOF
  Collaboration}),\ }\href {\doibase 10.1051/epjconf/201714608004} {\bibfield
  {journal} {\bibinfo  {journal} {EPJ Web Conf.}\ }\textbf {\bibinfo {volume}
  {146}},\ \bibinfo {pages} {08004} (\bibinfo {year}
  {2017}{\natexlab{b}})}\BibitemShut {NoStop}%
\bibitem [{\citenamefont {Damone}\ \emph {et~al.}(2018)\citenamefont {Damone}
  \emph {et~al.}}]{Damone_Be7_2018}%
  \BibitemOpen
  \bibfield  {author} {\bibinfo {author} {\bibfnamefont {L.}~\bibnamefont
  {Damone}} \emph {et~al.} (\bibinfo {collaboration} {The n\_TOF
  Collaboration}),\ }\href {\doibase 10.1103/PhysRevLett.121.042701} {\bibfield
   {journal} {\bibinfo  {journal} {Phys. Rev. Lett.}\ }\textbf {\bibinfo
  {volume} {121}},\ \bibinfo {pages} {042701} (\bibinfo {year}
  {2018})}\BibitemShut {NoStop}%
\bibitem [{\citenamefont {Stamatopoulos}\ \emph {et~al.}(2020)\citenamefont
  {Stamatopoulos} \emph {et~al.}}]{Stamatopoulos_Pu240_2020}%
  \BibitemOpen
  \bibfield  {author} {\bibinfo {author} {\bibfnamefont {A.}~\bibnamefont
  {Stamatopoulos}} \emph {et~al.} (\bibinfo {collaboration} {The n\_TOF
  Collaboration}),\ }\href {\doibase 10.1103/PhysRevC.102.014616} {\bibfield
  {journal} {\bibinfo  {journal} {Phys. Rev. C}\ }\textbf {\bibinfo {volume}
  {102}},\ \bibinfo {pages} {014616} (\bibinfo {year} {2020})}\BibitemShut
  {NoStop}%
\bibitem [{\citenamefont {Marrone}\ \emph {et~al.}(2006)\citenamefont {Marrone}
  \emph {et~al.}}]{Marrone_Sm151_2006}%
  \BibitemOpen
  \bibfield  {author} {\bibinfo {author} {\bibfnamefont {S.}~\bibnamefont
  {Marrone}} \emph {et~al.} (\bibinfo {collaboration} {The n\_TOF
  Collaboration}),\ }\href {\doibase 10.1103/PhysRevC.73.034604} {\bibfield
  {journal} {\bibinfo  {journal} {Phys. Rev. C}\ }\textbf {\bibinfo {volume}
  {73}},\ \bibinfo {pages} {034604} (\bibinfo {year} {2006})}\BibitemShut
  {NoStop}%
\bibitem [{\citenamefont {Domingo-Pardo}\ \emph {et~al.}(2006)\citenamefont
  {Domingo-Pardo} \emph {et~al.}}]{Domingo_Bi209_2006}%
  \BibitemOpen
  \bibfield  {author} {\bibinfo {author} {\bibfnamefont {C.}~\bibnamefont
  {Domingo-Pardo}} \emph {et~al.} (\bibinfo {collaboration} {The n\_TOF
  Collaboration}),\ }\href {\doibase 10.1103/PhysRevC.74.025807} {\bibfield
  {journal} {\bibinfo  {journal} {Phys. Rev. C}\ }\textbf {\bibinfo {volume}
  {74}},\ \bibinfo {pages} {025807} (\bibinfo {year} {2006})}\BibitemShut
  {NoStop}%
\bibitem [{\citenamefont {Casanovas}\ \emph {et~al.}(2020)\citenamefont
  {Casanovas} \emph {et~al.}}]{Casanovas_204Tl_2020}%
  \BibitemOpen
  \bibfield  {author} {\bibinfo {author} {\bibfnamefont {A.}~\bibnamefont
  {Casanovas}} \emph {et~al.} (\bibinfo {collaboration} {The n\_TOF
  Collaboration}),\ }\href {\doibase 10.1088/1742-6596/1668/1/012005}
  {\bibfield  {journal} {\bibinfo  {journal} {J. Phys. Conf. Series}\ }\textbf
  {\bibinfo {volume} {1668}},\ \bibinfo {pages} {012005} (\bibinfo {year}
  {2020})}\BibitemShut {NoStop}%
\bibitem [{\citenamefont {Gunsing}\ \emph {et~al.}(2012)\citenamefont {Gunsing}
  \emph {et~al.}}]{Gunsing_Th232_2012}%
  \BibitemOpen
  \bibfield  {author} {\bibinfo {author} {\bibfnamefont {F.}~\bibnamefont
  {Gunsing}} \emph {et~al.} (\bibinfo {collaboration} {The n\_TOF
  Collaboration}),\ }\href {\doibase 10.1103/PhysRevC.85.064601} {\bibfield
  {journal} {\bibinfo  {journal} {Phys. Rev. C}\ }\textbf {\bibinfo {volume}
  {85}},\ \bibinfo {pages} {064601} (\bibinfo {year} {2012})}\BibitemShut
  {NoStop}%
\bibitem [{\citenamefont {Fraval}\ \emph {et~al.}(2014)\citenamefont {Fraval}
  \emph {et~al.}}]{Fraval_Am241_2014}%
  \BibitemOpen
  \bibfield  {author} {\bibinfo {author} {\bibfnamefont {K.}~\bibnamefont
  {Fraval}} \emph {et~al.} (\bibinfo {collaboration} {The n\_TOF
  Collaboration}),\ }\href {\doibase 10.1103/PhysRevC.89.044609} {\bibfield
  {journal} {\bibinfo  {journal} {Phys. Rev. C}\ }\textbf {\bibinfo {volume}
  {89}},\ \bibinfo {pages} {044609} (\bibinfo {year} {2014})}\BibitemShut
  {NoStop}%
\bibitem [{\citenamefont {Mingrone}\ \emph {et~al.}(2017)\citenamefont
  {Mingrone} \emph {et~al.}}]{Mingrone_U238_2017}%
  \BibitemOpen
  \bibfield  {author} {\bibinfo {author} {\bibfnamefont {F.}~\bibnamefont
  {Mingrone}} \emph {et~al.} (\bibinfo {collaboration} {The n\_TOF
  Collaboration}),\ }\href {\doibase 10.1103/PhysRevC.95.034604} {\bibfield
  {journal} {\bibinfo  {journal} {Phys. Rev. C}\ }\textbf {\bibinfo {volume}
  {95}},\ \bibinfo {pages} {034604} (\bibinfo {year} {2017})}\BibitemShut
  {NoStop}%
\bibitem [{\citenamefont {Mastromarco}\ \emph {et~al.}(2017)\citenamefont
  {Mastromarco} \emph {et~al.}}]{Mastromarco_U236_2017}%
  \BibitemOpen
  \bibfield  {author} {\bibinfo {author} {\bibfnamefont {M.}~\bibnamefont
  {Mastromarco}} \emph {et~al.} (\bibinfo {collaboration} {The n\_TOF
  Collaboration}),\ }\href {\doibase 10.1051/epjconf/201714611054} {\bibfield
  {journal} {\bibinfo  {journal} {EPJ Web Conf.}\ }\textbf {\bibinfo {volume}
  {146}},\ \bibinfo {pages} {11054} (\bibinfo {year} {2017})}\BibitemShut
  {NoStop}%
\bibitem [{\citenamefont {Lerendegui-Marco}\ \emph {et~al.}(2018)\citenamefont
  {Lerendegui-Marco} \emph {et~al.}}]{Lerendegui_Pu242_2018}%
  \BibitemOpen
  \bibfield  {author} {\bibinfo {author} {\bibfnamefont {J.}~\bibnamefont
  {Lerendegui-Marco}} \emph {et~al.} (\bibinfo {collaboration} {The n\_TOF
  Collaboration}),\ }\href {\doibase 10.1103/PhysRevC.97.024605} {\bibfield
  {journal} {\bibinfo  {journal} {Phys. Rev. C}\ }\textbf {\bibinfo {volume}
  {97}},\ \bibinfo {pages} {024605} (\bibinfo {year} {2018})}\BibitemShut
  {NoStop}%
\bibitem [{\citenamefont {Agostinelli}\ \emph {et~al.}(2003)\citenamefont
  {Agostinelli} \emph {et~al.}}]{Agostinelli_GEANT4_2003}%
  \BibitemOpen
  \bibfield  {author} {\bibinfo {author} {\bibfnamefont {S.}~\bibnamefont
  {Agostinelli}} \emph {et~al.} (\bibinfo {collaboration} {the GEANT4
  Collaboration}),\ }\href {\doibase 10.1016/S0168-9002(03)01368-8} {\bibfield
  {journal} {\bibinfo  {journal} {Nucl. Instrum. Methods A}\ }\textbf {\bibinfo
  {volume} {506}},\ \bibinfo {pages} {250} (\bibinfo {year}
  {2003})}\BibitemShut {NoStop}%
\bibitem [{\citenamefont {Cosentino}\ and\ \citenamefont
  {others.}(2015)}]{Consentino_Simon_2015}%
  \BibitemOpen
  \bibfield  {author} {\bibinfo {author} {\bibfnamefont {L.}~\bibnamefont
  {Cosentino}}\ and\ \bibinfo {author} {\bibnamefont {others.}},\ }\href
  {\doibase 10.1063/1.4927073} {\bibfield  {journal} {\bibinfo  {journal} {Rev.
  Sci. Instrum.}\ }\textbf {\bibinfo {volume} {86}},\ \bibinfo {pages} {073509}
  (\bibinfo {year} {2015})}\BibitemShut {NoStop}%
\bibitem [{\citenamefont {Carlson}\ \emph {et~al.}(2009)\citenamefont {Carlson}
  \emph {et~al.}}]{iaea_StandartLibraires_2007}%
  \BibitemOpen
  \bibfield  {author} {\bibinfo {author} {\bibfnamefont {A.}~\bibnamefont
  {Carlson}} \emph {et~al.},\ }\href {\doibase 10.1016/j.nds.2009.11.001}
  {\bibfield  {journal} {\bibinfo  {journal} {Nucl. Data Sheets}\ }\textbf
  {\bibinfo {volume} {110}},\ \bibinfo {pages} {3215} (\bibinfo {year}
  {2009})}\BibitemShut {NoStop}%
\bibitem [{\citenamefont {{U. Abbondanno}}\ \emph {et~al.}(2005)\citenamefont
  {{U. Abbondanno}} \emph {et~al.}}]{Abbondano_DAQ_2005}%
  \BibitemOpen
  \bibfield  {author} {\bibinfo {author} {\bibnamefont {{U. Abbondanno}}} \emph
  {et~al.} (\bibinfo {collaboration} {The n\_TOF Collaboration}),\ }\href
  {\doibase 10.1016/j.nima.2004.09.002} {\bibfield  {journal} {\bibinfo
  {journal} {Nucl. Instrum. Methods A}\ }\textbf {\bibinfo {volume} {538}},\
  \bibinfo {pages} {692} (\bibinfo {year} {2005})}\BibitemShut {NoStop}%
\bibitem [{\citenamefont {{\v Zugec}}(2016)}]{Zugec_DataProcessing_2016}%
  \BibitemOpen
  \bibfield  {author} {\bibinfo {author} {\bibfnamefont {P.}~\bibnamefont {{\v
  Zugec}}} (\bibinfo {collaboration} {The n\_TOF Collaboration}),\ }\href
  {\doibase 10.1016/j.nima.2015.12.054} {\bibfield  {journal} {\bibinfo
  {journal} {Nucl. Instrum. Methods A}\ }\textbf {\bibinfo {volume} {812}},\
  \bibinfo {pages} {134} (\bibinfo {year} {2016})}\BibitemShut {NoStop}%
\bibitem [{\citenamefont {{C. Guerrero}}\ \emph {et~al.}(2008)\citenamefont
  {{C. Guerrero}}, \citenamefont {{D. Cano-Ott}}, \citenamefont
  {Fern{\'a}ndez-Ord{\'o}{\~n}ez}, \citenamefont {Gonz{\'a}lez-Romero},
  \citenamefont {Mart\'{\i}nez},\ and\ \citenamefont
  {Villamar\'{\i}n}}]{Guerrero_SignalBC501A_2008}%
  \BibitemOpen
  \bibfield  {author} {\bibinfo {author} {\bibnamefont {{C. Guerrero}}},
  \bibinfo {author} {\bibnamefont {{D. Cano-Ott}}}, \bibinfo {author}
  {\bibfnamefont {M.}~\bibnamefont {Fern{\'a}ndez-Ord{\'o}{\~n}ez}}, \bibinfo
  {author} {\bibfnamefont {E.}~\bibnamefont {Gonz{\'a}lez-Romero}}, \bibinfo
  {author} {\bibfnamefont {T.}~\bibnamefont {Mart\'{\i}nez}}, \ and\ \bibinfo
  {author} {\bibfnamefont {D.}~\bibnamefont {Villamar\'{\i}n}},\ }\href
  {\doibase 10.1016/j.nima.2008.09.017} {\bibfield  {journal} {\bibinfo
  {journal} {Nucl. Instrum. Methods A}\ }\textbf {\bibinfo {volume} {597}},\
  \bibinfo {pages} {212} (\bibinfo {year} {2008})}\BibitemShut {NoStop}%
\bibitem [{\citenamefont {Brun}\ and\ \citenamefont
  {Rademakers}(1997)}]{Brun_ROOT_1997}%
  \BibitemOpen
  \bibfield  {author} {\bibinfo {author} {\bibfnamefont {R.}~\bibnamefont
  {Brun}}\ and\ \bibinfo {author} {\bibfnamefont {F.}~\bibnamefont
  {Rademakers}},\ }\href {\doibase 10.1016/S0168-9002(97)00048-X} {\bibfield
  {journal} {\bibinfo  {journal} {Nucl. Instrum. Meth. A}\ }\textbf {\bibinfo
  {volume} {389}},\ \bibinfo {pages} {81} (\bibinfo {year} {1997})}\BibitemShut
  {NoStop}%
\bibitem [{\citenamefont {Chechev}(2006)}]{Chechev_Cm244HalfLife_2006}%
  \BibitemOpen
  \bibfield  {author} {\bibinfo {author} {\bibfnamefont {V.~P.}\ \bibnamefont
  {Chechev}},\ }\href {\doibase 10.1134/S1063778806070155} {\bibfield
  {journal} {\bibinfo  {journal} {Physics of Atomic Nuclei}\ }\textbf {\bibinfo
  {volume} {69}},\ \bibinfo {pages} {1188} (\bibinfo {year}
  {2006})}\BibitemShut {NoStop}%
\bibitem [{\citenamefont {Alcayne}\ \emph {et~al.}(2024)\citenamefont {Alcayne}
  \emph {et~al.}}]{Alcayne_sted_2024}%
  \BibitemOpen
  \bibfield  {author} {\bibinfo {author} {\bibfnamefont {V.}~\bibnamefont
  {Alcayne}} \emph {et~al.} (\bibinfo {collaboration} {The n\_TOF
  Collaboration}),\ }\href {\doibase
  https://doi.org/10.1016/j.radphyschem.2024.111525} {\bibfield  {journal}
  {\bibinfo  {journal} {Radiat. Phys. Chem.}\ }\textbf {\bibinfo {volume}
  {217}},\ \bibinfo {pages} {111525} (\bibinfo {year} {2024})}\BibitemShut
  {NoStop}%
\bibitem [{\citenamefont {Alcayne}(2022)}]{Alcayne_Thesis_2022}%
  \BibitemOpen
  \bibfield  {author} {\bibinfo {author} {\bibfnamefont {V.}~\bibnamefont
  {Alcayne}},\ }\emph {\bibinfo {title} {{Measurement of the Cm-244, Cm-246 and
  Cm-248 neutron-induced capture cross sections at the CERN n TOF facility}}},\
  \href {https://cds.cern.ch/record/2852035?ln=en} {Ph.D. thesis} (\bibinfo
  {year} {2022})\BibitemShut {NoStop}%
\bibitem [{\citenamefont {Knoll}(2000)}]{Knoll_2000}%
  \BibitemOpen
  \bibfield  {author} {\bibinfo {author} {\bibfnamefont {G.}~\bibnamefont
  {Knoll}},\ }\href@noop {} {\emph {\bibinfo {title} {{Radiation Detection and
  Measurement}}}},\ \bibinfo {edition} {3rd}\ ed.\ (\bibinfo  {publisher}
  {Wiley},\ \bibinfo {year} {2000})\BibitemShut {NoStop}%
\bibitem [{JEF(2017)}]{JEFF3.3_2017}%
  \BibitemOpen
  \href@noop {} {\emph {\bibinfo {title} {{Data Bank. JEFF-3.3}}}},\ \bibinfo
  {type} {Tech. Rep.}\ (\bibinfo  {institution} {OECD/NEA},\ \bibinfo {year}
  {2017})\BibitemShut {NoStop}%
\bibitem [{\citenamefont {Iwamoto}\ \emph {et~al.}(2023)\citenamefont {Iwamoto}
  \emph {et~al.}}]{JENDL5}%
  \BibitemOpen
  \bibfield  {author} {\bibinfo {author} {\bibfnamefont {O.}~\bibnamefont
  {Iwamoto}} \emph {et~al.},\ }\href {\doibase 10.1080/00223131.2022.2141903}
  {\bibfield  {journal} {\bibinfo  {journal} {J. Nucl. Sci. Technol.}\ }\textbf
  {\bibinfo {volume} {60}},\ \bibinfo {pages} {1} (\bibinfo {year}
  {2023})}\BibitemShut {NoStop}%
\bibitem [{\citenamefont {Brown}\ \emph {et~al.}(2018)\citenamefont {Brown}
  \emph {et~al.}}]{Brown_ENDF8_2018}%
  \BibitemOpen
  \bibfield  {author} {\bibinfo {author} {\bibfnamefont {D.}~\bibnamefont
  {Brown}} \emph {et~al.},\ }\href {\doibase 10.1016/j.nds.2018.02.001}
  {\bibfield  {journal} {\bibinfo  {journal} {Nucl. Data Sheets}\ }\textbf
  {\bibinfo {volume} {148}},\ \bibinfo {pages} {1} (\bibinfo {year}
  {2018})}\BibitemShut {NoStop}%
\bibitem [{\citenamefont {Macklin}\ and\ \citenamefont
  {Gibbons}(1967)}]{Mackin_PHWT_1967}%
  \BibitemOpen
  \bibfield  {author} {\bibinfo {author} {\bibfnamefont {R.~L.}\ \bibnamefont
  {Macklin}}\ and\ \bibinfo {author} {\bibfnamefont {J.~H.}\ \bibnamefont
  {Gibbons}},\ }\href {\doibase 10.1103/PhysRev.159.1007} {\bibfield  {journal}
  {\bibinfo  {journal} {Phys. Rev. C}\ }\textbf {\bibinfo {volume} {159}},\
  \bibinfo {pages} {1007} (\bibinfo {year} {1967})}\BibitemShut {NoStop}%
\bibitem [{\citenamefont {Abbondanno}\ \emph {et~al.}(2004)\citenamefont
  {Abbondanno} \emph {et~al.}}]{Tain_PHWT_2004}%
  \BibitemOpen
  \bibfield  {author} {\bibinfo {author} {\bibfnamefont {U.}~\bibnamefont
  {Abbondanno}} \emph {et~al.} (\bibinfo {collaboration} {The n\_TOF
  Collaboration}),\ }\href {\doibase 10.1016/j.nima.2003.09.066} {\bibfield
  {journal} {\bibinfo  {journal} {Nucl. Instrum. Methods A}\ }\textbf {\bibinfo
  {volume} {521}},\ \bibinfo {pages} {454} (\bibinfo {year}
  {2004})}\BibitemShut {NoStop}%
\bibitem [{\citenamefont {{E. Mendoza}}\ \emph {et~al.}(2023)\citenamefont {{E.
  Mendoza}} \emph {et~al.}}]{Mendoza_PHWT_23}%
  \BibitemOpen
  \bibfield  {author} {\bibinfo {author} {\bibnamefont {{E. Mendoza}}} \emph
  {et~al.},\ }\href {\doibase 10.1016/j.nima.2022.167894} {\bibfield  {journal}
  {\bibinfo  {journal} {Nucl. Instrum. Methods A}\ }\textbf {\bibinfo {volume}
  {1047}},\ \bibinfo {pages} {167894} (\bibinfo {year} {2023})}\BibitemShut
  {NoStop}%
\bibitem [{\citenamefont {Capote}\ \emph {et~al.}(2009)\citenamefont {Capote}
  \emph {et~al.}}]{Capote_RIPL_2007}%
  \BibitemOpen
  \bibfield  {author} {\bibinfo {author} {\bibfnamefont {R.}~\bibnamefont
  {Capote}} \emph {et~al.},\ }\href {\doibase 10.1016/j.nds.2009.10.004}
  {\bibfield  {journal} {\bibinfo  {journal} {Nucl. Data Sheets}\ }\textbf
  {\bibinfo {volume} {110}},\ \bibinfo {pages} {3107} (\bibinfo {year}
  {2009})}\BibitemShut {NoStop}%
\bibitem [{ENS()}]{ENSDF}%
  \BibitemOpen
  \href {{http://www.nndc.bnl.gov/ensarchivals/}} {\enquote {\bibinfo {title}
  {{From ENSDF database as of 2023.Version available at
  http://www.nndc.bnl.gov/ensarchivals/}},}\ }\BibitemShut {NoStop}%
\bibitem [{\citenamefont {Guerrero}\ \emph {et~al.}(2009)\citenamefont
  {Guerrero} \emph {et~al.}}]{Guerrero_TAC_2009}%
  \BibitemOpen
  \bibfield  {author} {\bibinfo {author} {\bibfnamefont {C.}~\bibnamefont
  {Guerrero}} \emph {et~al.} (\bibinfo {collaboration} {The n\_TOF
  Collaboration}),\ }\href {\doibase 10.1016/j.nima.2009.07.025} {\bibfield
  {journal} {\bibinfo  {journal} {Nucl. Instrum. Methods A}\ }\textbf {\bibinfo
  {volume} {608}},\ \bibinfo {pages} {424} (\bibinfo {year}
  {2009})}\BibitemShut {NoStop}%
\bibitem [{\citenamefont {Mendoza}\ \emph {et~al.}(2020)\citenamefont {Mendoza}
  \emph {et~al.}}]{Mendoza_PSF_2020}%
  \BibitemOpen
  \bibfield  {author} {\bibinfo {author} {\bibfnamefont {E.}~\bibnamefont
  {Mendoza}} \emph {et~al.} (\bibinfo {collaboration} {The n\_TOF
  Collaboration}),\ }\href {\doibase 10.1051/epjconf/202023901015} {\bibfield
  {journal} {\bibinfo  {journal} {EPJ Web Conf.}\ }\textbf {\bibinfo {volume}
  {239}},\ \bibinfo {pages} {01015} (\bibinfo {year} {2020})}\BibitemShut
  {NoStop}%
\bibitem [{\citenamefont {Lerendegui-Marco}(2018)}]{Lerendegui_Thesis_2018}%
  \BibitemOpen
  \bibfield  {author} {\bibinfo {author} {\bibfnamefont {J.}~\bibnamefont
  {Lerendegui-Marco}},\ }\emph {\bibinfo {title} {{Neutron radiative capture on
  Pu-242: addressing the target accuracies for innovative nuclear systems}}},\
  \href {https://cds.cern.ch/record/2661485?ln=es} {Ph.D. thesis} (\bibinfo
  {year} {2018})\BibitemShut {NoStop}%
\bibitem [{\citenamefont {Macklin}\ \emph {et~al.}(1979)\citenamefont
  {Macklin}, \citenamefont {Halperin},\ and\ \citenamefont
  {Winters}}]{Macklin_Saturated_1979}%
  \BibitemOpen
  \bibfield  {author} {\bibinfo {author} {\bibfnamefont {R.}~\bibnamefont
  {Macklin}}, \bibinfo {author} {\bibfnamefont {J.}~\bibnamefont {Halperin}}, \
  and\ \bibinfo {author} {\bibfnamefont {R.}~\bibnamefont {Winters}},\ }\href
  {\doibase 10.1016/0029-554X(79)90457-9} {\bibfield  {journal} {\bibinfo
  {journal} {Nucl. Instrum. Methods}\ }\textbf {\bibinfo {volume} {164}},\
  \bibinfo {pages} {213} (\bibinfo {year} {1979})}\BibitemShut {NoStop}%
\bibitem [{\citenamefont {{A. Borella}}\ \emph {et~al.}(2007)\citenamefont {{A.
  Borella}}, \citenamefont {Aerts}, \citenamefont {Gunsing}, \citenamefont
  {Moxon}, \citenamefont {Schillebeeckx},\ and\ \citenamefont
  {Wynants}}]{Borella_C6D6_2007}%
  \BibitemOpen
  \bibfield  {author} {\bibinfo {author} {\bibnamefont {{A. Borella}}},
  \bibinfo {author} {\bibfnamefont {G.}~\bibnamefont {Aerts}}, \bibinfo
  {author} {\bibfnamefont {F.}~\bibnamefont {Gunsing}}, \bibinfo {author}
  {\bibfnamefont {M.}~\bibnamefont {Moxon}}, \bibinfo {author} {\bibfnamefont
  {P.}~\bibnamefont {Schillebeeckx}}, \ and\ \bibinfo {author} {\bibfnamefont
  {R.}~\bibnamefont {Wynants}},\ }\href {\doibase 10.1016/j.nima.2007.03.034}
  {\bibfield  {journal} {\bibinfo  {journal} {Nucl. Instrum. Methods A}\
  }\textbf {\bibinfo {volume} {577}},\ \bibinfo {pages} {626} (\bibinfo {year}
  {2007})}\BibitemShut {NoStop}%
\bibitem [{\citenamefont {Larsson}(2006)}]{Larsson_SAMMY_2006}%
  \BibitemOpen
  \bibfield  {author} {\bibinfo {author} {\bibfnamefont {N.}~\bibnamefont
  {Larsson}},\ }\href@noop {} {\emph {\bibinfo {title} {{Updated User's Guide
  for SAMMY: Multilevel R-matrix Fits to Neutron Data Using Bayes
  Equations}}}},\ \bibinfo {type} {Tech. Rep.}\ (\bibinfo  {institution}
  {ORNL},\ \bibinfo {year} {2006})\BibitemShut {NoStop}%
\bibitem [{\citenamefont {Frohner}(2000)}]{Frohner_Res_2000}%
  \BibitemOpen
  \bibfield  {author} {\bibinfo {author} {\bibfnamefont {F.~H.}\ \bibnamefont
  {Frohner}},\ }\href
  {http://inis.iaea.org/search/search.aspx?orig_q=RN:32011631} {}\ (\bibinfo
  {publisher} {Organisation for Economic Co-Operation and Development},\
  \bibinfo {address} {Nuclear Energy Agency of the OECD (NEA)},\ \bibinfo
  {year} {2000})\ \bibinfo {note} {{Evaluation and analysis of nuclear
  resonance data}}\BibitemShut {NoStop}%
\bibitem [{\citenamefont {Lorusso}\ \emph {et~al.}(2004)\citenamefont {Lorusso}
  \emph {et~al.}}]{Lorusso_RFEAR1_2004}%
  \BibitemOpen
  \bibfield  {author} {\bibinfo {author} {\bibfnamefont {G.}~\bibnamefont
  {Lorusso}} \emph {et~al.} (\bibinfo {collaboration} {The n\_TOF
  Collaboration}),\ }\href {\doibase 10.1016/j.nima.2004.04.247} {\bibfield
  {journal} {\bibinfo  {journal} {Nucl. Instrum. Methods A}\ }\textbf {\bibinfo
  {volume} {532}},\ \bibinfo {pages} {622} (\bibinfo {year}
  {2004})}\BibitemShut {NoStop}%
\bibitem [{\citenamefont {Vlachoudis}\ \emph {et~al.}(2021)\citenamefont
  {Vlachoudis} \emph {et~al.}}]{Vlachoudis_RFEAR2_2021}%
  \BibitemOpen
  \bibfield  {author} {\bibinfo {author} {\bibfnamefont {V.}~\bibnamefont
  {Vlachoudis}} \emph {et~al.},\ }\href {https://cds.cern.ch/record/2764434}
  {\emph {\bibinfo {title} {{On the resolution function of the n\_TOF facility:
  a comprehensive study and user guide}}}},\ \bibinfo {type} {Tech. Rep.}\
  (\bibinfo {year} {2021})\BibitemShut {NoStop}%
\bibitem [{\citenamefont {Ahdida}\ \emph {et~al.}(2022)\citenamefont {Ahdida}
  \emph {et~al.}}]{Ahdida_22}%
  \BibitemOpen
  \bibfield  {author} {\bibinfo {author} {\bibfnamefont {C.}~\bibnamefont
  {Ahdida}} \emph {et~al.},\ }\href {\doibase 10.3389/fphy.2021.788253}
  {\bibfield  {journal} {\bibinfo  {journal} {Frontiers in Physics}\ ,\
  \bibinfo {pages} {9}} (\bibinfo {year} {2022})}\BibitemShut {NoStop}%
\bibitem [{\citenamefont {Maslov}\ \emph {et~al.}(1996)\citenamefont {Maslov}
  \emph {et~al.}}]{Maslov_EvalCm246_1996}%
  \BibitemOpen
  \bibfield  {author} {\bibinfo {author} {\bibfnamefont {V.}~\bibnamefont
  {Maslov}} \emph {et~al.},\ }\href
  {https://inis.iaea.org/collection/NCLCollectionStore/_Public/27/078/27078190.pdf}
  {\bibfield  {journal} {\bibinfo  {journal} {INDC(BLR)-004/L}\ } (\bibinfo
  {year} {1996})}\BibitemShut {NoStop}%
\bibitem [{\citenamefont {{R. E Cot{\'e}}}\ \emph {et~al.}(1964)\citenamefont
  {{R. E Cot{\'e}}}, \citenamefont {Barnes},\ and\ \citenamefont
  {Diamond}}]{Cote_Cm244_1964}%
  \BibitemOpen
  \bibfield  {author} {\bibinfo {author} {\bibnamefont {{R. E Cot{\'e}}}},
  \bibinfo {author} {\bibfnamefont {R.~F.}\ \bibnamefont {Barnes}}, \ and\
  \bibinfo {author} {\bibfnamefont {H.}~\bibnamefont {Diamond}},\ }\href
  {\doibase 10.1103/PhysRev.134.B1281} {\bibfield  {journal} {\bibinfo
  {journal} {Phys. Rev.}\ }\textbf {\bibinfo {volume} {134}},\ \bibinfo {pages}
  {B1281} (\bibinfo {year} {1964})}\BibitemShut {NoStop}%
\bibitem [{\citenamefont {{J. R. Berreth}}\ \emph {et~al.}(1972)\citenamefont
  {{J. R. Berreth}}, \citenamefont {Simpson},\ and\ \citenamefont
  {Rusche}}]{Berreth_Cm244_1972}%
  \BibitemOpen
  \bibfield  {author} {\bibinfo {author} {\bibnamefont {{J. R. Berreth}}},
  \bibinfo {author} {\bibfnamefont {F.}~\bibnamefont {Simpson}}, \ and\
  \bibinfo {author} {\bibfnamefont {B.~C.}\ \bibnamefont {Rusche}},\ }\href
  {\doibase 10.13182/NSE72-A35502} {\bibfield  {journal} {\bibinfo  {journal}
  {Nucl. Sci. Eng.}\ }\textbf {\bibinfo {volume} {49}},\ \bibinfo {pages} {145}
  (\bibinfo {year} {1972})}\BibitemShut {NoStop}%
\bibitem [{\citenamefont {{R. W. Benjamin}}\ \emph {et~al.}(1974)\citenamefont
  {{R. W. Benjamin}}, \citenamefont {Ahlfeld}, \citenamefont {Harvey},\ and\
  \citenamefont {Hill}}]{Benjamin_Cm248_1974}%
  \BibitemOpen
  \bibfield  {author} {\bibinfo {author} {\bibnamefont {{R. W. Benjamin}}},
  \bibinfo {author} {\bibfnamefont {C.}~\bibnamefont {Ahlfeld}}, \bibinfo
  {author} {\bibfnamefont {J.}~\bibnamefont {Harvey}}, \ and\ \bibinfo {author}
  {\bibfnamefont {N.}~\bibnamefont {Hill}},\ }\href@noop {} {\bibfield
  {journal} {\bibinfo  {journal} {Nucl. Sci. Eng.}\ }\textbf {\bibinfo {volume}
  {55}},\ \bibinfo {pages} {440} (\bibinfo {year} {1974})}\BibitemShut
  {NoStop}%
\bibitem [{\citenamefont {Belanova}\ \emph {et~al.}(1975)\citenamefont
  {Belanova} \emph {et~al.}}]{Belanova_Cm_1975}%
  \BibitemOpen
  \bibfield  {author} {\bibinfo {author} {\bibfnamefont {T.~S.}\ \bibnamefont
  {Belanova}} \emph {et~al.},\ }\href {\doibase 10.1007/BF01126376} {\bibfield
  {journal} {\bibinfo  {journal} {Soviet Atomic Energy}\ }\textbf {\bibinfo
  {volume} {39}},\ \bibinfo {pages} {1020} (\bibinfo {year}
  {1975})}\BibitemShut {NoStop}%
\bibitem [{\citenamefont {Kin}\ \emph {et~al.}(2009)\citenamefont {Kin} \emph
  {et~al.}}]{Tin_ANNRI_2009}%
  \BibitemOpen
  \bibfield  {author} {\bibinfo {author} {\bibfnamefont {T.}~\bibnamefont
  {Kin}} \emph {et~al.},\ }in\ \href {\doibase 10.1109/NSSMIC.2009.5402387}
  {\emph {\bibinfo {booktitle} {{IEEE Nucl. Sci. Symp. Conf. Rec. }}}}\
  (\bibinfo {year} {2009})\ pp.\ \bibinfo {pages} {1194--1197}\BibitemShut
  {NoStop}%
\bibitem [{\citenamefont {Kikuchi}\ and\ \citenamefont
  {Nakagawa}(1984)}]{Kikuchi_EvalCm248_1984}%
  \BibitemOpen
  \bibfield  {author} {\bibinfo {author} {\bibfnamefont {Y.}~\bibnamefont
  {Kikuchi}}\ and\ \bibinfo {author} {\bibfnamefont {T.}~\bibnamefont
  {Nakagawa}},\ }\href
  {https://www-nds.iaea.org/publications/indc/indc-jpn-0087} {\emph {\bibinfo
  {title} {{Evaluation of neutron data for 248Cm and 249Cm }}}},\ \bibinfo
  {type} {Tech. Rep.}\ (\bibinfo {address} {Japan},\ \bibinfo {year} {1984})\
  \bibinfo {note} {jAERI-M--84-116}\BibitemShut {NoStop}%
\end{thebibliography}%

\end{document}